\documentclass[twocolumn]{aastex63}

\usepackage{graphicx}
\usepackage{natbib}
\usepackage{verbatim}
\usepackage{color}

\usepackage{txfonts}
\usepackage{booktabs}   

\usepackage{lineno}            

\received{27 February 2021}
\revised{30 September 2021}
\accepted{19 October 2021}     
\submitjournal{ApJ}

\shorttitle{Connection between radio and $\gamma$-ray activity in 0716$+$714}
\shortauthors{Kim et al.}
\graphicspath{{./}{figures/}}

\begin{document}

\title{Radio and $\gamma$-ray activity in the jet of the blazar S5 0716$+$714}


\author[0000-0003-4997-2153]{Dae-Won Kim}
\affiliation{Department of Physics and Astronomy, Seoul National University, Gwanak-gu, Seoul 08826, Korea; \url{dwkimastro@gmail.com}}
\affiliation{Max-Planck-Institut f\"{u}r Radioastronomie, Auf dem H\"{u}gel 69, D-53121, Bonn, Germany}

\author[0000-0003-4540-4095]{Evgeniya V. Kravchenko} 
\affiliation{Moscow Institute of Physics and Technology, Institutsky per. 9, Moscow region, Dolgoprudny, 141700, Russia}
\affiliation{Astro Space Center, Lebedev Physical Institute, Russian Academy of Sciences, Profsouznaya st., 84/32, Moscow, 117997, Russia}
\affiliation{INAF Istituto di Radioastronomia, Via P. Gobetti, 101, Bologna, 40129, Italy}

\author[0000-0002-1123-7498]{Alexander M. Kutkin} 
\affiliation{ASTRON, Netherlands Institute for Radio Astronomy, Oude Hoogeveensedijk 4, 7991PD, Dwingeloo, The Netherlands}
\affiliation{Astro Space Center, Lebedev Physical Institute, Russian Academy of Sciences, Profsouznaya st., 84/32, Moscow, 117997, Russia}

\author[0000-0002-8434-5692]{Markus B\"{o}ttcher} 
\affiliation{Centre for Space Research, North-West University, Potchefstroom, 2520, South Africa}


\author[0000-0003-4190-7613]{Jos\'e L. G\'omez} 
\affiliation{Instituto de Astrof\'isica de Andaluc\'ia, CSIC, Glorieta de la Astronom\'ia s/n, Granada, 18008, Spain}

\author[0000-0003-0685-3621]{Mark Gurwell} 
\affiliation{Center for Astrophysics $\vert$ Harvard $\&$ Smithsonian, 60 Garden Street, Cambridge, MA 02138, USA}

\author[0000-0001-6158-1708]{Svetlana G. Jorstad} 
\affiliation{Institute for Astrophysical Research, Boston University, 725 Commonwealth Avenue, Boston, MA 02215, USA}
\affiliation{Astronomical Institute, St. Petersburg State University, Universitetskij Pr. 28, Petrodvorets, 198504 St. Petersburg, Russia}

\author[0000-0002-0393-0647]{Anne L\"{a}hteenm\"{a}ki} 
\affiliation{Aalto University Mets\"{a}hovi Radio Observatory, Mets\"{a}hovintie 114, FI-02540 Kylm\"{a}l\"{a}, Finland}
\affiliation{Aalto University Department of Electronic and Nanoengineering, PL15500, FI-00076 Aalto, Finland} 

\author[0000-0001-7396-3332]{Alan P. Marscher} 
\affiliation{Institute for Astrophysical Research, Boston University, 725 Commonwealth Avenue, Boston, MA 02215, USA}

\author[0000-0002-9248-086X]{Venkatessh Ramakrishnan} 
\affiliation{Astronomy Department, Universidad de Concepci\'{o}n, Casilla 160-C, 4030000 Concepci\'{o}n, Chile}

\author[0000-0003-1249-6026]{Merja Tornikoski} 
\affiliation{Aalto University Mets\"{a}hovi Radio Observatory, Mets\"{a}hovintie 114, FI-02540 Kylm\"{a}l\"{a}, Finland}

\author[0000-0003-0465-1559]{Sascha Trippe}
\affiliation{Department of Physics and Astronomy, Seoul National University, Gwanak-gu, Seoul 08826, Korea; \url{dwkimastro@gmail.com}}
\affiliation{SNU Astronomy Research Center, Seoul National University, 1 Gwanak-ro, Gwanak-gu, Seoul 08826, Korea}

\author[0000-0001-6314-0690]{Zachary Weaver} 
\affiliation{Institute for Astrophysical Research, Boston University, 725 Commonwealth Avenue, Boston, MA 02215, USA}

\author[0000-0003-1318-8535]{Karen E. Williamson} 
\affiliation{Institute for Astrophysical Research, Boston University, 725 Commonwealth Avenue, Boston, MA 02215, USA}



\begin{abstract}
We explore the connection between the $\gamma$-ray and radio emission in the jet of the blazar 0716$+$714 by using 15, 37, and 230\,GHz radio and 0.1--200\,GeV $\gamma$-ray light curves spanning 10.5\,years (2008--2019). We find significant positive and negative correlations between radio and $\gamma$-ray fluxes in different time ranges. The time delays between radio and $\gamma$-ray emission suggest that the observed $\gamma$-ray flares originated from multiple regions upstream of the radio core, within a few parsecs from the central engine. Using time-resolved 43\,GHz VLBA maps we identified 14 jet components moving downstream along the jet. Their apparent speeds range from 6 to 26\,$c$, showing notable variations in their position angles upstream the stationary component ($\sim$0.53\,mas from the core). The brightness temperature declines as function of distance from the core according to a power-law which becomes shallower at the location of the stationary component. We also find that the periods at which significant correlations between radio and $\gamma$-ray emission occur overlap with the times when the jet was oriented to the north. Our results indicate that the passage of a propagating disturbance (or shock) through the radio core and the orientation of the jet might be responsible for the observed correlation between the radio and $\gamma$-ray variability. We present a scenario that connects the positive correlation and the unusual anti-correlation by combining the production of a flare and a dip at $\gamma$-rays by a strong moving shock at different distances from the jet apex.
\end{abstract}


\section{Introduction} \label{sec:int}

Blazars \citep{1980ARA&A..18..321A} are a class of active galactic nuclei (AGN) that show extreme variability on various timescales (e.g. \citealt{2013A&A...558A..92B}).
It is widely believed that they are powered by accretion onto supermassive black holes, accompanied by the formation of two-sided relativistic outflows or jets.
Due to their preferential orientation close to the line of sight ($\lesssim5^{\circ}$, \citealt{2009A&A...494..527H, 2017ApJ...846...98J, 2018ApJ...866..137L}), blazar jet components exhibit superluminal speeds up to a few tens of the speed of light \citep[e.g.,][]{2018ApJ...866..137L}, as well as significant enhancement of their emission due to Doppler boosting.
The vast majority of sources detected by the Large Area Telescope (LAT) on board the \textit{Fermi Gamma-ray Space Telescope} in the GeV band are blazars \citep{2010ApJ...715..429A}.
Detection of a significant correlation between the radio flux density and $\gamma$-ray flux \citep{2009ApJ...696L..17K, 2010ApJ...722L...7P, 2014MNRAS.441.1899F} indicates a common emission mechanism.
The common view is that the jet radio emission is produced by synchrotron radiation from electrons, while the high energy emission can either be generated through inverse Compton scattering (IC) of seed photons by the same electrons (e.g. \citealt{1994ApJ...421..153S}) or by hadronic processes (e.g. \citealt{1998Sci...279..684M}). 
Most likely, the bulk of the $\gamma$-ray emission is produced within the parsec-scale jet \citep{2014MNRAS.445.1636R, 2016MNRAS.462.2747K, 2017MNRAS.468.4478L, 2018MNRAS.480.2324K} and may be associated with stationary structures \citep[e.g.][]{2019MNRAS.482.2336P, 2020A&A...636A..62K}.

S5~0716$+$714 (hereafter 0716$+$714) is one of the most active BL~Lacertae objects and also shows intra-day variability (e.g. \citealt{2014ApJ...783...83L}).
An early high angular resolution study performed with the Very Long Baseline Interferometry Space Observatory Program in 2000 (VSOP, \citealt{2006AA...452...83B}) located the short term variability of the blazar within 100\,$\mu$as from the radio core (surface of the optical depth $\tau_{\nu}\,\sim\,1$).
Later, on 2015 January 3--4, 0716$+$714 was observed by the \textit{RadioAstron} space VLBI mission \citep{2013ARep...57..153K}. 
These observations, conducted at 22\,GHz \citep{2020ApJ...893...68K}, revealed complex structure within the innermost $100~\mu$as of the jet: a core size of $\leq 8\, \mu$as extended toward the southeast with limits of $<12\times5 \, \mu$as along the major and minor beam axis, respectively, and a change of the jet position angle by about 95\degr{} toward the northeast.
A compact linearly polarized component that might be associated with the stationary feature which is considered to be a recollimation shock, is located at a position of $58~\mu$as downstream from the core \citep{2015A&A...578A.123R, 2017ApJ...846...98J}. 

Observations constrain the 0716$+$714 jet viewing angle to $\leq5{\degr}$ \citep{2005AA...433..815B,2009AA...494..527H,2017ApJ...846...98J}, and a stacked-epoch analysis showed that the intrinsic opening angle of the outflow is about $2\degr$ \citep{2017MNRAS.468.4992P}. 
Alongside the significant curvature in the innermost $100\,\mu$as \citep{2020ApJ...893...68K}, this suggests that the jet has been viewed at an angle smaller than the opening angle itself, meaning the line of sight is located inside the outflow.

Analysis of the correlation between light curves obtained at different wavelengths is crucial to explore the physics of the long-term flux variations, especially the radiative processes and the location of the emission region \citep[e.g.,][]{2018MNRAS.480.5517L}. Previous studies of 0716$+$714 revealed a tight correlation between optical and $\gamma$-ray fluxes \citep{2013ApJ...768...40L, 2013A&A...552A..11R, 2016MNRAS.456..171R}. Variability of the emission in 0716$+$714 at these frequencies usually occurs contemporaneously; this supports leptonic emission models \citep{1997A&A...320...19M, 2000ApJ...536..729L, 2002ApJ...581..127B}. The connection becomes much more complicated when comparing $\gamma$-ray and radio emission. On the one hand, it has been reported that the $\gamma$-ray flux shows no significant correlation with the radio emission \citep{2014MNRAS.445..428M, 2015MNRAS.452.1280R}. On the other hand, \citet{2013A&A...552A..11R, 2014A&A...571L...2R} reported not only significant correlations, but also anti-correlated behavior.
\citet{2018ARep...62..654B} explains these changes from positive to negative correlations with a helical jet model of 0716+714, where emission observed at different wavelengths originates from spatially separate regions and at different fixed distances from the jet base. 
Moreover, different mechanisms might be responsible for the generation of high-energy emission at different locations along the jet. Furthermore, as \citet{2015MNRAS.452.1280R} noted, the rapid variability of 0716$+$714 complicates correlation studies of light curves observed in different frequency bands.
In this paper, we further investigate the radio and $\gamma$-ray emission of 0716$+$714 and to pinpoint their production sites in the jet.

0716$+$714 has no spectroscopic redshift measurement because of its featureless optical continuum and bright optical nucleus.
Here we adopt the value of $z=0.31\pm0.08$, derived from the photometric detection of the blazar host galaxy \citep{2008A&A...487L..29N}, which is compatible with other estimates \citep{2013ApJ...764...57D, 2018A&A...619A..45M}.
Throughout the paper, we assume a flat $\Lambda$CDM cosmology with a matter density parameter $\Omega_\mathrm{m}=0.3$, cosmological constant parameter $\Omega_{\Lambda}=0.7$, and Hubble constant $H_0 = 70$~km\,s$^{-1}$\,Mpc$^{-1}$, \citep{2014ApJ...794..135B,2016AA...594A..13P}.
This corresponds to a luminosity distance $D_\mathrm{L}$ of 1.6~Gpc and a scale of 4.56~pc/mas at a redshift of 0.31. 

\section{Observations} \label{sec:obs}

\begin{figure*}
\centering
\includegraphics[angle=0, height=12cm, keepaspectratio]{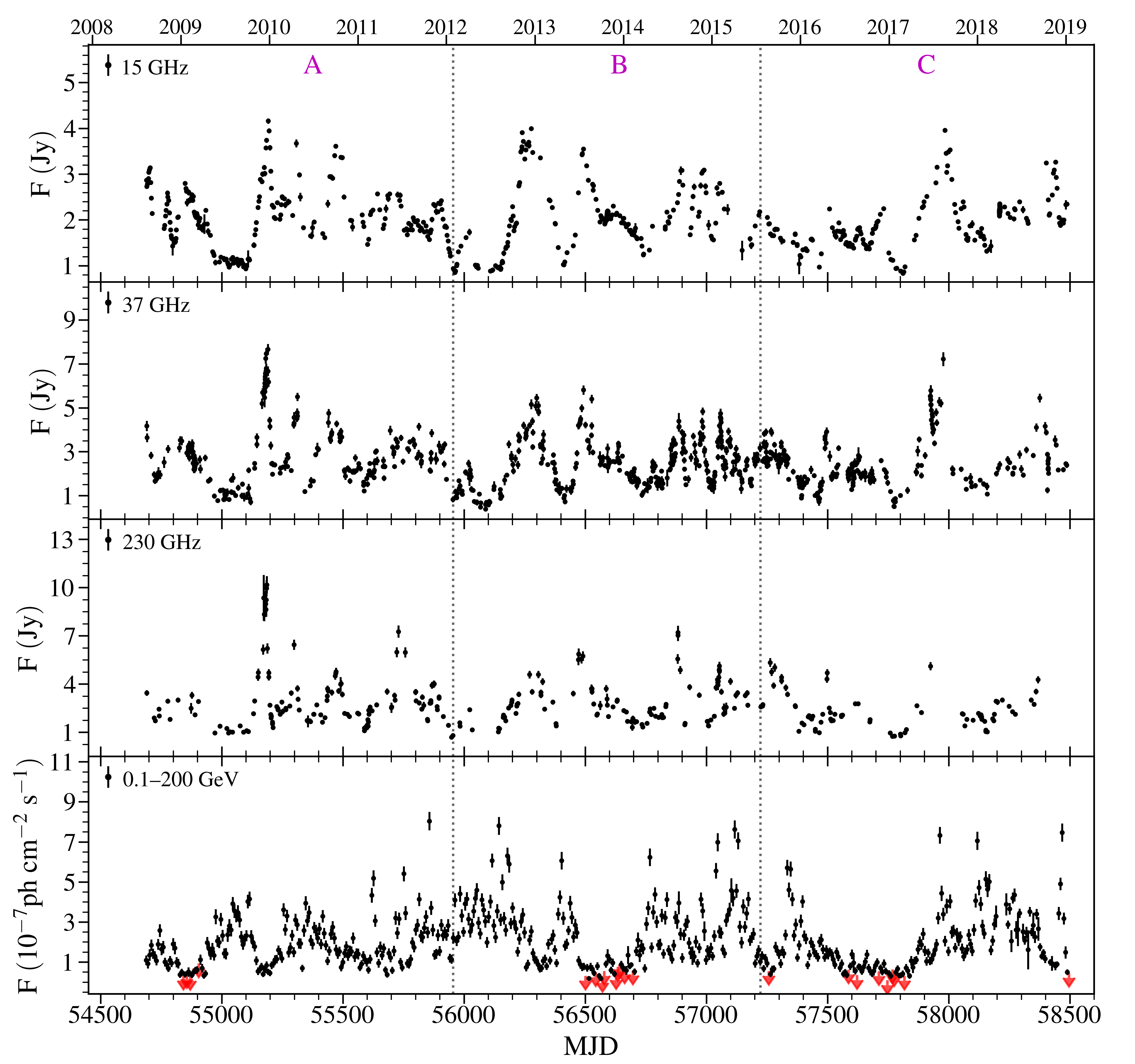}
\caption{From top to bottom: light curves of the blazar 0716+714 at radio (15, 37 and 230\,GHz) and $\gamma$-ray (0.1--200\,GeV) bands. The $\gamma$-ray light curve was binned to one value per week, the red arrows indicate 3$\sigma$ upper limits. The vertical dashed lines distinguish three 3.5-year long periods ($A$, $B$, and $C$), thus dividing the whole time range into three equal segments (see Section~\ref{sec:cor1}). Observing times are given in years along the top and in MJD along the bottom horizontal axis.
}
\label{fig:lightcurve0}
\end{figure*}

\subsection{15, 37, and 43 GHz data}

15\,GHz and 37\,GHz continuum single-dish monitoring data were obtained from the Owens Valley Radio Observatory\footnote{\url{https://sites.astro.caltech.edu/ovroblazars/data.php}} (OVRO) and Mets\"{a}hovi Radio Observatory\footnote{\url{http://www.metsahovi.fi/AGN/data/}} operated by Aalto University in Finland, respectively. 0716$+$714 is one of the target sources in their AGN monitoring programs. A detailed description of the observations and data calibration can be found in \citet{richards_etal11} for OVRO data and \citet{1998AAS..132..305T} for Mets\"{a}hovi data.

Interferometric multi-epoch observations were obtained through the VLBA-BU-BLAZAR (BU) monitoring program\footnote{\url{https://www.bu.edu/blazars/index.html}} at 43\,GHz. This program observes $\sim$33 $\gamma$-ray bright blazars monthly with the Very Long Baseline Array (VLBA). We analyzed 118 BU-epochs in total, which span about 10.5\,years (from June 2008 to December 2018). Details of the data calibration are given in \citep{2017ApJ...846...98J}.

\subsection{SMA 230 GHz (1.3 mm)}

The 230\,GHz (1.3\,mm) flux density data were obtained at the Submillimeter Array (SMA) on Mauna Kea (Hawaii).
0716$+$714 is included in an ongoing monitoring program at the SMA which monitors the fluxes of compact extragalactic radio sources that can be used as calibrators at mm wavelengths \citep{2007ASPC..375..234G}.
Potential calibrators are occasionally observed for 3 to 5 minutes, and the measured source signal strength is calibrated against known standards, typically solar system objects (Titan, Uranus, Neptune, or Callisto).  Data from this program are updated regularly and are available at the SMA website\footnote{\url{http://sma1.sma.hawaii.edu/callist/callist.html}}.

\subsection{\texorpdfstring{$\gamma$}{}-ray flux}

We obtained $\gamma$-ray fluxes in the energy range 0.1--200 GeV from observations by the \textit{Fermi}-LAT, by compiling data from 2008 through the beginning of 2019. 
The analysis was done using the \textit{Fermi} {\tt Science Tools} software package\footnote{\url{http://fermi.gsfc.nasa.gov/ssc/}} version v10r0p5 and Pass 8 data, including the instrument response functions gll\_iem\_v07 and the iso\_P8R3\_SOURCE\_V2\_v1 diffuse source models. Photons were selected within a 15$^{\circ}$ radius centered on the source, using the {\tt gtselect} tool with evclass of 128, evtype of 3 and maximum zenith angle of 90$^{\circ}$.
Sources from the 4FGL catalog \citep{2015ApJS..218...23A} with high test statistic values \citep[TS\,$\geq$\,25,][]{1996ApJ...461..396M} were used to generate the background source model. 
The emission from 0716$+$714 was modelled assuming a log-parabolic photon spectrum 
($dN/dE=N_0(E/E_\mathrm{b})^{-(\alpha+\beta\mathrm{log}(E/E_\mathrm{b}))}$, where $E_\mathrm{b}=428.655$\,MeV, 
$\alpha$=2.01 and $\beta$=0.0375).
We produced a light curve binned into weekly intervals.
For each time bin, the integrated flux values were computed using the maximum-likelihood algorithm implemented in the science tool {\tt gtlike} to estimate the significance of a detection. 
We used a detection criterion which corresponds to a maximum-likelihood test statistic TS $>10$ \citep{2011ApJ...730..101A}. 
If no significant flux was detected, we computed a 3$\sigma$ upper limit. 
More details on the $\gamma$-ray data reduction can be found in \citet{2014ApJ...789..135W}.

\section{Results} \label{sec:res}
\subsection{Radio and \texorpdfstring{$\gamma$}{}-ray light curves} \label{sec:res1}
Figure~\ref{fig:lightcurve0} shows the OVRO (15\,GHz), Mets\"{a}hovi (37\,GHz), SMA (230\,GHz), and LAT (0.1--200\,GeV) light curves of 0716$+$714 from MJD 54687 to MJD 58490 (August 2008 to January 2019). 0716$+$716 is highly variable at all bands, while the average flux density and variability become stronger and faster at shorter wavelengths. The radio light curves were sampled at irregular intervals, the average sampling intervals are $\sim$6, 5, and 12 days at 15, 37, and 230\,GHz, respectively.
From the weekly $\gamma$-ray light curve, two distinct groups of flaring states can be distinguished. First, there are three minor, short-term (on scales of months) flares in an otherwise quiescent period until 2011 (see the top and bottom time axes in Figure~\ref{fig:lightcurve0}). Afterwards, three major, long-term (on scales of years) flares can be seen in 2011--2013, 2014--2016, and 2017--2019, respectively. The radio light curves show frequent flaring: there are roughly 20--30 individual events throughout our observations, with an average duration of a radio flare of about one month.
This is untypical for blazars, which usually show characteristic radio variability time scales of on the order of years \citep[e.g.,][]{2007A&A...469..899H, 2011A&A...533A..97T}. 

\subsection{Correlation analysis} \label{sec:cor}

\subsubsection{Long-term \texorpdfstring{$\gamma$}{}-ray--37 GHz correlation} \label{sec:cor1}

We search for correlations between the observed radio and $\gamma$-ray light curves over the entire observing time range using the discrete cross-correlation function (DCF; \citealt{1988ApJ...333..646E}). 
We computed confidence levels for DCF curves by simulating $\gamma$-ray light curves in the manner of \citet{2013MNRAS.433..907E}. In order to obtain input parameters for the simulation, we calculated a periodogram and measured the power-law slope of the spectral density (powerlaw spectral density, PSD) which follows PSD $\propto$ 1/$\nu^{\beta}$. We also calculated the probability density function (PDF) of the $\gamma$-ray light curve, using log-normal and gamma ($\Gamma$) functions.
To calculate the periodogram of the $\gamma$-ray light curve, we linearly interpolated missing flux points (i.e., upper limits); the difference between the upper limits and the resultant fluxes is on average $\pm 0.1\times10^{-7}$\,ph\,cm$^{-2}$\,s$^{-1}$.
For consistency, this interpolated $\gamma$-ray light curve was used in the simulation and DCF analysis throughout this study. Bin sizes for DCFs were the larger of the two sampling intervals for any pair of time series.

We generated 10,000 artificial $\gamma$-ray light curves spanning time ranges 10 times longer than the length of the input light curve to minimize the impact of red-noise leakage \citep[][]{2002MNRAS.332..231U}. From each artificial light curve we drew a sub-sample having the length of the observed light curve and re-scaled it to the mean and standard deviation of the input data. We sampled the artificial light curves into time bins spanning one week to match the sampling of the observed $\gamma$-ray data. Therefore, we expect that aliasing is negligible \citep{2013MNRAS.433..907E}. 
Since the DCF coefficient values do not always follow Gaussian distributions, we calculated the cumulative distribution function (CDF) of the DCF coefficients to find the 95, 99, and 99.9\% confidence levels \citep{2015MNRAS.453.3455R}; these three levels correspond to 2\,$\sigma$, 2.6\,$\sigma$, and 3.3\,$\sigma$, respectively for a normal distribution. Since the DCF coefficient values frequently deviate from a normal distribution, we used the more conservative significance level of  
3.3\,$\sigma$ ($\sim$99.9\%) instead of 3\,$\sigma$ ($\sim$99.7\%) as the limit for a significant detection.

\begin{figure}
\centering
\includegraphics[angle=0, width=0.45\textwidth, keepaspectratio]{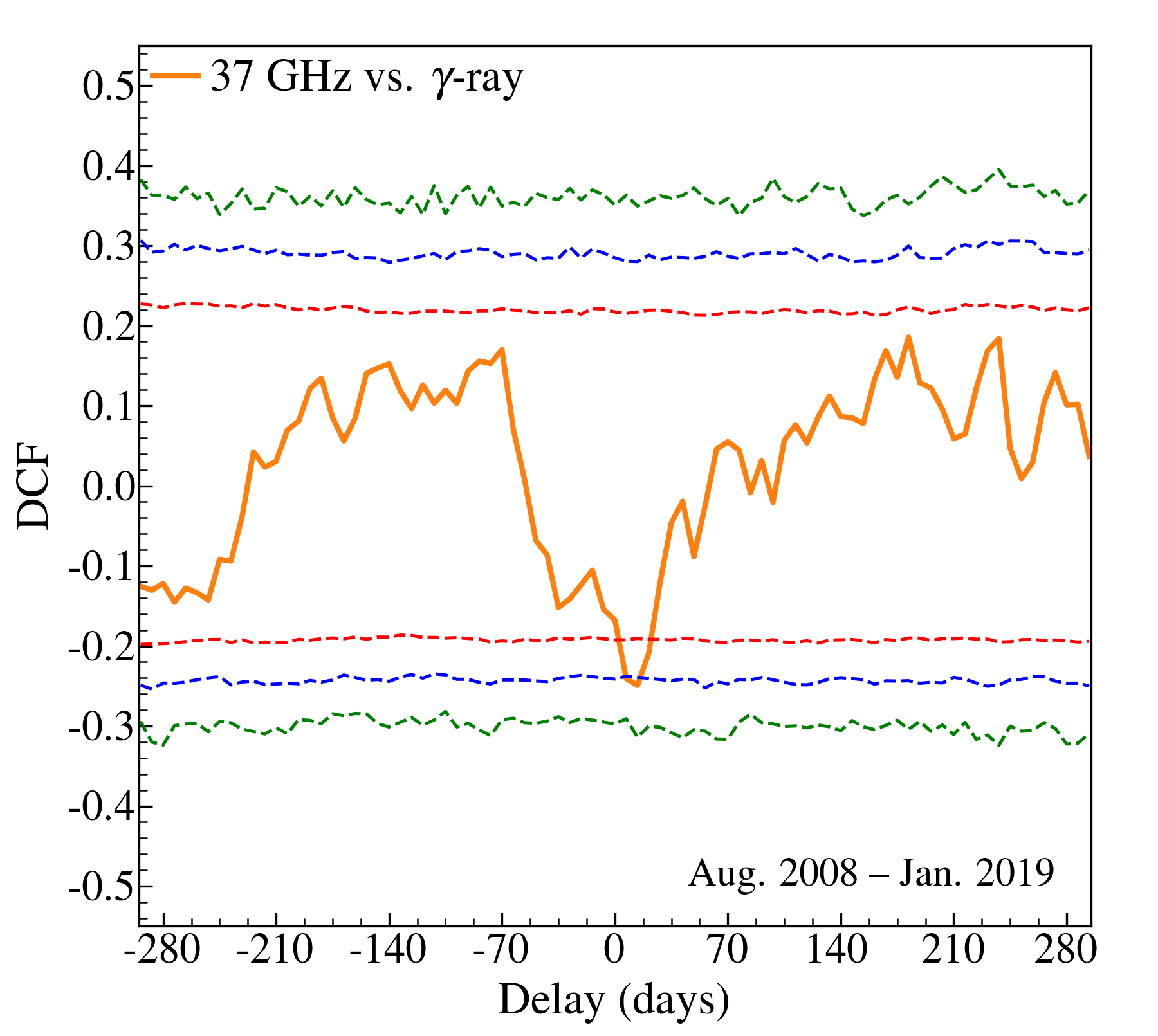}
\caption{Discrete correlation function (DCF) as function of relative time comparing the 37\,GHz and $\gamma$-ray light curves over the full 10.5 year observing time range. The red, blue, and green dashed lines denote the 95, 99, and 99.9 \% confidence levels, respectively.
}
\label{fig:dcf_full}
\end{figure}

\begin{figure*}
\centering
\includegraphics[angle=0,width=0.33\textwidth]{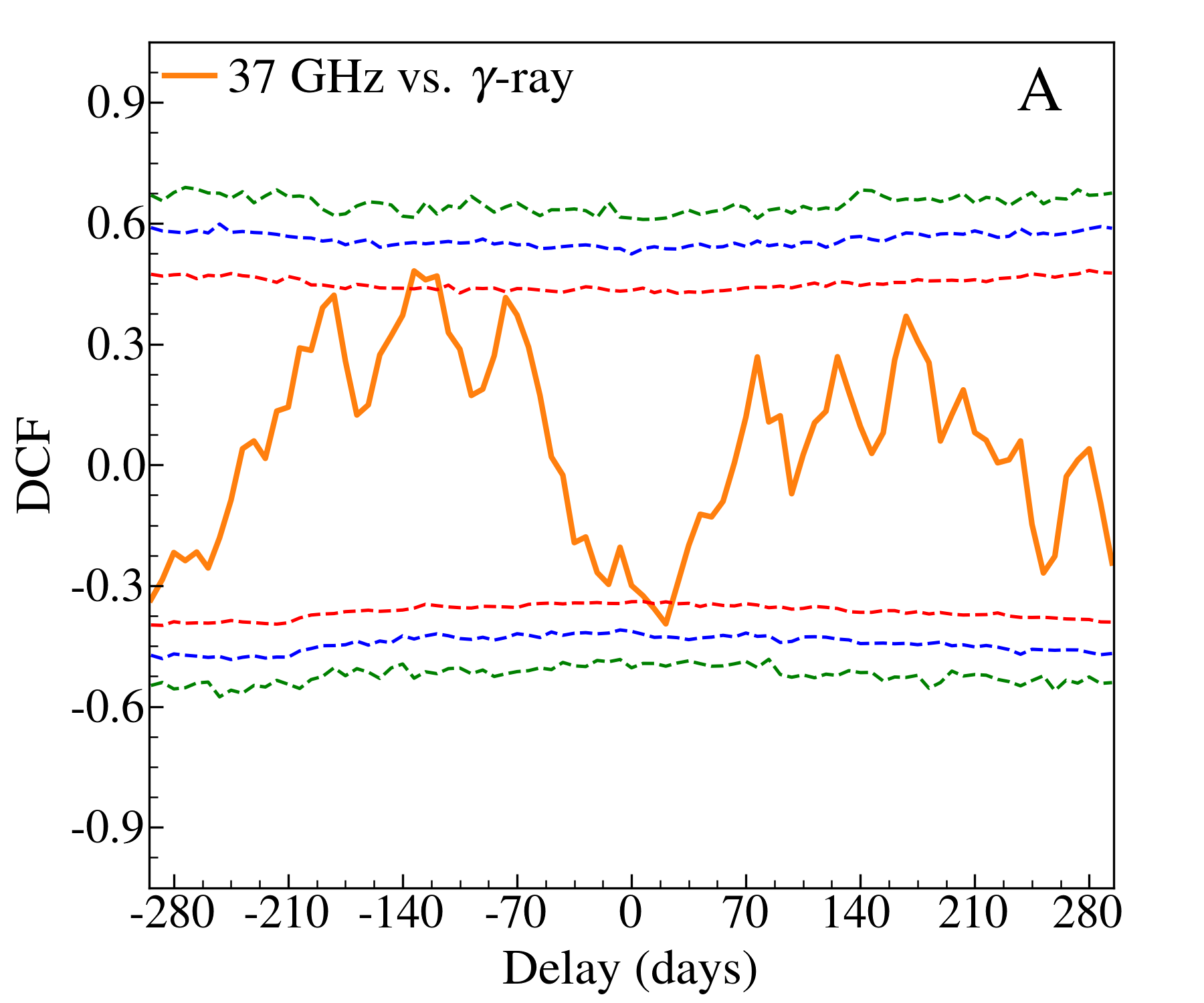}
\includegraphics[angle=0,width=0.33\textwidth]{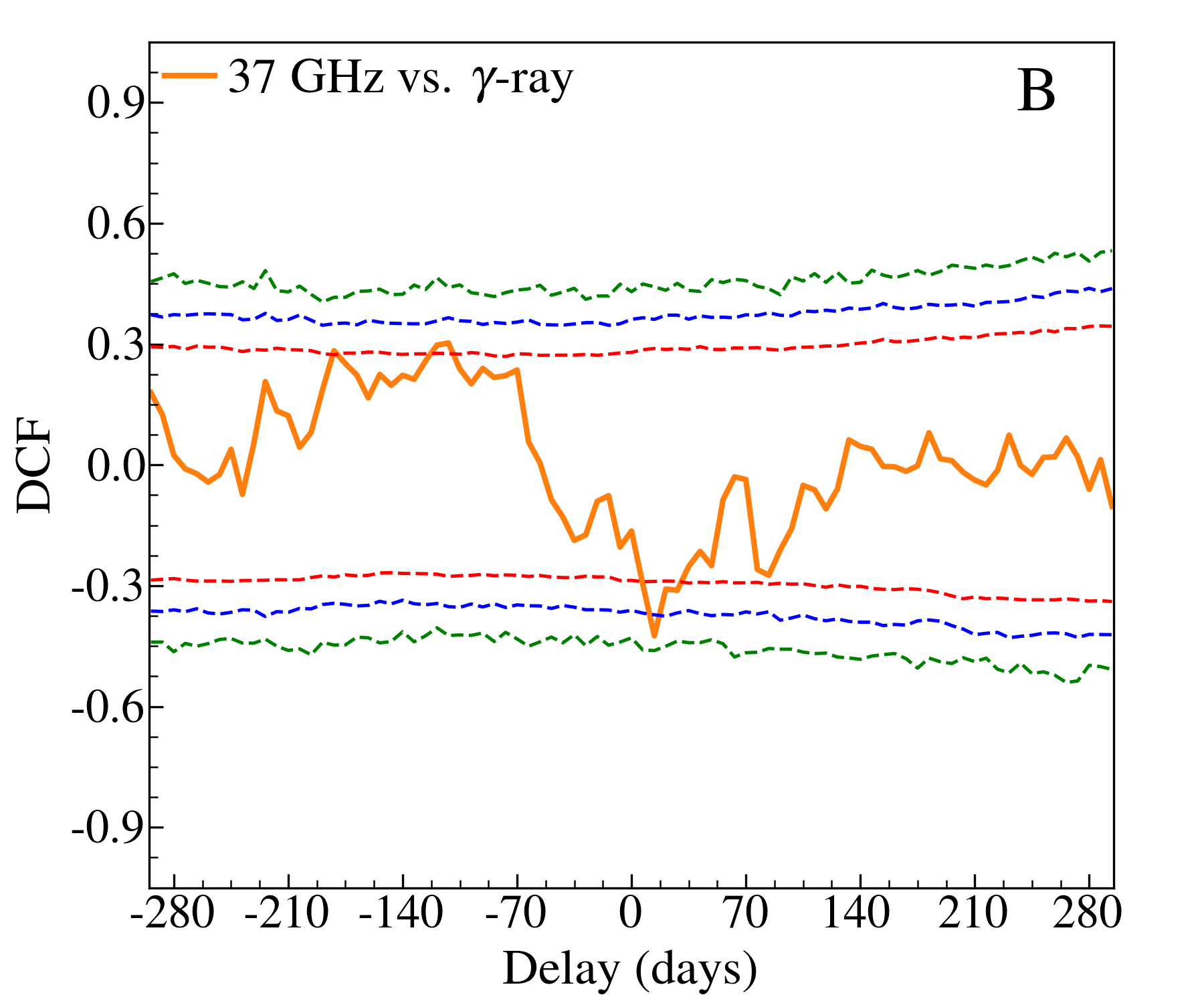}
\includegraphics[angle=0,width=0.33\textwidth]{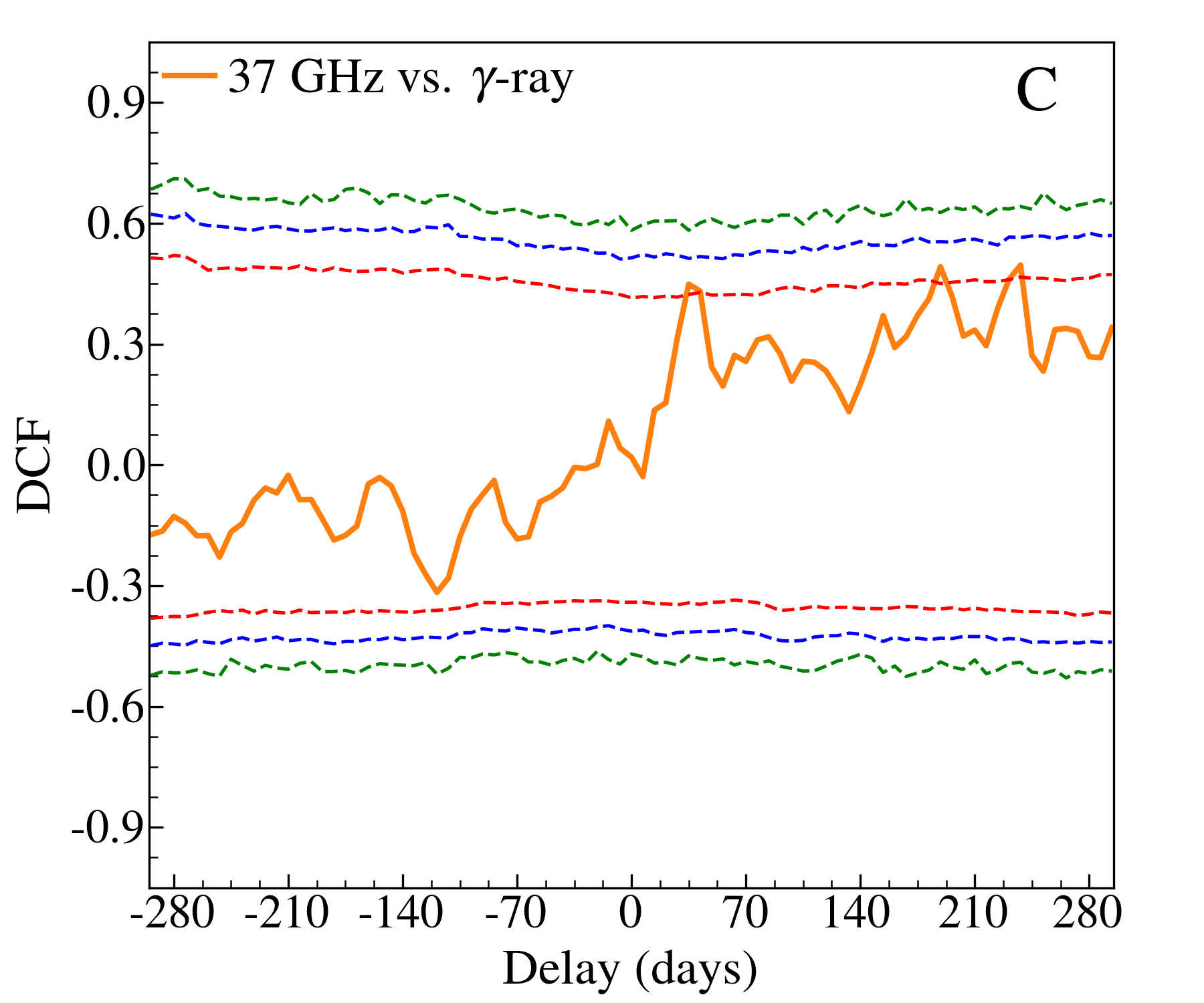}\\
\caption{Discrete correlation function (DCF) as function of time delay comparing 37\,GHz and $\gamma$-ray light curves within three 3.5\,yr periods: $A$ (MJD 54687--55954.7), $B$ (MJD 55954.7--57222.3), and $C$ (MJD 57222.3--58490). The red, blue, and green dashed lines denote the 95, 99, and 99.9 \% confidence levels, respectively.
}
\label{fig:dcf_abc}
\end{figure*}

First, we investigated the overall radio-to-$\gamma$-ray correlation using the 37\,GHz radio data owing to their denser sampling and because of non-zero contribution of interstellar scintillation up to 15\,GHz \citep{2019MNRAS.489.5365K}.
Figure~\ref{fig:dcf_full} shows the resultant DCF over the whole observed 10.5-year time range adopting a seven-day binning. We find a significant negative correlation exceeding the 99\% confidence level. The minimum DCF value of $-$0.25 is located at a delay of around 15 days, with the radio leading the $\gamma$-ray emission.

To explore the origin of this unusual anticorrelation more deeply, we split the whole observing time range into three equal 3.5-year long intervals: $A$ (MJD 54687--55954.7), $B$ (MJD 55954.7--57222.3), and $C$ (MJD 57222.3--58490). The correlation analysis was performed again within each of these time ranges. The results are shown in Figure~\ref{fig:dcf_abc}. In interval $A$, we found two DCF peaks exceeding the 95\% confidence levels. A negative correlation was again found at 21 days with a coefficient of $-$0.39. In addition, there is a positive correlation at about $-$125 days with a coefficient of $\sim$0.48. A similar trend can be seen in region $B$. Here, the negative correlation is more significant than the one in $A$, exceeding the 99\% confidence level. It is located at 14 days with the coefficient being $-$0.43. In addition, positive correlation values exceed the 95\% confidence level at two time delays. Given their values and locations, the one located at $-$112 days with the coefficient of 0.3 appears more probable. In interval $C$, positive correlation values exceed the 95\% confidence level at multiple delays. Given the positive delays (i.e., radio-leading) and the relatively sparse sampling of the radio data in this period, these correlation peaks seem to be artifacts (see Section~\ref{sec:cor2}).

Our correlation analysis suggests the presence of both positive and negative significant correlations between the radio and $\gamma$-ray light curves in each of the $A$ (2008--2012) and $B$ (2012--2016) periods, plus a potential positive correlation in $C$. The correlation coefficients are larger in each of the individual intervals than for the entire observing time. This is consistent with rapid variability, including multiple flaring events, weakening any underlying intrinsic correlation. Another natural explanation is a time dependence of the delays, which washes out the correlation if too long time intervals are analyzed.

\begin{figure*}
\centering
\includegraphics[angle=0,width=0.275\textwidth]{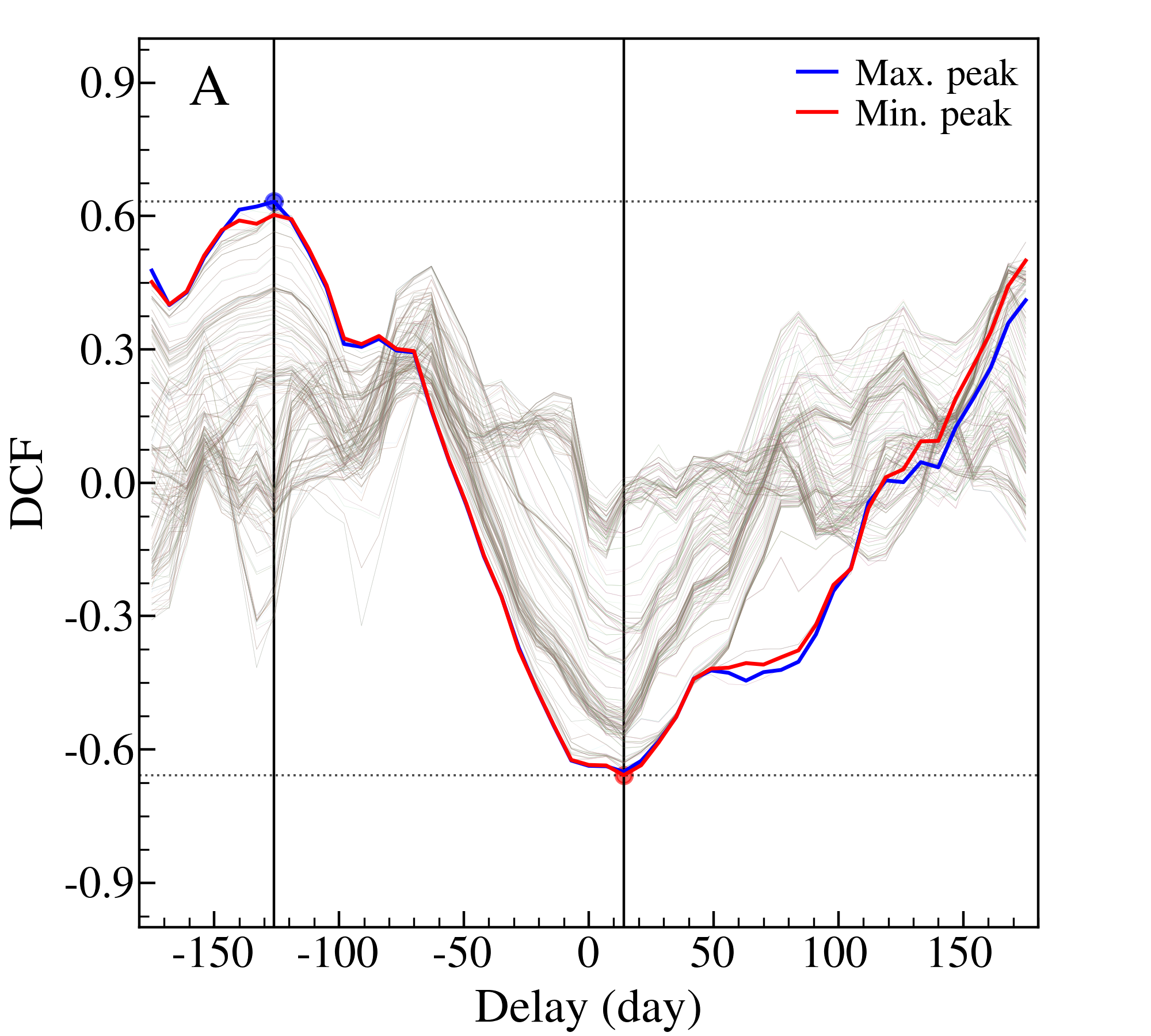}
\includegraphics[angle=0,width=0.275\textwidth]{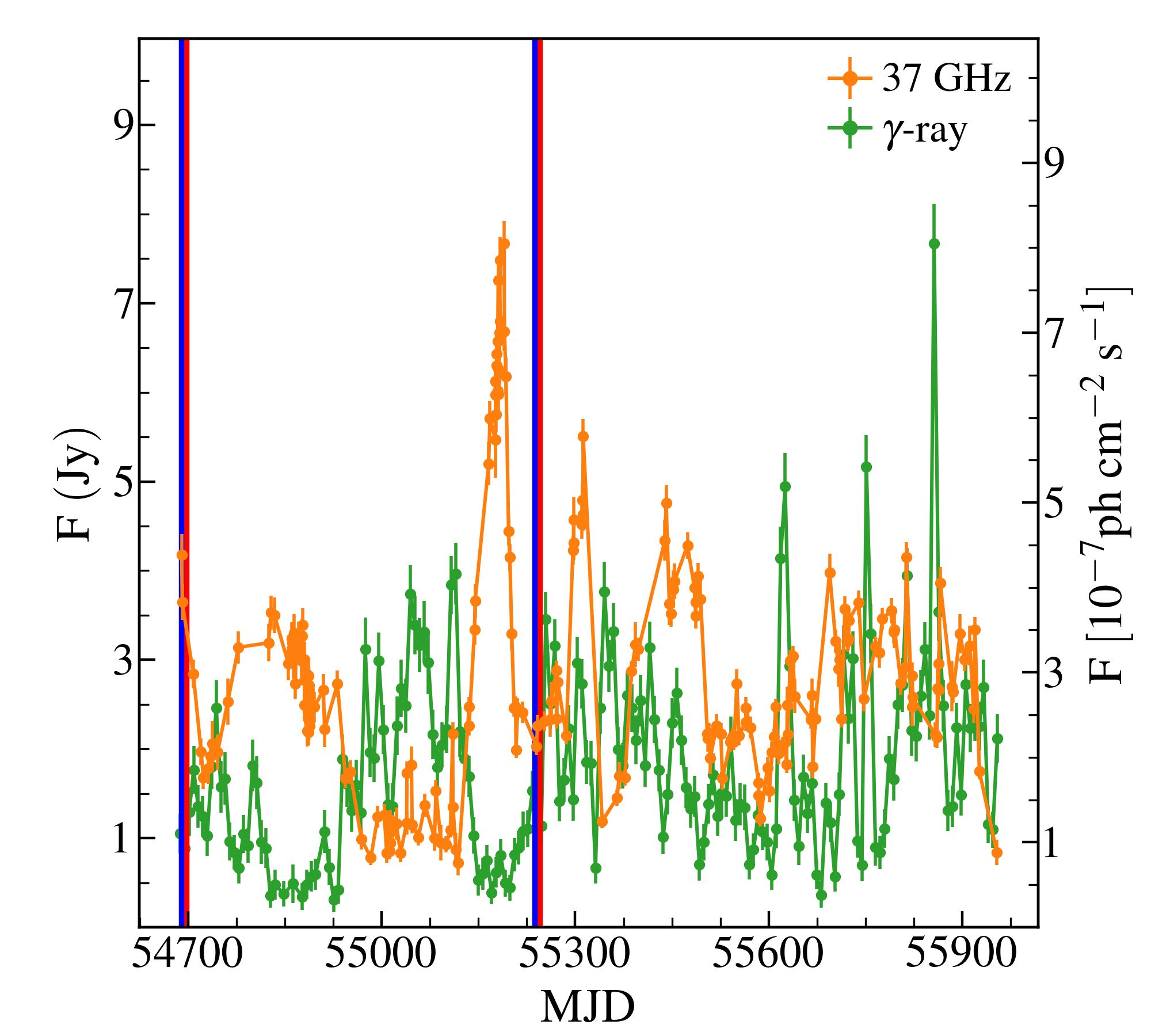}
\includegraphics[angle=0,width=0.275\textwidth]{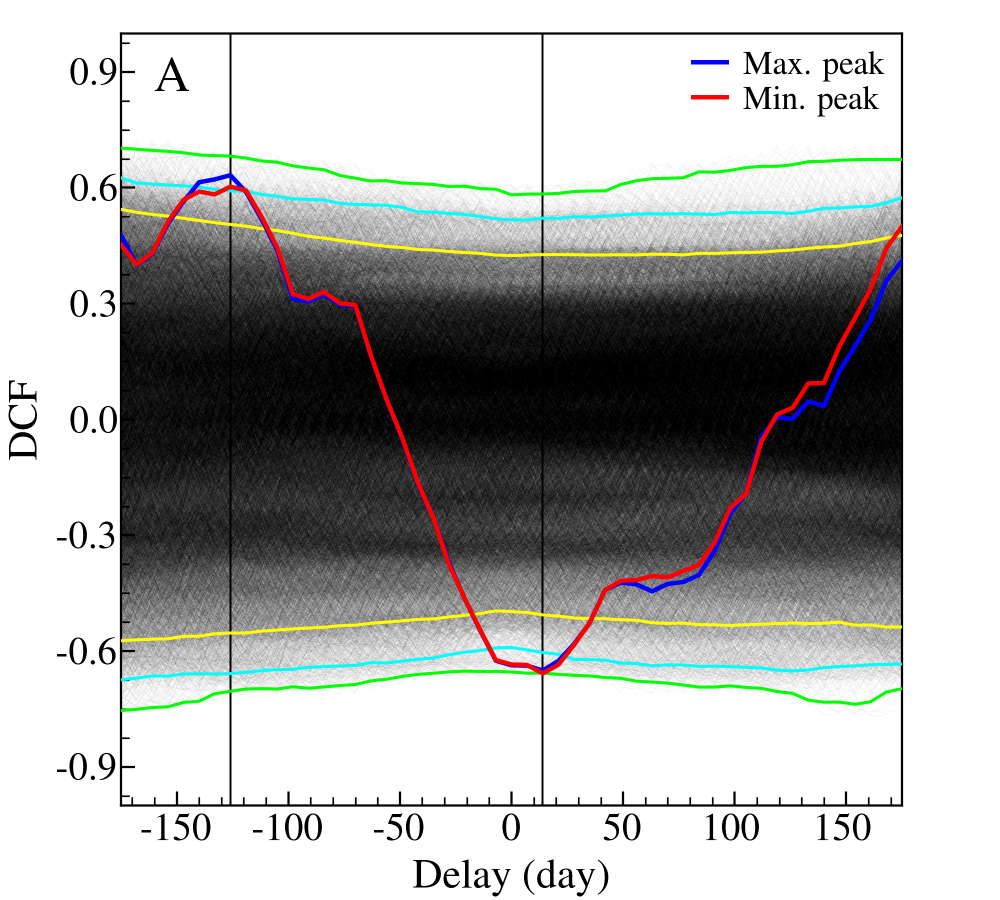}\\
\includegraphics[angle=0,width=0.275\textwidth]{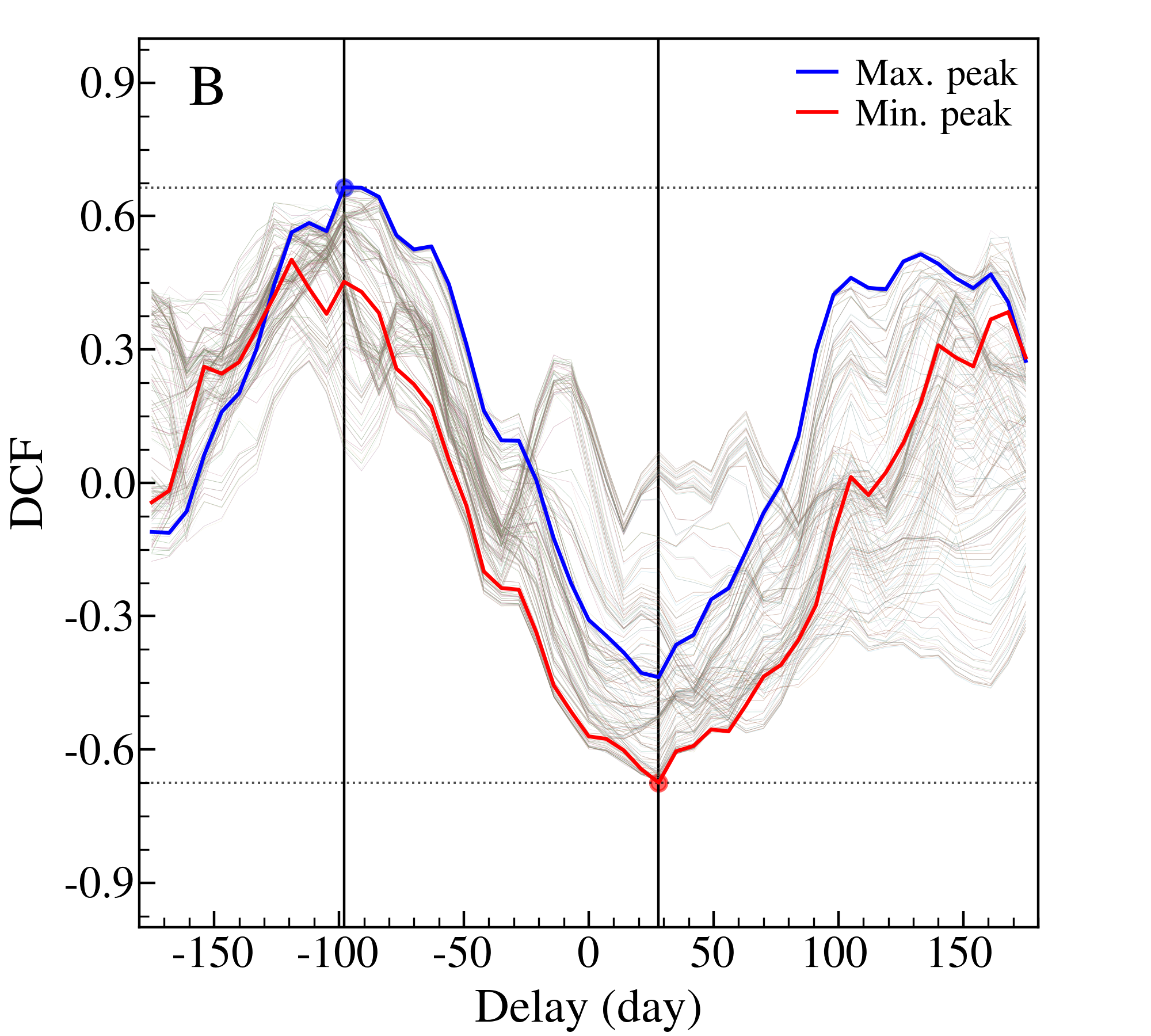}
\includegraphics[angle=0,width=0.275\textwidth]{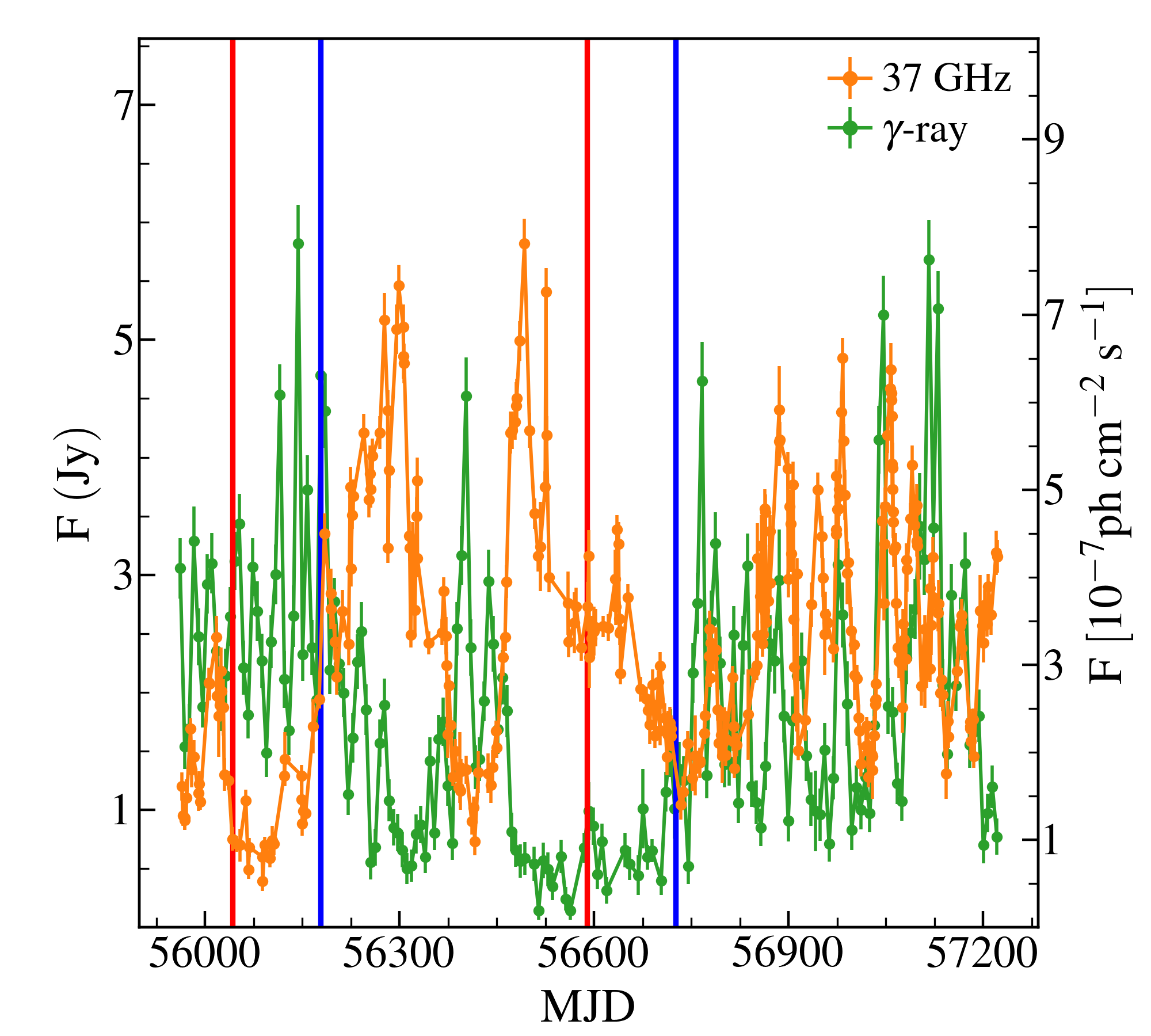}
\includegraphics[angle=0,width=0.275\textwidth]{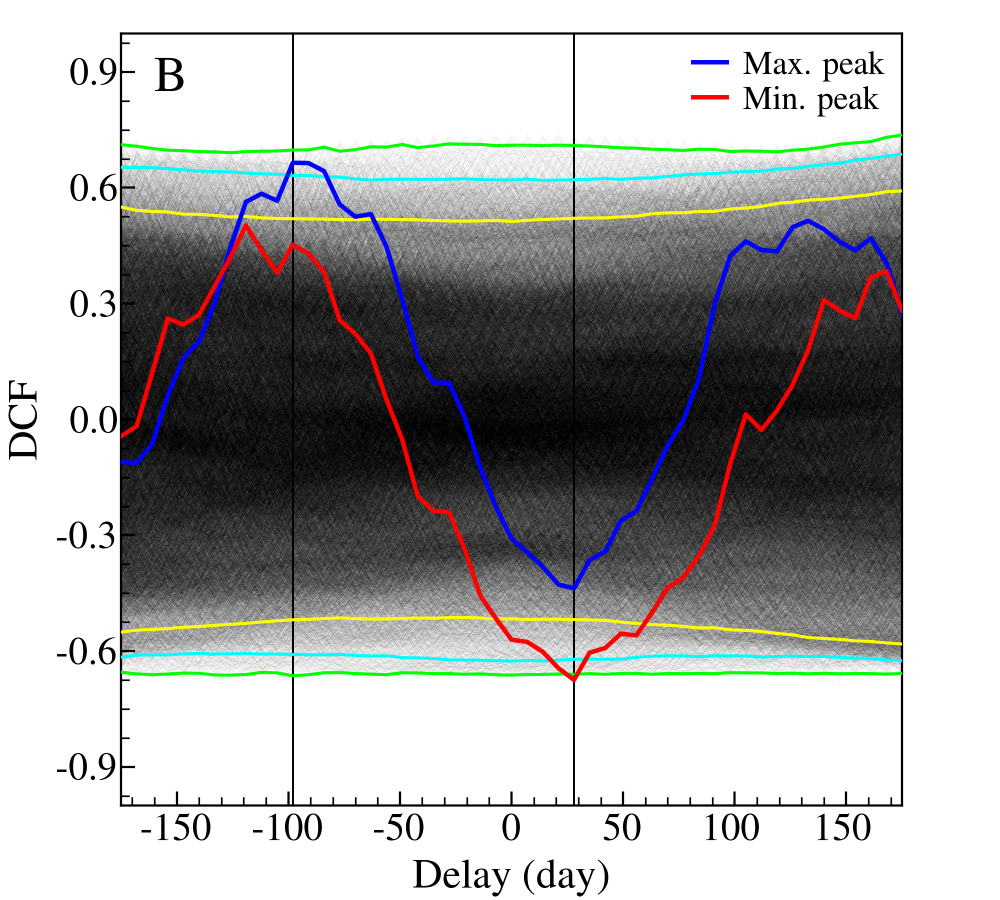}\\
\includegraphics[angle=0,width=0.275\textwidth]{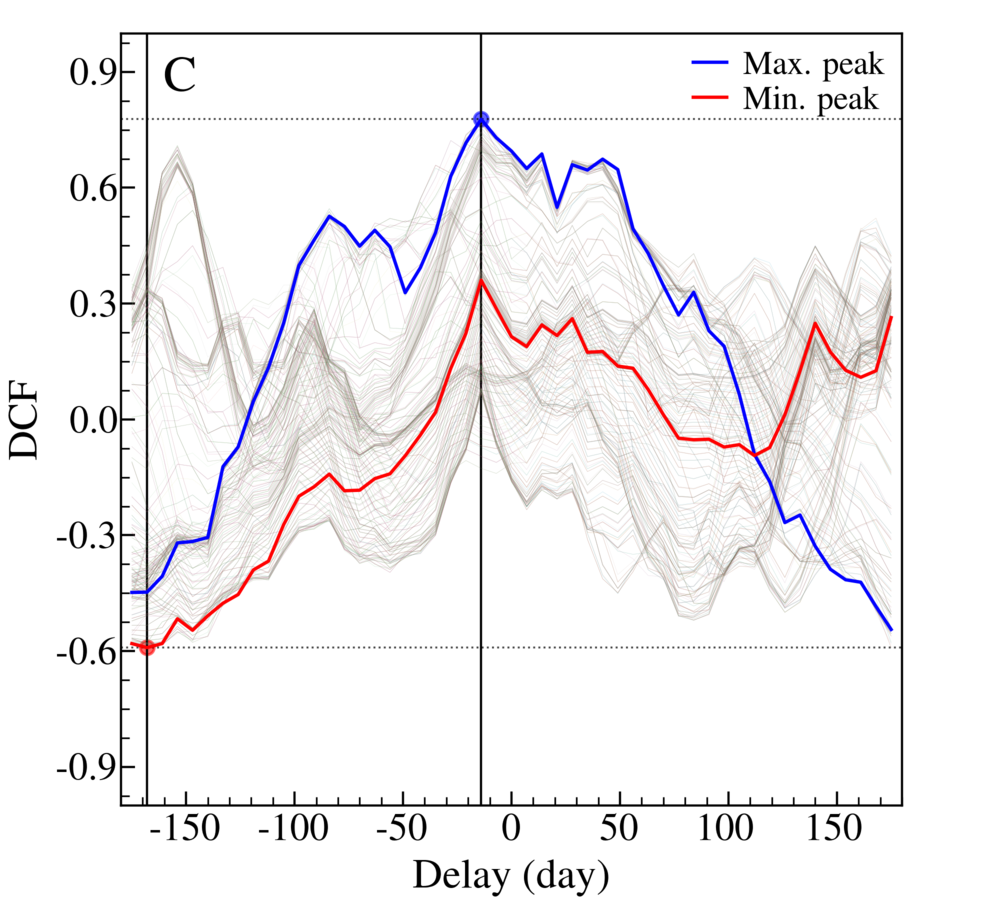}
\includegraphics[angle=0,width=0.275\textwidth]{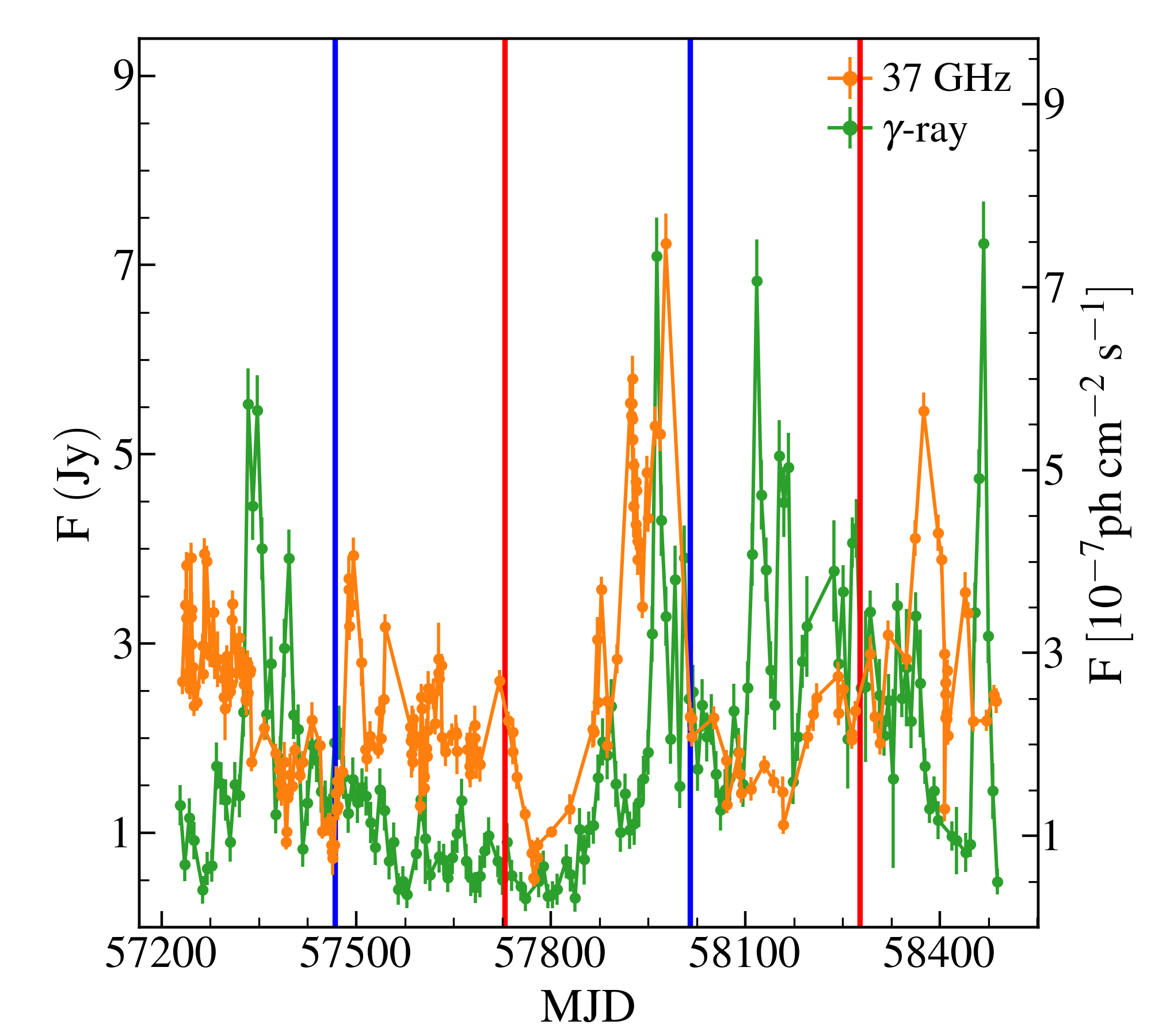}
\includegraphics[angle=0,width=0.275\textwidth]{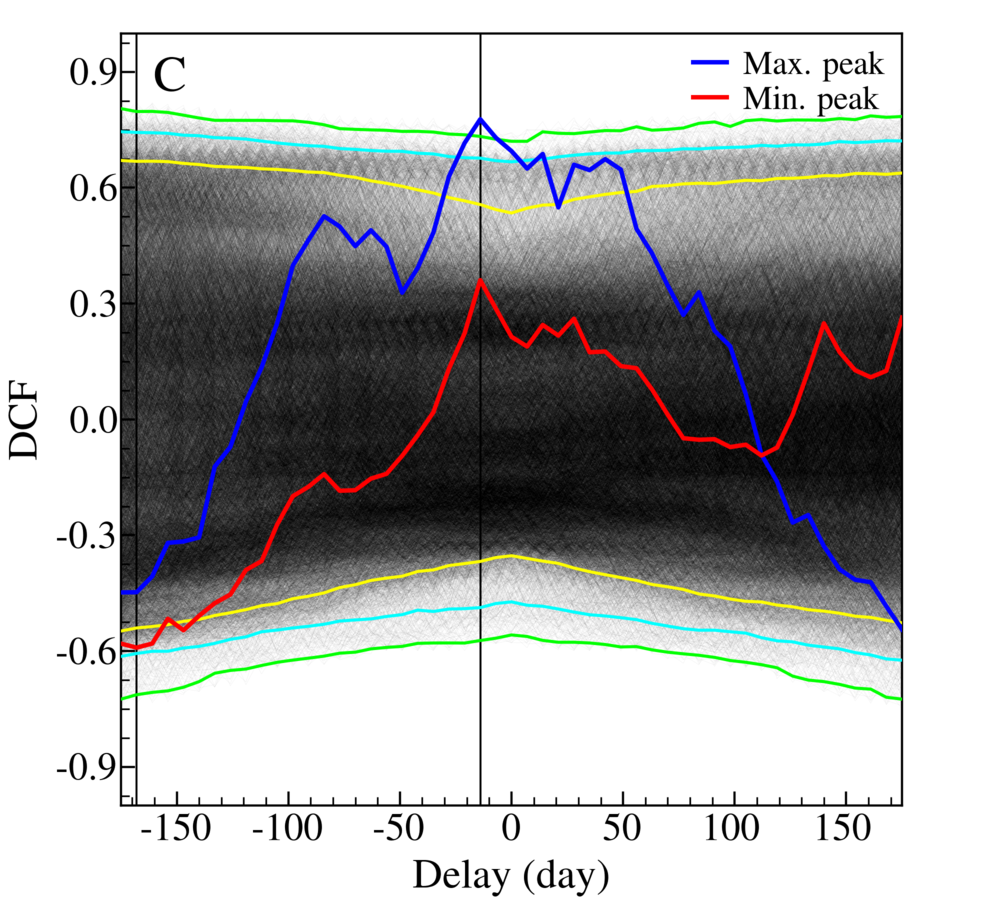}\\
\caption{From left to right: DCF curves derived from 1.5-year windows, observed 37\,GHz and $\gamma$-ray light curves, and the ``extremal'' DCF curves containing the highest (blue curves) and lowest DCF (red curves) values plotted on top of $\sim$15000 DCF curves obtained from random selection. From top to bottom, each row shows the $A$, $B$, and $C$ periods, respectively. The 1.5-year time ranges corresponding to the DCF curves with maximum (``Max.'') and minimum (``Min.'') DCF values are marked by blue and red vertical lines, respectively, in the 3.5\,year light curves. In the third column, the yellow, cyan, and green curves denote the 95, 99, and 99.9\% confidence levels, respectively. The positions of extremal coefficients are denoted by vertical black lines. 
}
\label{fig:searching}
\end{figure*}

\subsubsection{Optimization of time ranges} \label{sec:cor2}
We further explored the main source of the significant radio-to-$\gamma$-ray correlations by narrowing down the time range. During the correlation analysis for the $\gamma$-ray and the unevenly sampled radio light curves, we noted that some of the extended time gaps between radio data points (corresponding to empty bins in an evenly sampled light curve) result in spurious DCF peaks. This effect was particularly severe during the $C$ period for both the 37\,GHz and 230\,GHz data. To reduce this effect, we modeled the observed radio light curves using interpolation and Hanning windowing \citep[e.g.,][]{2014MNRAS.445..437M, 2016MNRAS.456..171R}. The model light curves were then sampled into regular bins with durations corresponding to their observed average sampling intervals. Since the radio and $\gamma$-ray light curves were sampled differently, we employed the DCF, rather than a simpler direct cross-correlation method, for our analysis.

We analyzed the $A$, $B$, and $C$ periods again, using again the 37\,GHz light curve as proxy for radio emission in general. For each 3.5-year period, we defined a sliding window shifting from left to right along the time axis. Due to the frequent and short radio flares, DCF peaks decrease considerably with increasing window size \citep[see also][for the effects of the rapid variability on the correlation]{2015MNRAS.452.1280R}. \citet[][]{2008A&A...488..897H} reported an average AGN flare timescale of one year at 37\,GHz. Considering the 27 individual 37\,GHz flares identified by \citet{2020ApJ...893...68K}, however, the typical flare duration \citep[e.g.,][]{2010ApJ...722..520A} of 0716$+$714 is closer to approximately 4 months \citep[see also][for similar estimates in the source]{2017ApJ...841..119L}. Furthermore, the jet of 0716$+$714 is known to frequently eject new radio jet components (about every half year; \citealt{2017ApJ...846...98J}), which coincide with the numerous radio flares shown in Figure~\ref{fig:lightcurve0}. Hence, we settled on 1.5 years as a reasonable size for the sliding window.

\begin{table}[b]
\caption{The optimum 1.5-year regions found for the $\gamma$-ray--37-GHz correlation analysis}
\label{tab:searching_result}
\centering
\begin{tabular}{l c c}   
\toprule
Time  &  Delay$^{1}$  &  DCF$^{2}$  \\
(MJD)  &  (days)  &    \\
\midrule
$T1$: 54697.5 -- 55245.0  &  14  &  $-$0.66  \\
$T2$: 56042.0 -- 56589.5  &  28  &  $-$0.67  \\
$T3$: 57467.0 -- 58014.5  &  $-$14  &  0.78  \\
\bottomrule
\multicolumn{3}{l}{$^1$ $+$: radio leading, $-$: $\gamma$-ray leading.}\\
\multicolumn{3}{l}{$^2$ DCF coefficient values.}\\
\end{tabular}
\end{table}

While the window passes through the 37\,GHz and $\gamma$-ray light curves in parallel, we generated and stored DCF curves ($\sim$720 in total) at each time step. We then identified two ``extreme'' DCF curves in each of the $A$/$B$/$C$ periods: the one with the highest (positive) and the one with the lowest (most negative) DCF value.
If an extremal (maximum or minimum) coefficient value appears in several consecutive steps, the time of the middle step was used as the time of the strongest correlation. The panels in the first column of Figure~\ref{fig:searching} show the results of this search. The 1.5-year time ranges (labeled T1, T2, and T3 for periods A, B, and C, respectively) of the selected DCF curves are displayed in the second column of Figure~\ref{fig:searching}.

\begin{figure*}[!th]
\centering
\includegraphics[angle=0,width=0.27\textwidth]{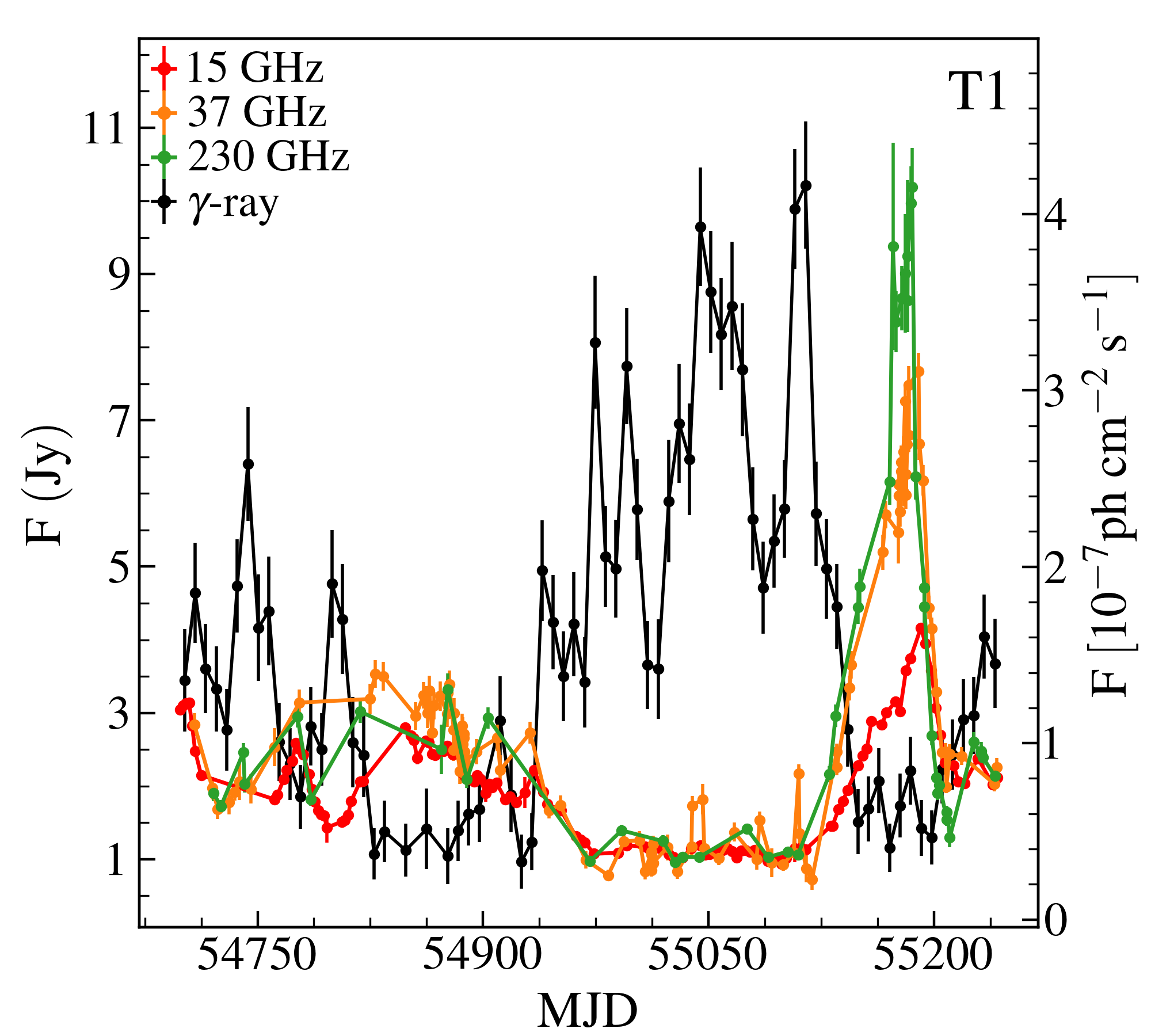}
\includegraphics[angle=0,width=0.27\textwidth]{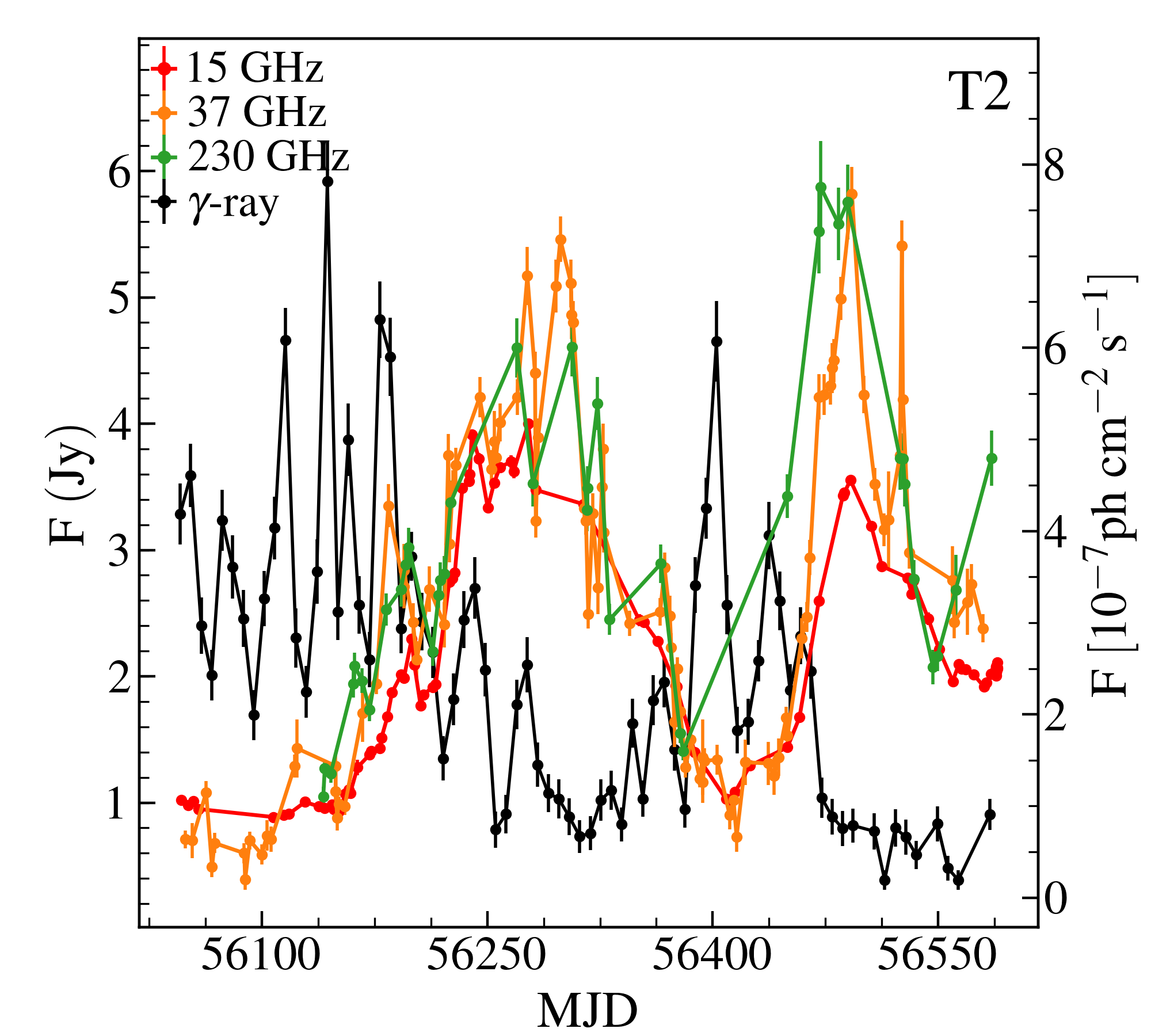}
\includegraphics[angle=0,width=0.27\textwidth]{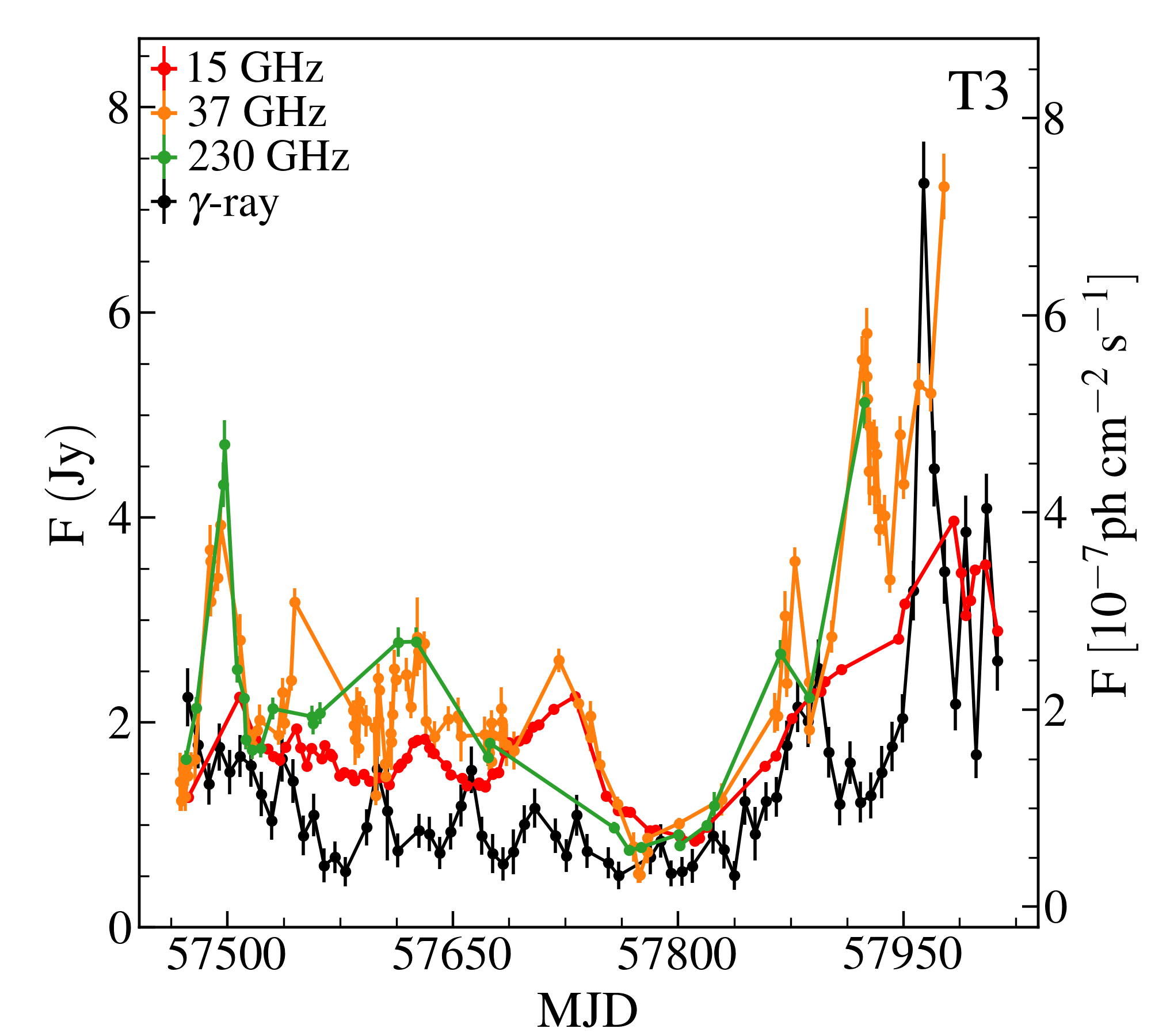}\\
\includegraphics[angle=0,width=0.27\textwidth]{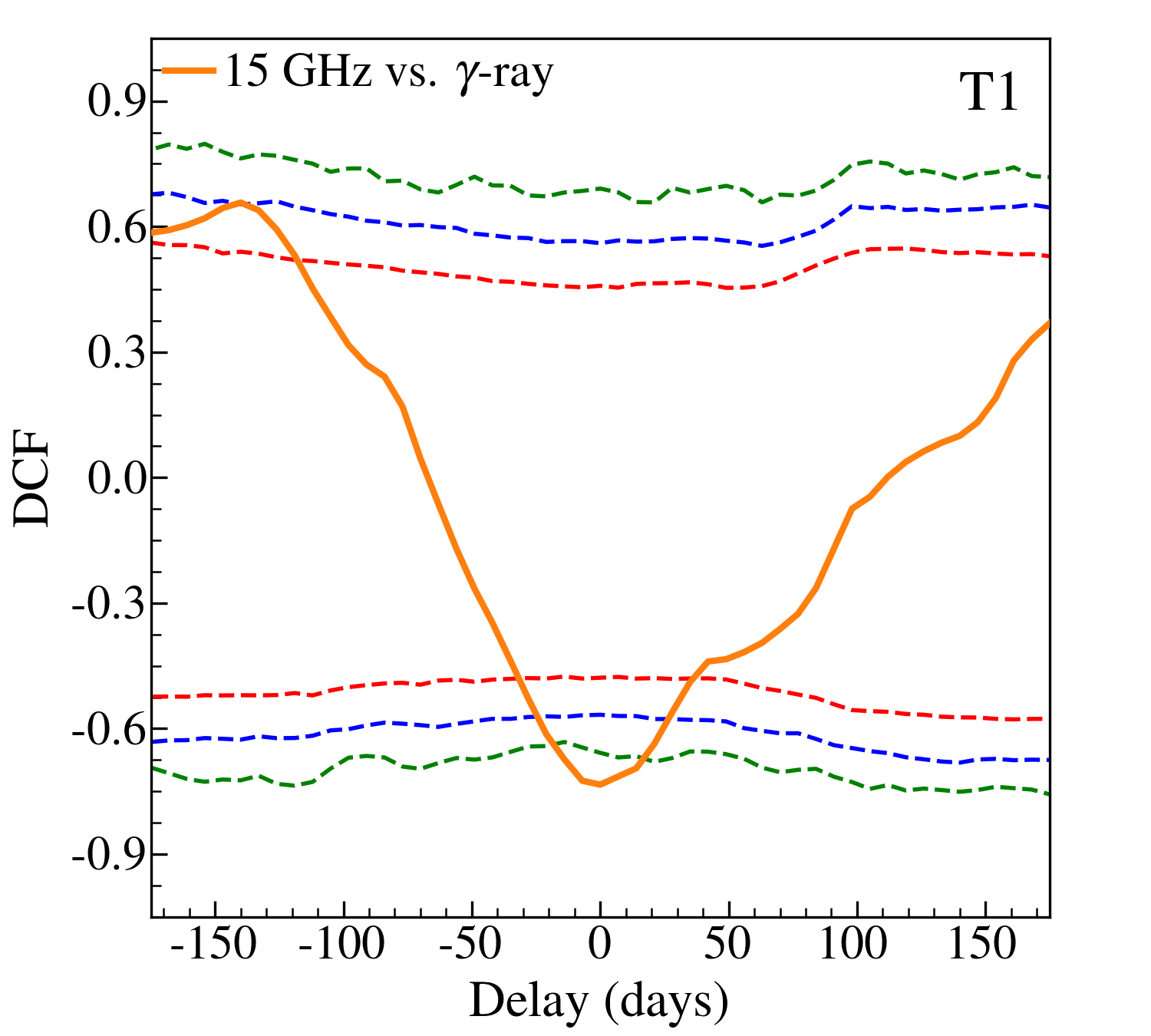}
\includegraphics[angle=0,width=0.27\textwidth]{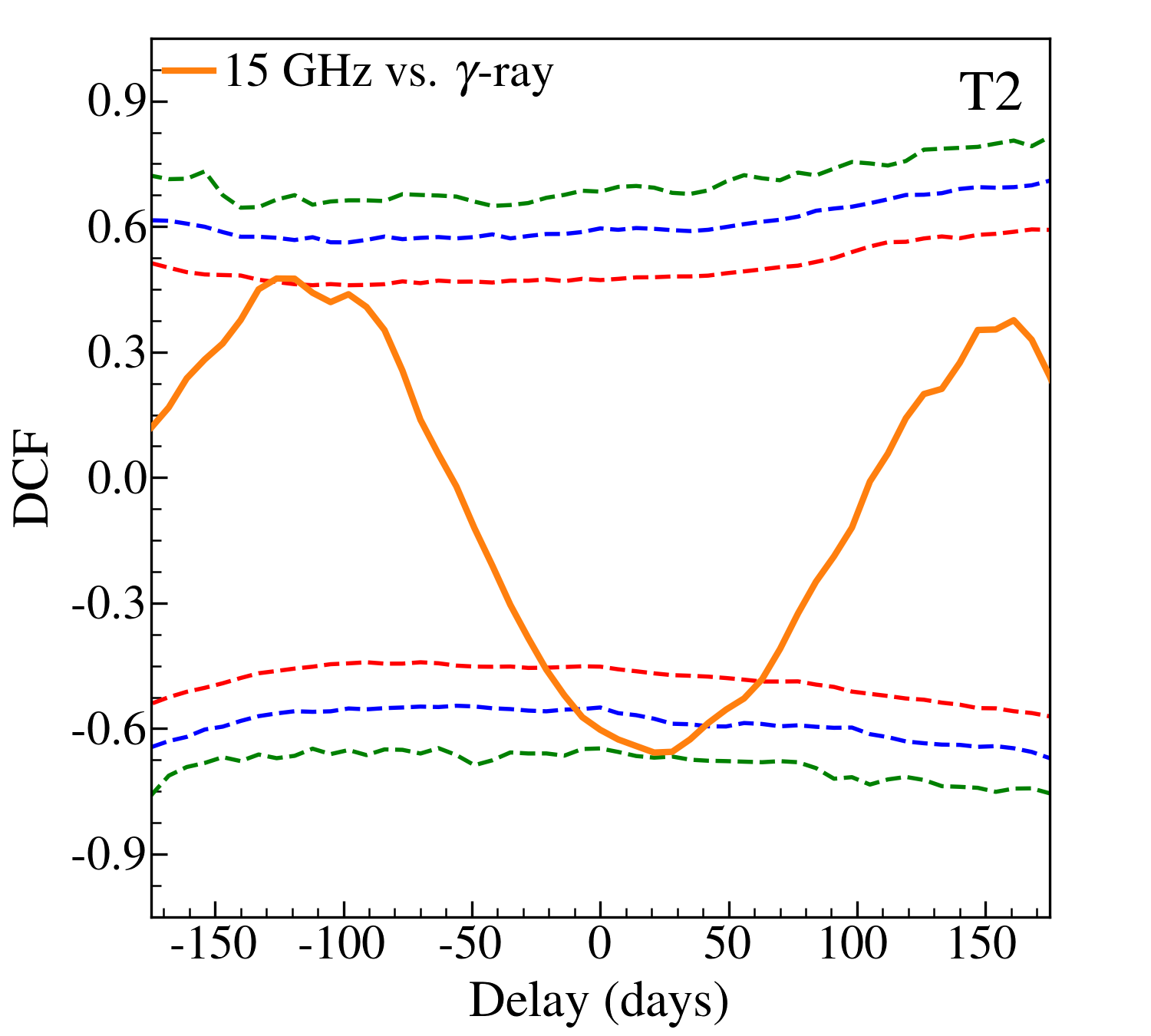}
\includegraphics[angle=0,width=0.27\textwidth]{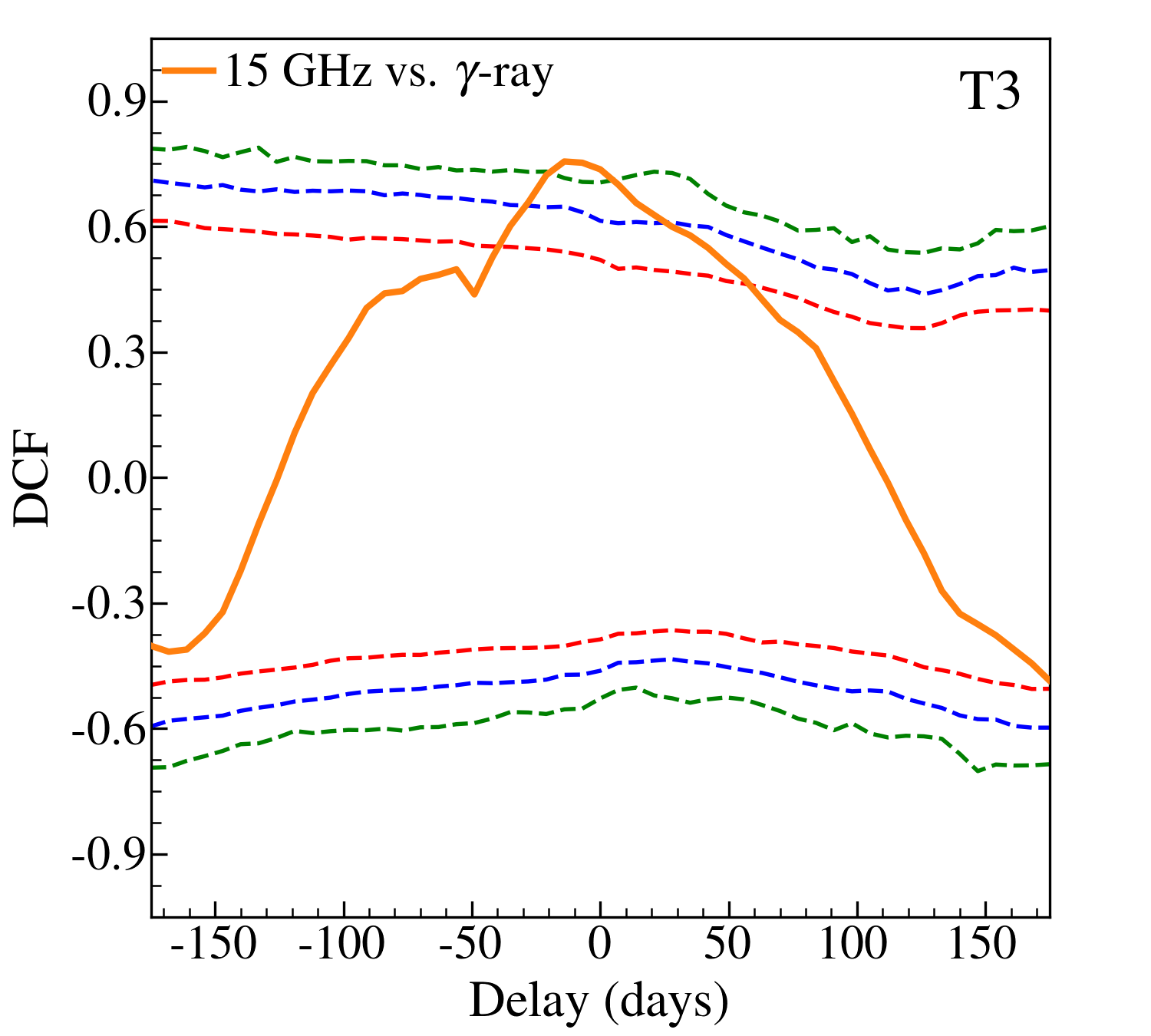}\\
\includegraphics[angle=0,width=0.27\textwidth]{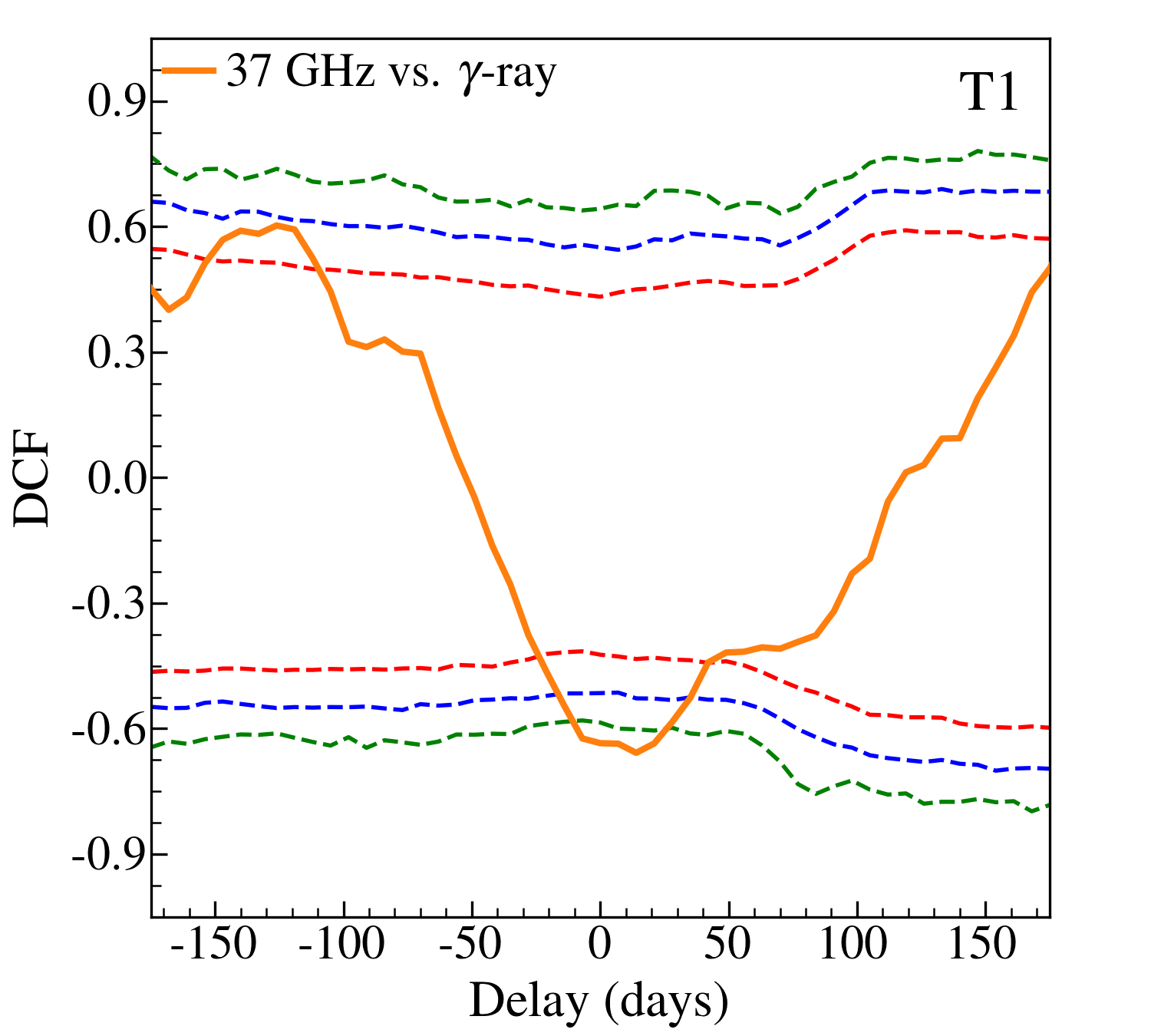}
\includegraphics[angle=0,width=0.27\textwidth]{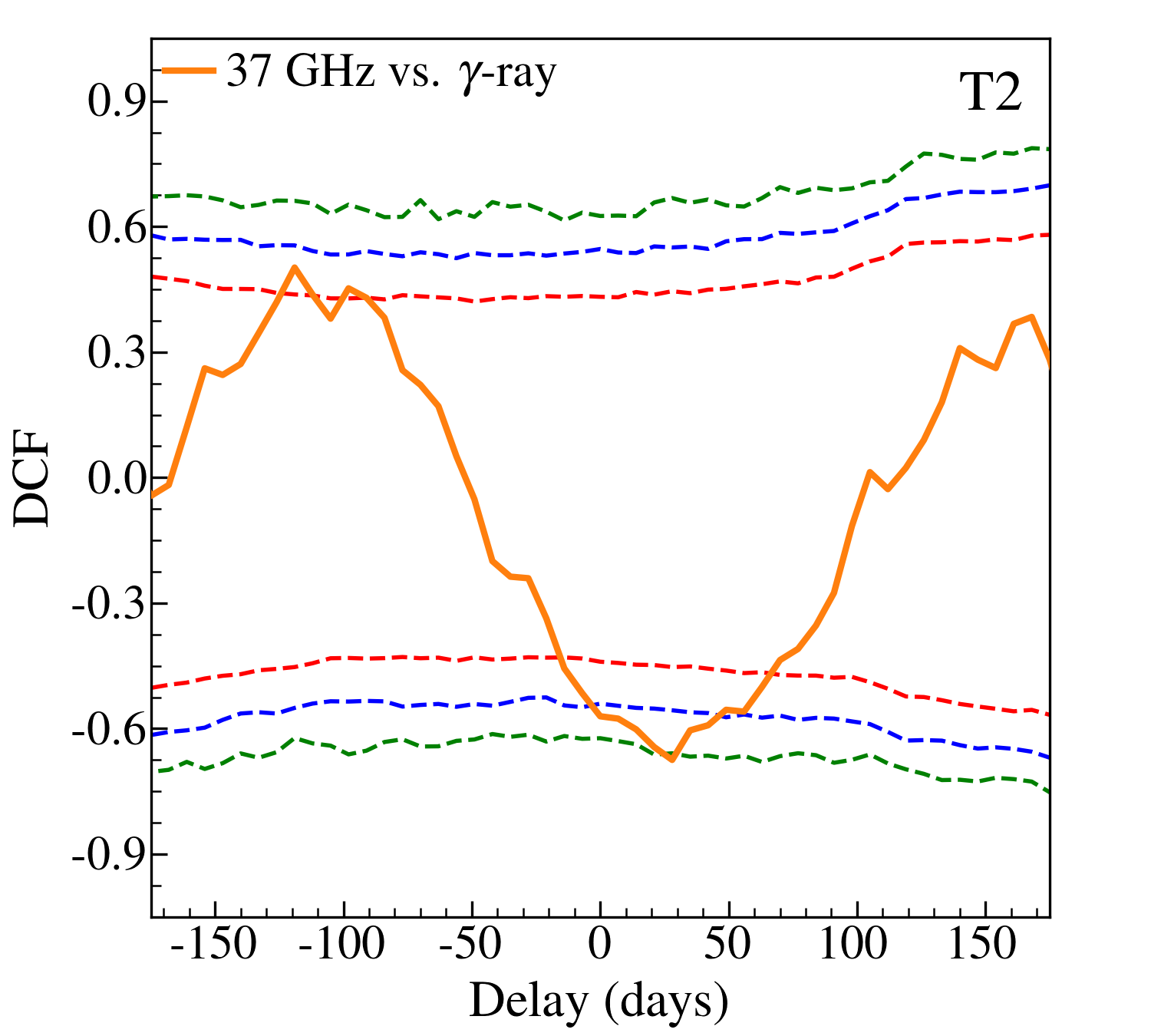}
\includegraphics[angle=0,width=0.27\textwidth]{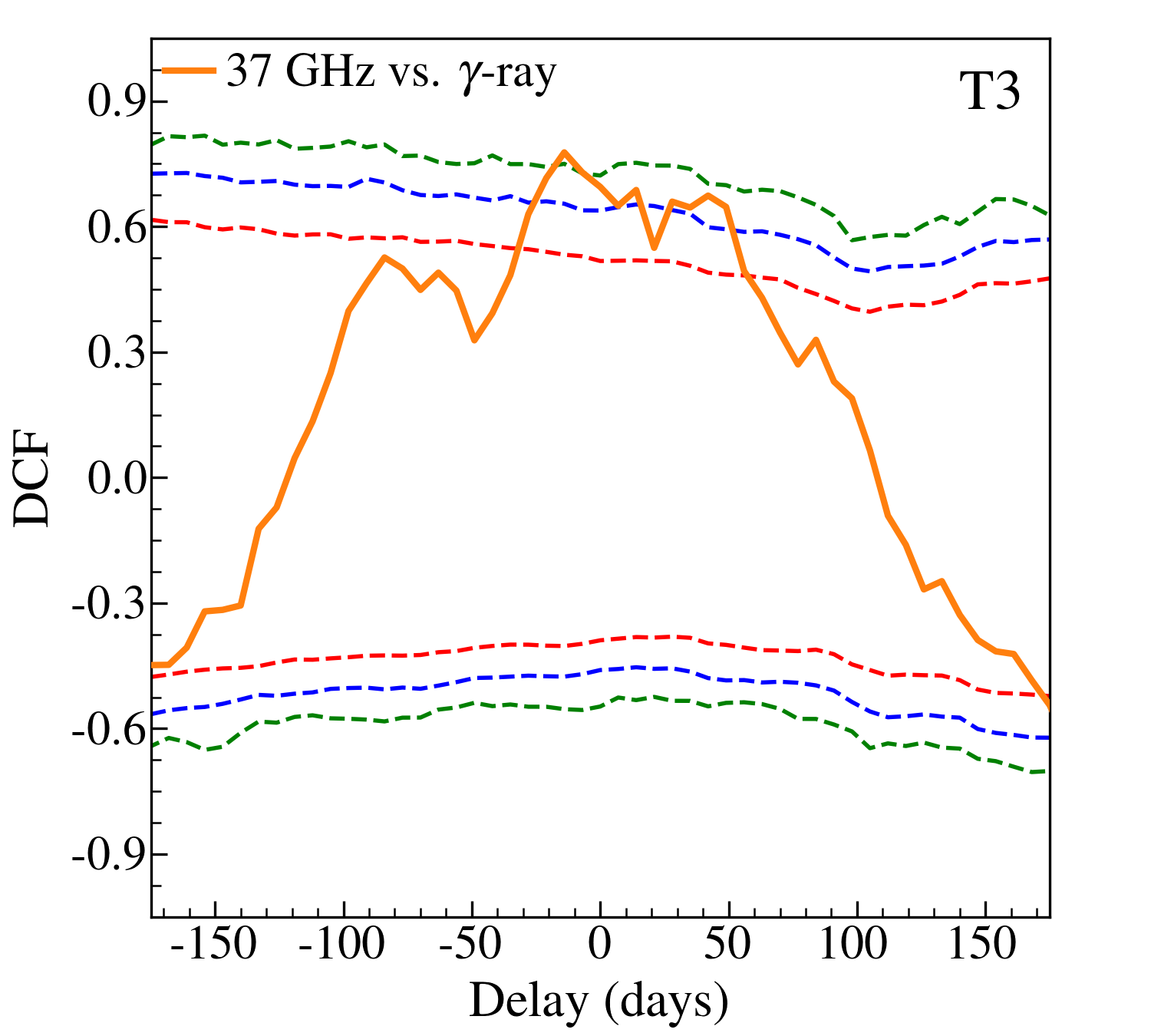}\\
\includegraphics[angle=0,width=0.27\textwidth]{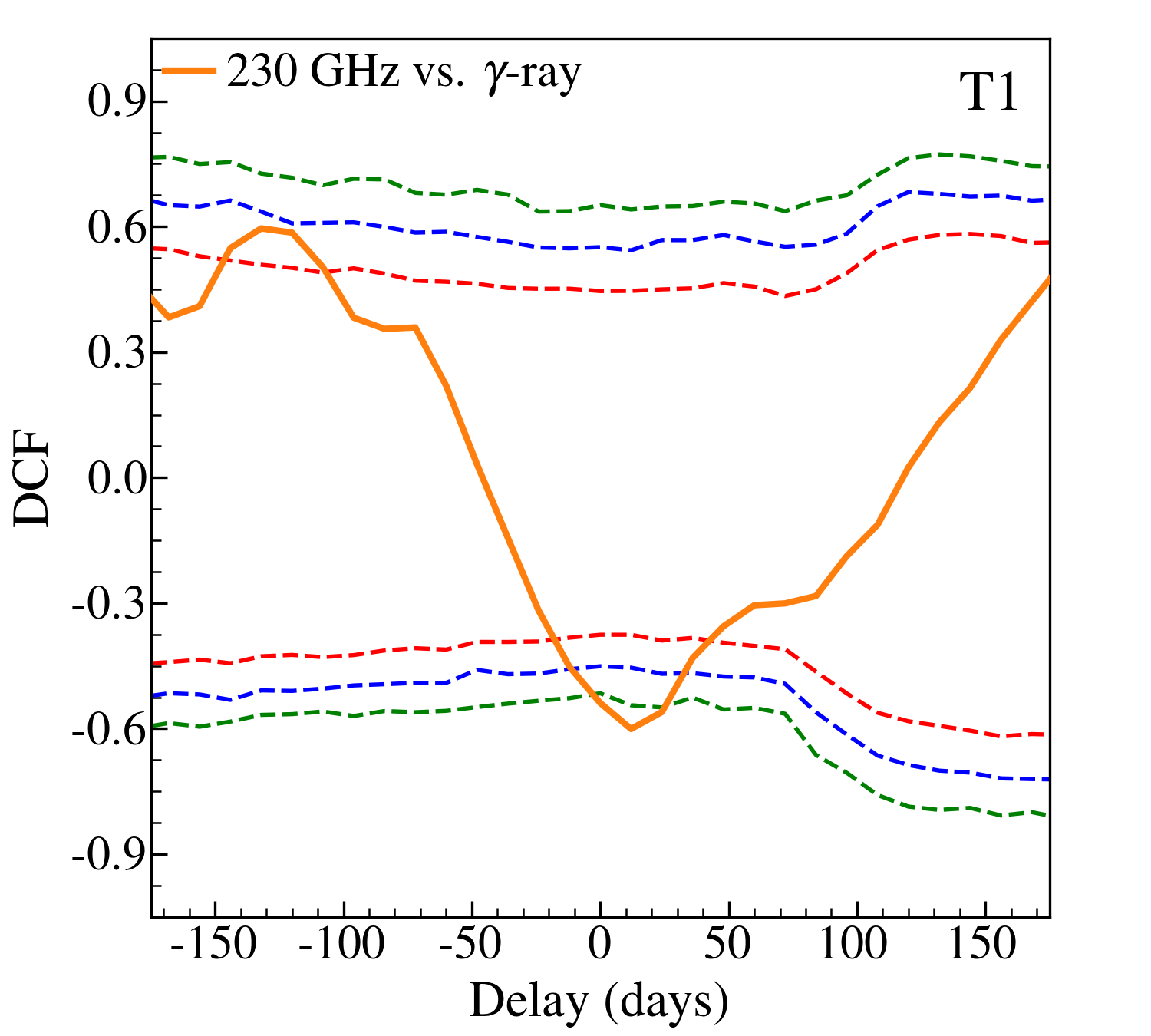}
\includegraphics[angle=0,width=0.27\textwidth]{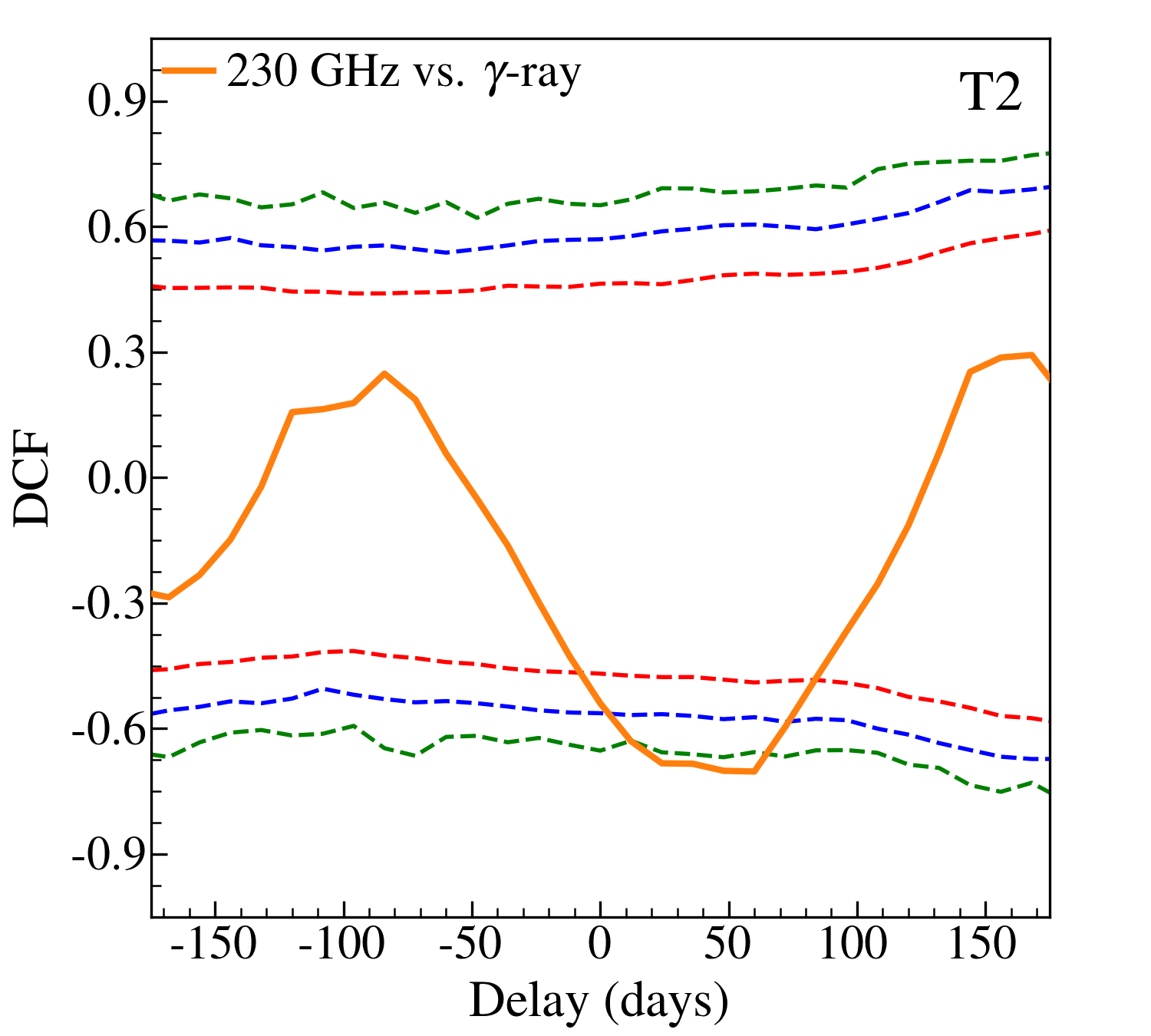}
\includegraphics[angle=0,width=0.27\textwidth]{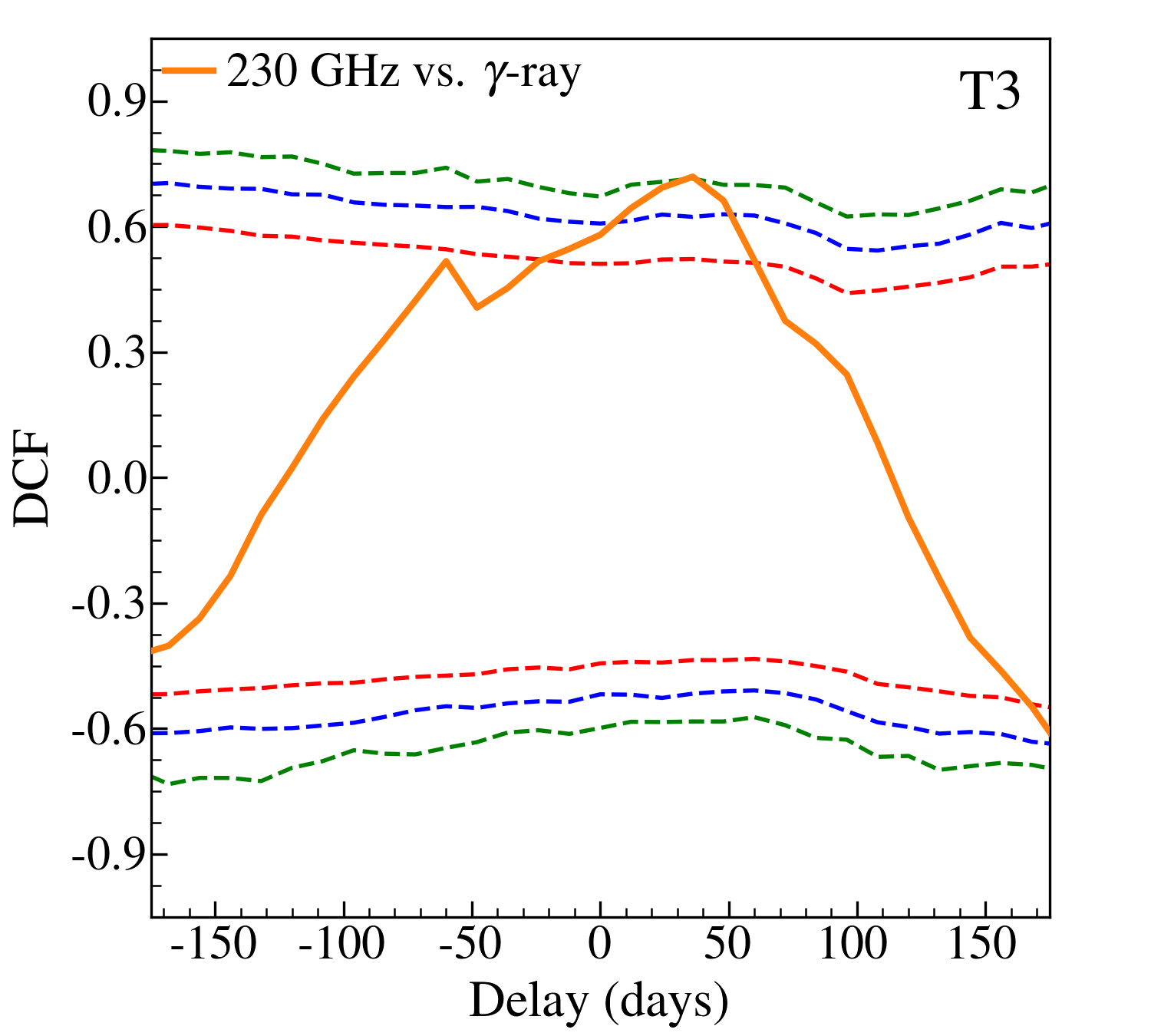}\\
\caption{From top to bottom: the radio and $\gamma$-ray light curves and the DCF curves from using 15, 37, and 230\,GHz radio data, respectively. Columns from left to right correspond to periods $T1$, $T2$, and $T3$, respectively. In each DCF plots, the orange curve indicates the DCF curve. The red, blue, and green solid lines denote the 95, 99, 99.9 \% confidence levels, respectively.
}
\label{fig:dcf_1.5y}
\end{figure*}

To construct significance levels, we generated DCF curves for many possible pairs of 1.5-year radio/$\gamma$-ray intervals (in total $\sim$15,000 DCF curves) for each of the $A$, $B$, and $C$ periods by random selection (any duplication was rejected). In the third column of Figure~\ref{fig:searching}, the two ``extremal'' DCF curves containing the highest and lowest DCF values (shown in the first column of Figure~\ref{fig:searching}) are overlaid onto the $\sim$15,000 DCF curves produced by the random pair test.

Interestingly, we found that the most negative DCF values exceeded the 99.9\% confidence level in periods $A$ and $B$ only, while this was the case only for the most positive DCF value in $C$. This suggests that the three 1.5-year time ranges ($T1$, $T2$, and $T3$) yielding these extreme DCF values are the origin of the significant (anti-)correlations between the radio and $\gamma$-ray light curves in the $A$, $B$, and $C$ periods, respectively. In each of the periods $A$, $B$, and $C$, extreme DCF values tend to appear at similar time delays. This is an actual measurement and not an artifact of external constraints. The light curves vary rapidly; correlated and anti-correlated segments are only a small displacement of the sliding 1.5-year window away from each other. We summarize the results of our test in Table~\ref{tab:searching_result}.

\begin{figure*}[!ht]
\centering
\includegraphics[angle=0,width=0.27\textwidth]{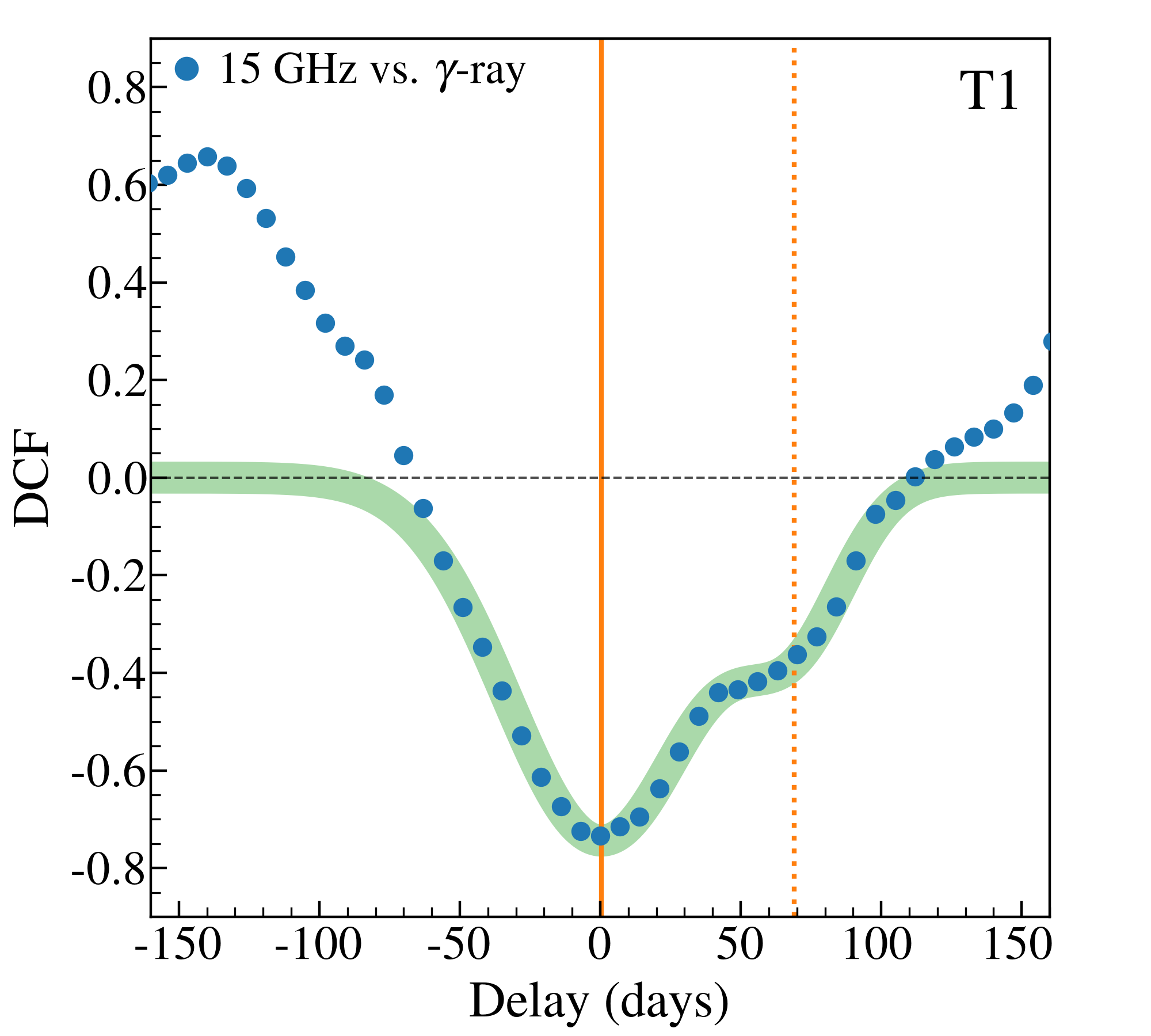}\quad
\includegraphics[angle=0,width=0.27\textwidth]{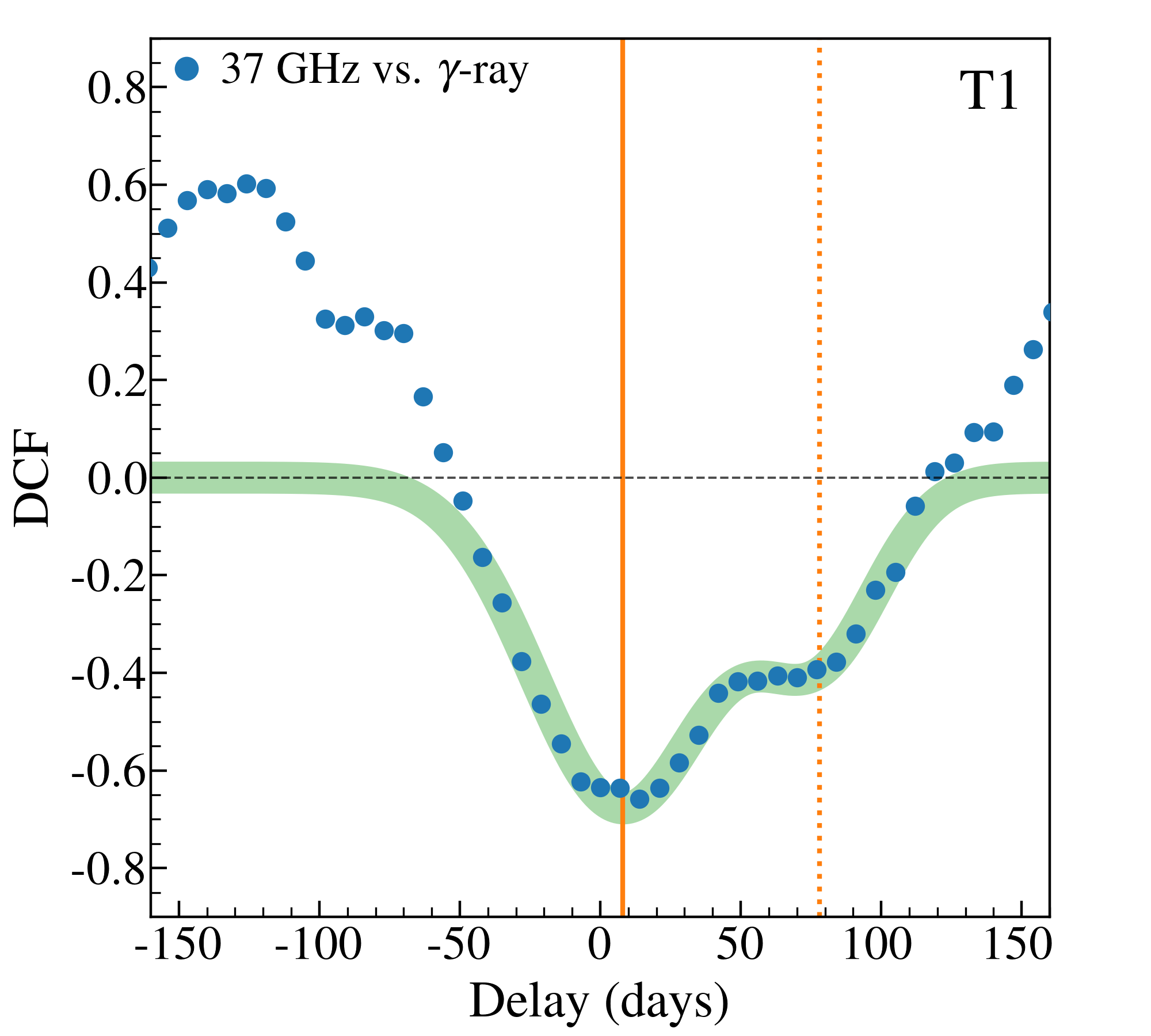}\quad
\includegraphics[angle=0,width=0.27\textwidth]{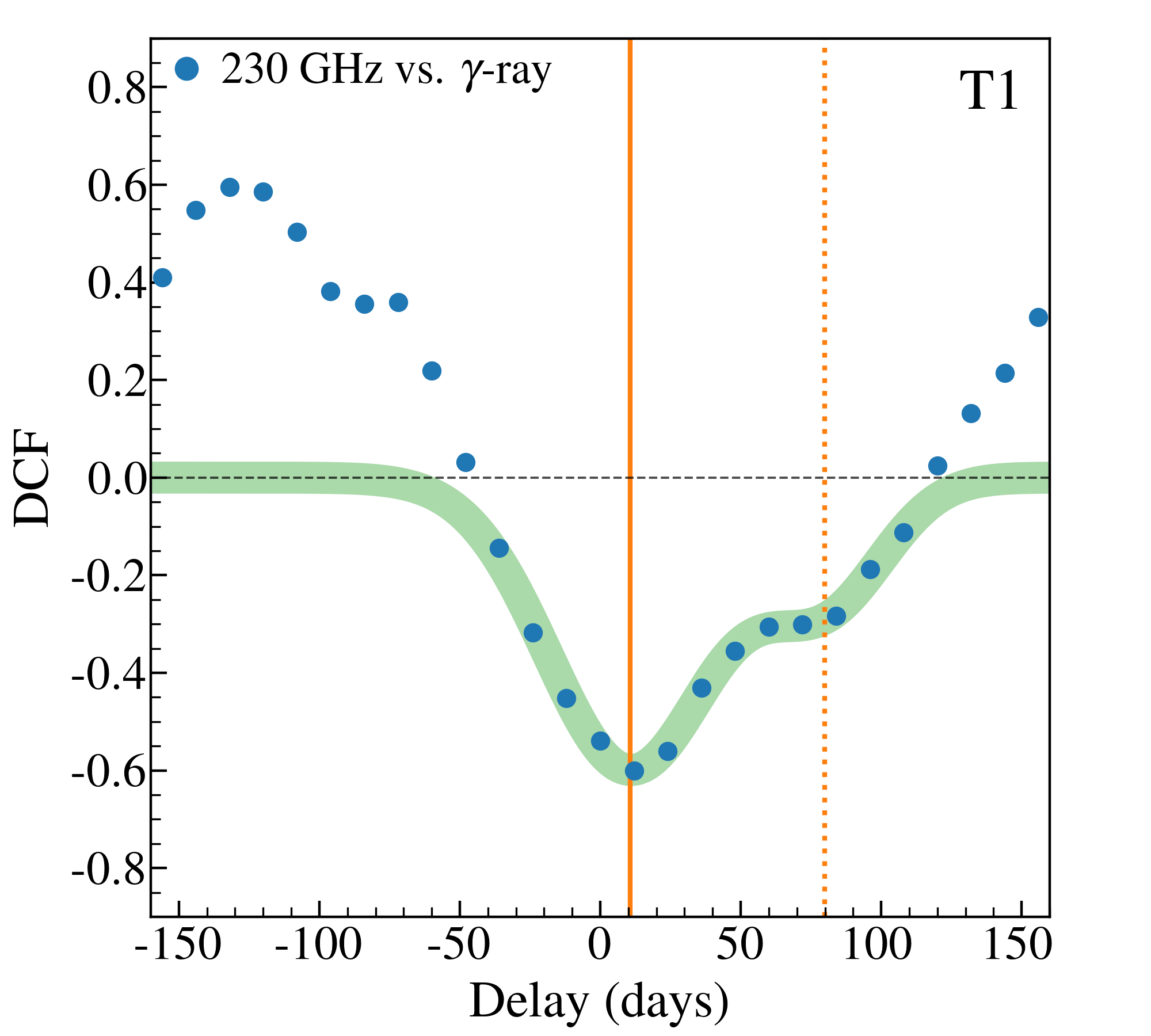}\\
\includegraphics[angle=0,width=0.27\textwidth]{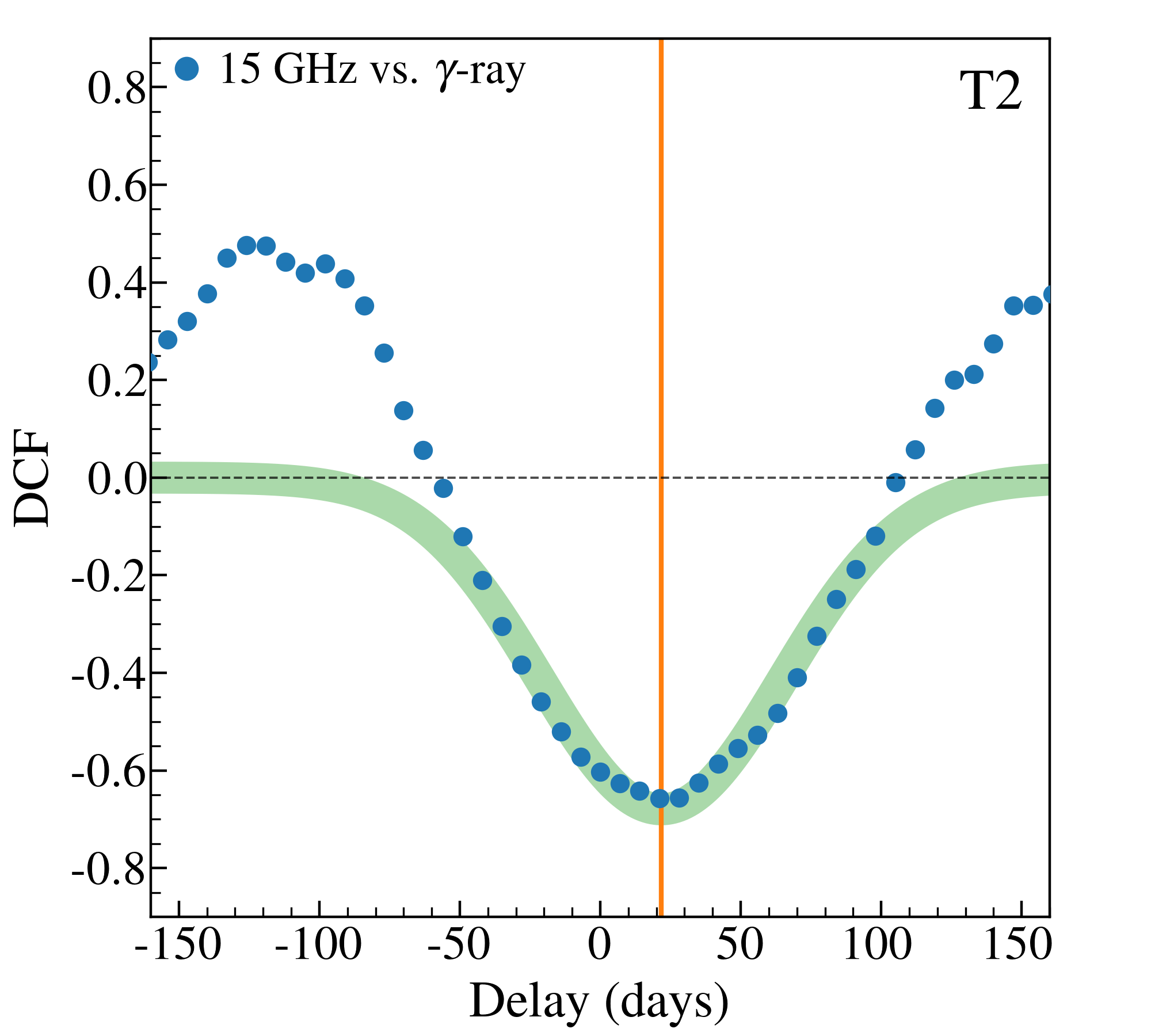}\quad
\includegraphics[angle=0,width=0.27\textwidth]{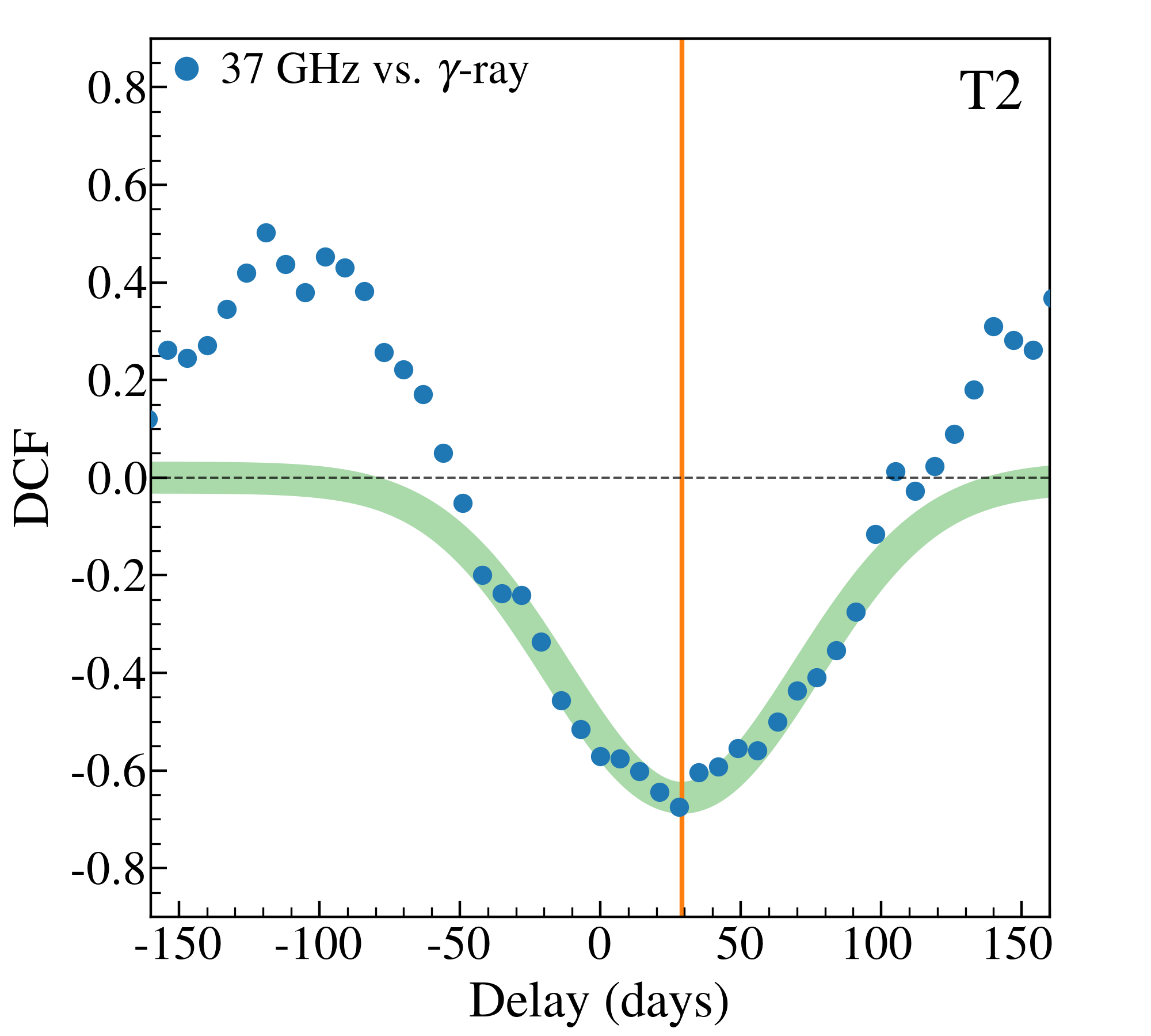}\quad
\includegraphics[angle=0,width=0.27\textwidth]{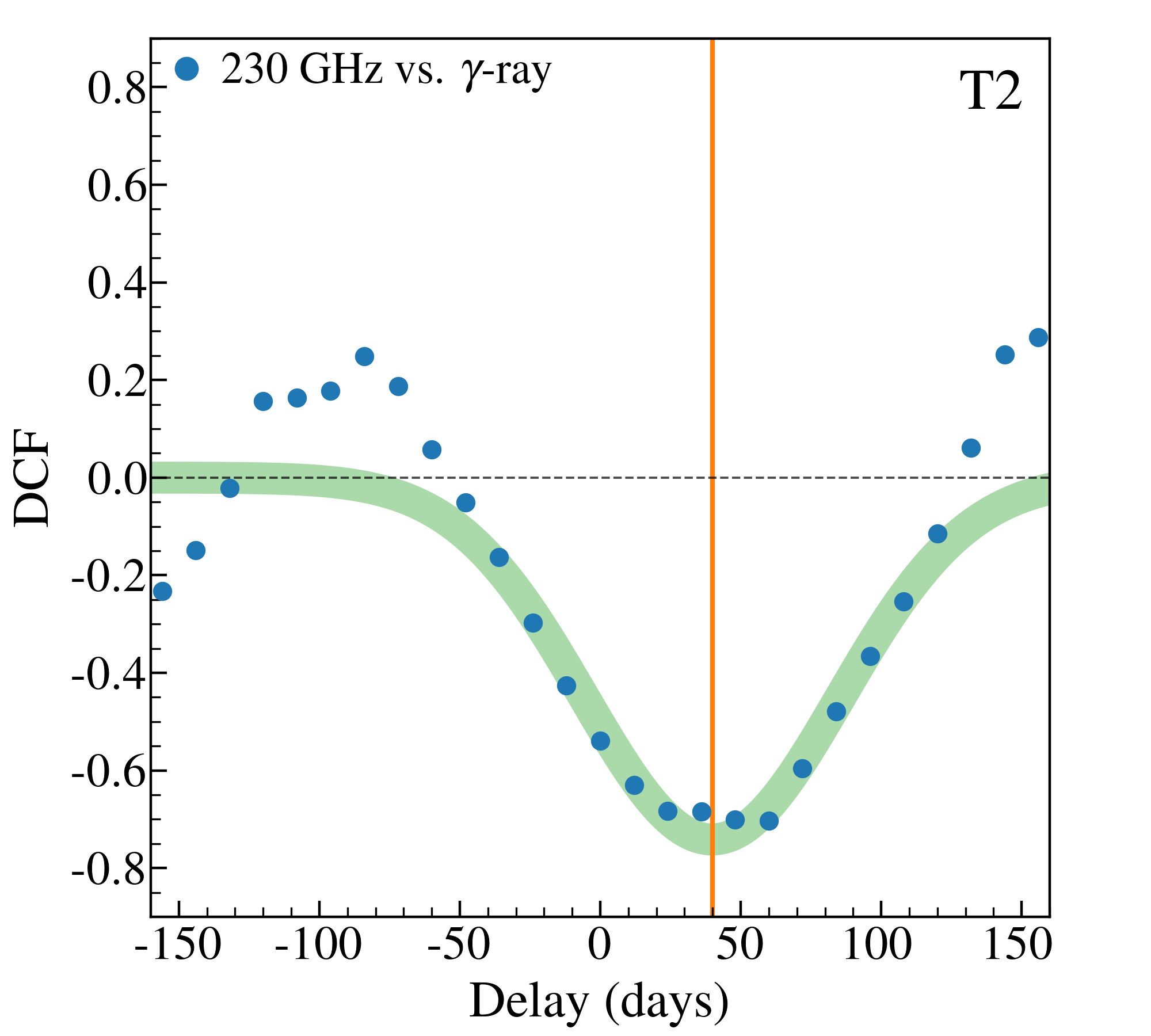}\\
\includegraphics[angle=0,width=0.27\textwidth]{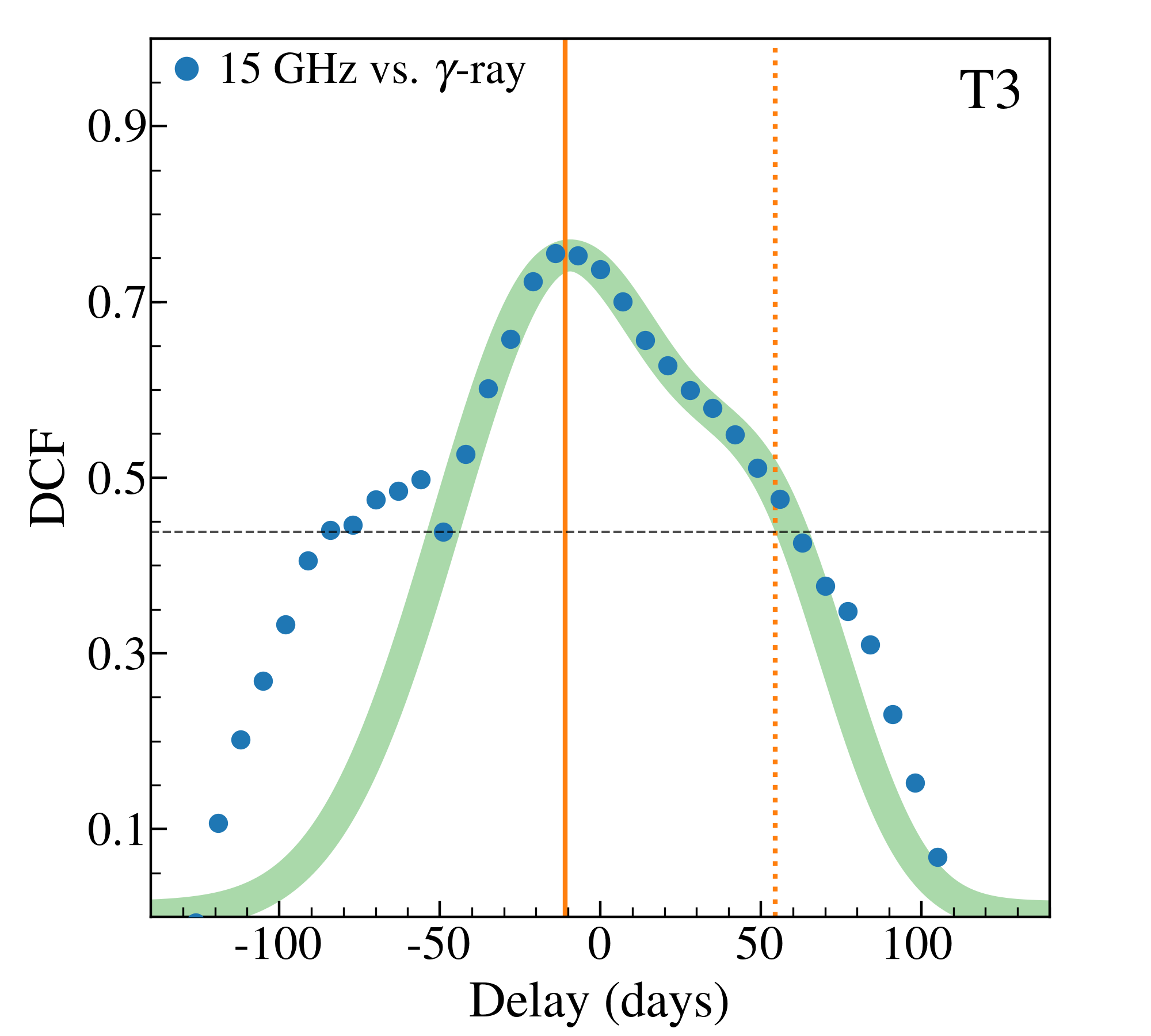}\quad
\includegraphics[angle=0,width=0.27\textwidth]{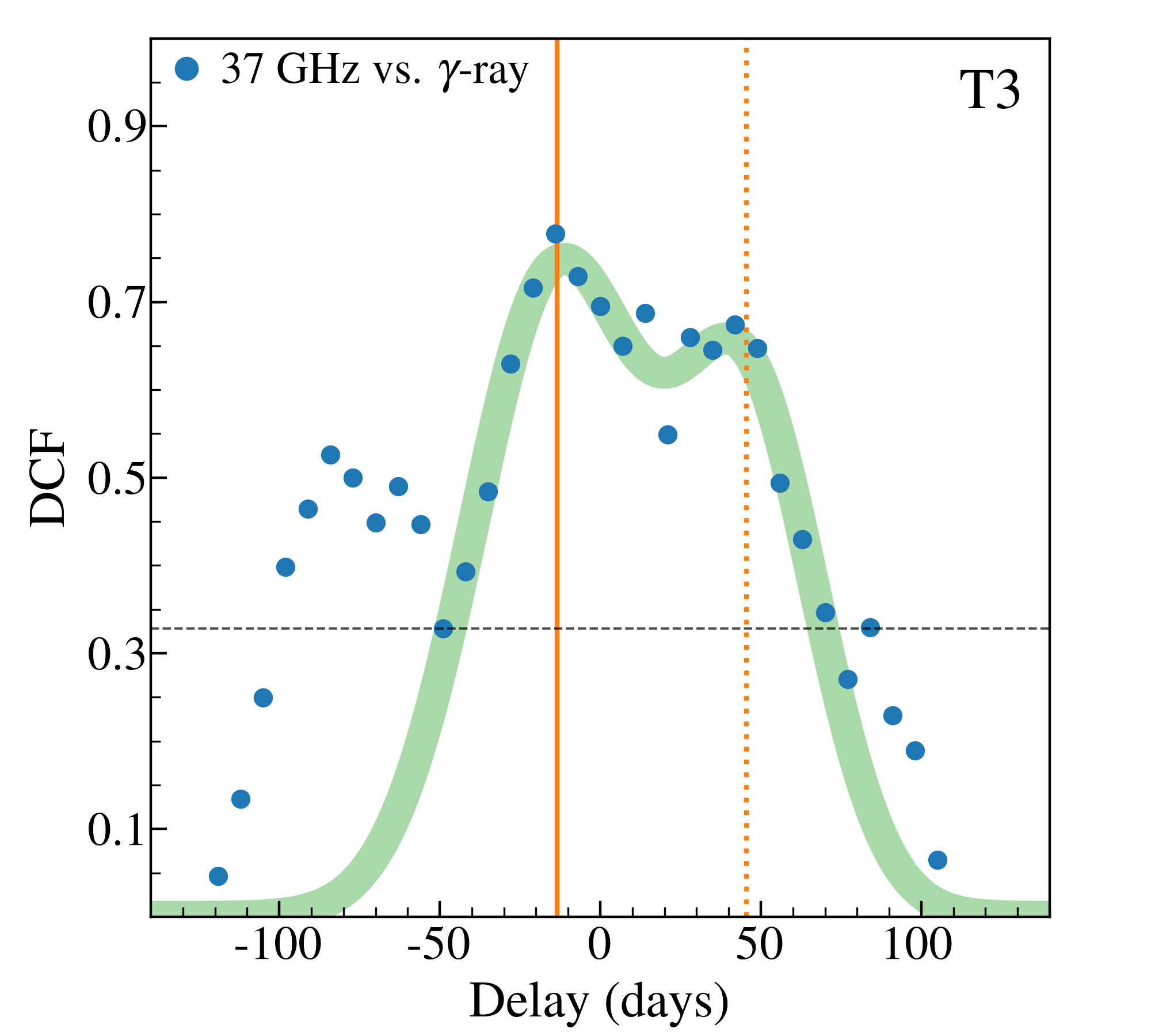}\\
\caption{Multi-component Gaussian fits to the DCF curves shown in Figure~\ref{fig:dcf_1.5y}. Thick green curve indicate the best-fit models. DCF peaks are marked with yellow vertical lines: absolute maxima or minima (\textit{solid}) and side-lobe peaks (\textit{dotted}). The fits used only the DCF values exceeding selected thresholds (\textit{horizontal black dashed lines}).
}
\label{fig:dcf_gauss}
\end{figure*}

\begin{table}[!hb]
\caption{Parameters of the best-fit Gaussian models shown in Figure~\ref{fig:dcf_gauss}}
\label{tab:dcf_gauss}
\centering
\begin{tabular}{l c c c}
\toprule
Data  &  $a^{1}$  &  $c^{2}$  &  $\mathrm{\left|w\right|}^{3}$  \\
  &    &  (days)  &  (days)  \\
\midrule
$T1$ at 15\,GHz  &  $-0.74\pm0.01$  &  $0.4\pm0.7$  &  $33.6\pm0.8$  \\
$T1$ at 37\,GHz  &  $-0.68\pm0.01$  &  $7.9\pm1.3$  &  $31.2\pm1.4$  \\
$T1$ at 230\,GHz  &  $-0.60\pm0.01$  &  $10.7\pm1.2$  &  $29.3\pm1.3$  \\
$T2$ at 15\,GHz  &  $-0.68\pm0.01$  &  $21.8\pm1.2$  &  $43.3\pm1.3$  \\
$T2$ at 37\,GHz  &  $-0.66\pm0.02$  &  $29.0\pm1.3$  &  $44.2\pm1.6$  \\
$T2$ at 230\,GHz  &  $-0.74\pm0.02$  &  $39.9\pm1.4$  &  $45.7\pm1.5$  \\
$T3$ at 15\,GHz  &  $0.75\pm0.01$  &  $-11.0\pm1.5$  &  $36.7\pm1.4$  \\
$T3$ at 37\,GHz  &  $0.73\pm0.03$  &  $-13.5\pm3.3$  &  $26.1\pm2.9$ \\
\bottomrule 
\multicolumn{4}{l}{$^1$ Peak amplitude of the Gaussian component.}\\
\multicolumn{4}{l}{$^2$ Peak location of the Gaussian component.}\\
\multicolumn{4}{l}{$^3$ Width of the Gaussian component.}\\
\end{tabular}
\end{table}

\subsubsection{DCF curves over the T1, T2, and T3 periods} \label{sec:cor3}
For the periods $T1$, $T2$, and $T3$, we performed the statistical test in the same manner as in Section~\ref{sec:cor1}, but using all radio bands. Figure~\ref{fig:dcf_1.5y} shows the results of the DCF analysis between the radio (15, 37, and 230\,GHz) and $\gamma$-ray light curves. Unsurprisingly, we find significant anti-correlations in both the $T1$ and $T2$ periods for all radio frequencies. In addition, we find DCF maxima (positive correlations) at delays between about $-$150 and $-$100 days, though these are less significant than the anti-correlations. Thus, our DCF results are consistent with the trends in the long-term correlation analysis presented in Section~\ref{sec:cor1}. We also find a significant positive correlation in the $T3$ period.


The location of the negative DCF minima indicates that the radio emission leads the $\gamma$-rays. Since the light curves vary rapidly `from peak to valley', a misidentification of corresponding flares can introduce negative correlations at positive time delays. Indeed, the negative delays for the positive correlation peaks in $T1$ and $T2$ indicate that the $\gamma$-ray emission leads the radio emission, which is typical for  blazars \citep[e.g.][]{2010ApJ...722L...7P}. However, the positive correlation between the $\gamma$-rays and the 230\,GHz data in $T3$ shows the opposite trend. 
The relatively poor sampling of the 230\,GHz data during this period missed the major radio flare that was observed in the other two radio frequencies around MJD\,57980 completely. In view of this, we were unable to estimate an accurate DCF curve for the 230\,GHz data in the $T3$ period and do not consider it in further analysis.

To identify the precise location and time lag of each DCF peak, we fit one or two Gaussian functions of the form ${\rm DCF}(t) = a\,\times$\,exp$\left[-(t-c)^{2}/2w^{2}\right]$, with $a$, $c$, and $w$ being the amplitude, delay, and width of the Gaussian profile, respectively, to the data \citep[e.g.,][]{2013A&A...552A..11R, 2018A&A...614A.148B}. Figure~\ref{fig:dcf_gauss} shows the results. A single Gaussian for $T2$ and double Gaussian for $T1$ and $T3$ were used to better describe the DCF extrema.
Table~\ref{tab:dcf_gauss} shows the parameters of the best-fit Gaussian models. In all three periods ($T1$, $T2$, and $T3$), the centimeter to millimeter radio light curves are highly correlated/anti-correlated with the $\gamma$-ray flux and show extremal coefficient values of very similar magnitude, thus suggesting the same physical processes behind the 15, 37, and 230\,GHz fluxes \citep[e.g.,][]{2013A&A...552A..11R, 2019A&A...626A..60A}. We also noticed that there is a frequency-dependent delay in the DCF peaks. For both the positive and negative correlations, the positions of the maxima and minima are located at larger absolute delay values for higher radio frequencies.
%

\subsection{Jet kinematics} \label{sec:res6}

\begin{figure}[!t]
\centering
\includegraphics[angle=0, width=0.45\textwidth, keepaspectratio]{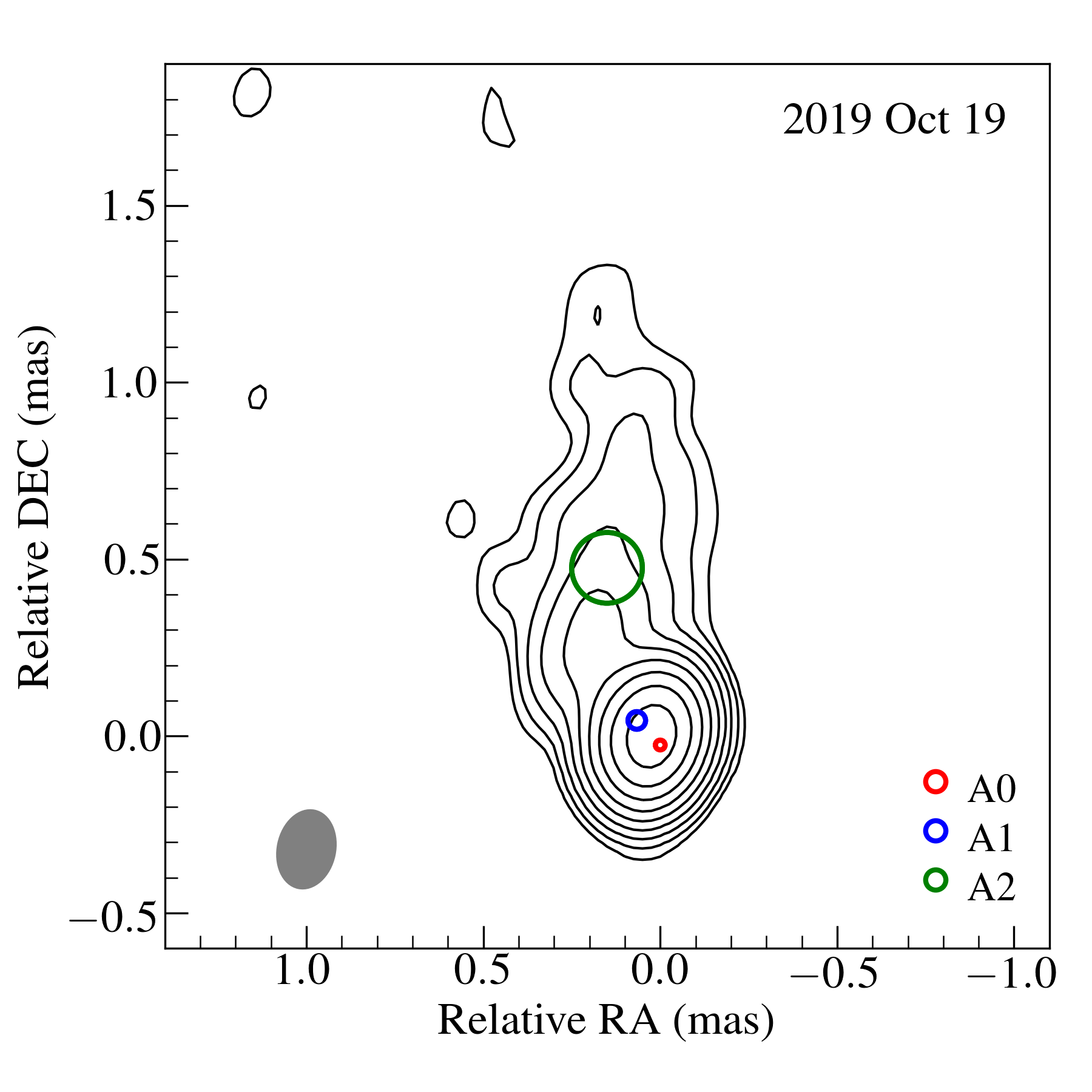} 
\caption{Total intensity image of the 0716$+$714 jet observed by the VLBA at 43\,GHz on 2019 October 19. The contour levels increase by a factor of 2, from 0.25\% to 64\% of the peak flux (i.e., 0.55\,Jy). The synthesized beam size is shown by the ellipse in the \textit{bottom-left} corner. The blue and green circles indicate the mean positions and sizes of $A1$ and $A2$ measured in 2008--2019, respectively. The red circle ($A0$; the core) is located at the map peak and indicates the mean core size (i.e., a FWHM of 0.03\,mas).
}
\label{fig:f6}
\end{figure}

\begin{figure}[!t]
\centering
\includegraphics[angle=0, width=0.45\textwidth, keepaspectratio]{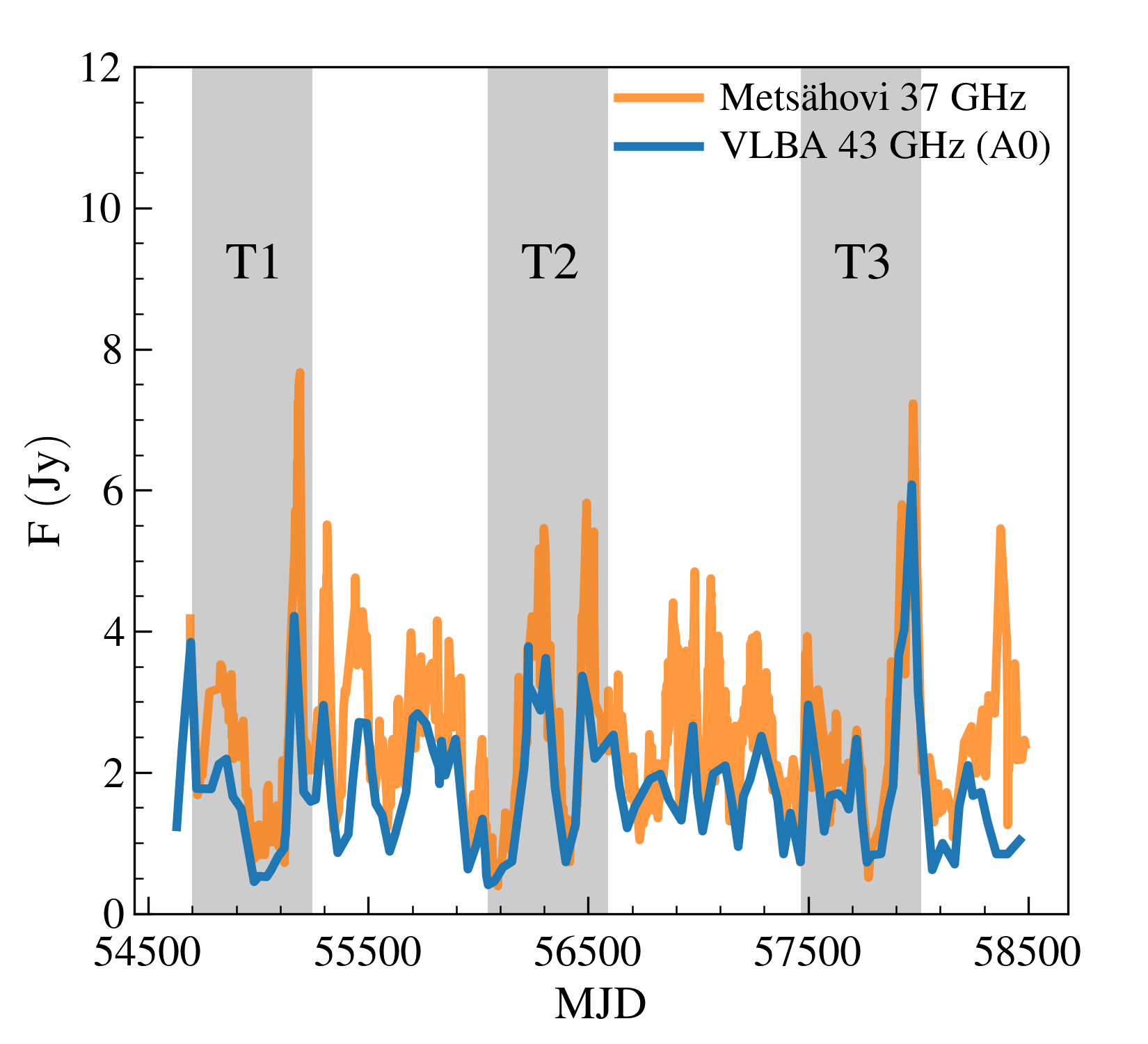} 
\caption{Radio light curves of 0716$+$714: total flux density at 37\,GHz obtained from Mets\"{a}hovi (\textit{orange}) and core ($A0$) flux density at 43\,GHz obtained from VLBA (\textit{blue}).
}
\label{fig:new8}
\end{figure}

\begin{figure*}[!t]
\centering
\includegraphics[angle=0, height=9cm, keepaspectratio]{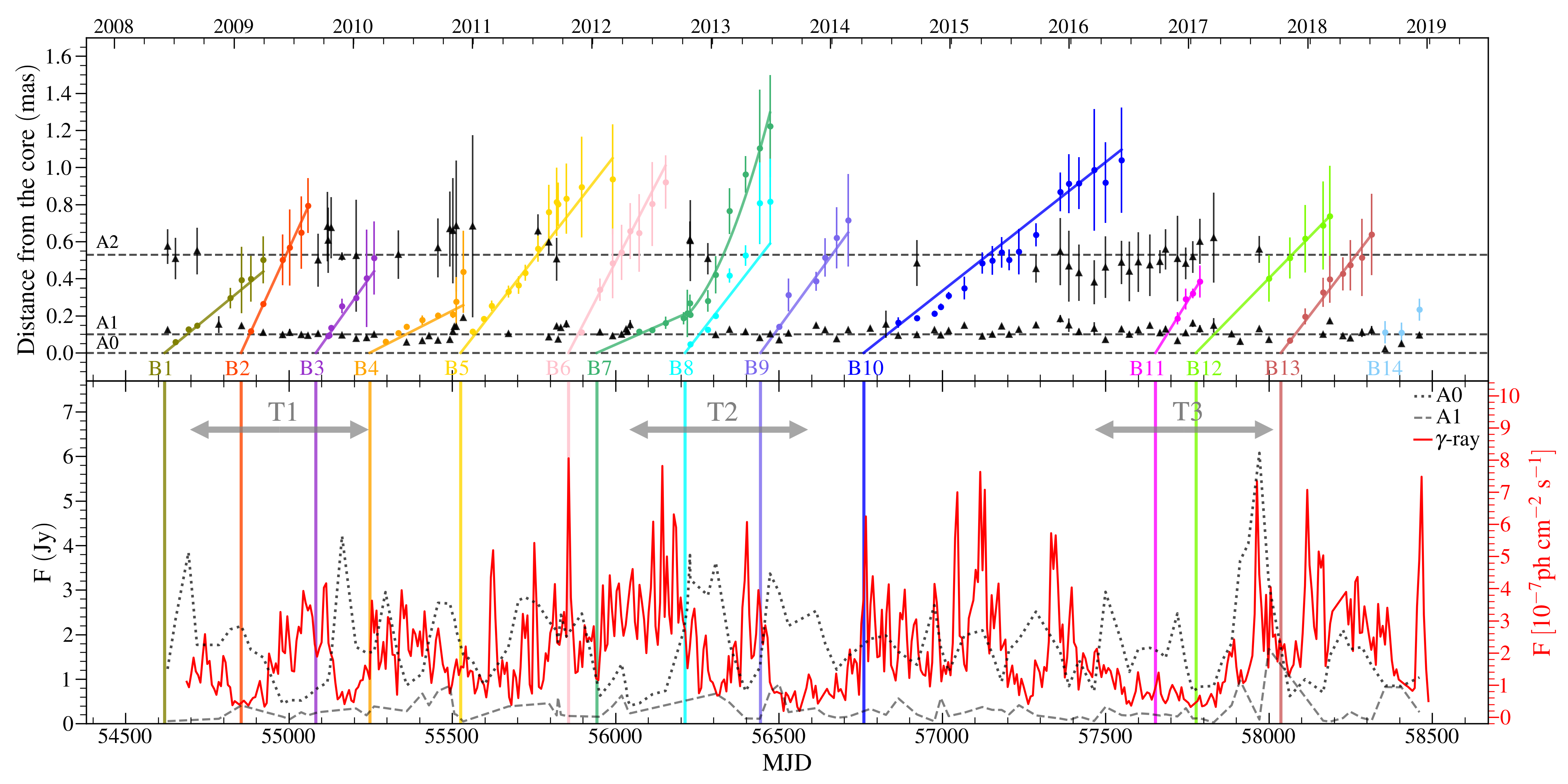} 
\caption{Top: Separation of jet components from the core as function of time. The solid lines represent linear fits to the component positions, dashed lines indicate the positions of the core $A0$ and the stationary features $A1$ and $A2$.
Bottom: The 43\,GHz light curves of $A0$ and $A1$ overlaid on the $\gamma$-ray light curve. The component $A2$ is omitted due to its weak emission. Vertical solid lines mark the epochs of ejection of the newborn jet components.
}
\label{fig:f7}
\end{figure*}

The activity in the parsec scale jet of 0716$+$714 has been monitored at 7\,mm (43\,GHz) by the VLBA-BU-BLAZAR program. To reconstruct the jet structure, we model the Fourier transform of the sky brightness distribution in the ($u$,$v$) visibility domain with a number of two-dimensional Gaussian intensity profiles using the \textit{modelfit} task in the \texttt{Difmap} software package \citep{1997ASPC..125...77S}.
Figure~\ref{fig:f6} shows the radio jet structure of the blazar on 2019 October 19.
We designate the 7\,mm core as $A0$, which is assumed to be the brightest region at the upstream end of the jet flow.
Apart from the core, we identify two stationary features, $A1$ and $A2$, located at average distances of 0.10$\pm$0.02 and 0.53$\pm$0.10\,mas from the core, respectively, and 14 individual moving knots ($B1$--$B14$); here the errors of the mean distances of $A1$ and $A2$ from the core are their average sizes. Comparison of the total and the core light curves (see Figure~\ref{fig:new8}) indicates that the core is nearly always the dominant source of the millimeter emission. Table~\ref{app:buparam} summarizes the parameters of all jet components.
Figure~\ref{fig:f7} shows the knot propagation and the radio and $\gamma$-ray light curves.

The $B1$--$B14$ jet components show a linear motion except for $B7$ which was better fit by a quadratic polynomial. 
We exclude $B14$ from the further kinematics analysis because it is detected only at three epochs.
The ejection epoch $T_0$ for each of the components $B1$--$B13$ was calculated from fits to the RA and DEC coordinate using a linear regression. Due to large errors in $T_0$ exceeding 0.5 years for the $B3$, $B5$, $B7$, and $B10$ knots, we calculated the ejection epoch from a linear fit to the radial distance from the core versus time. A complete description of the observations and data calibration will be provided in Weaver et al. (in prep.). The kinematics of the knots is summarized in Table~\ref{tab:tb6}.


\begin{table}[b]
\caption{Newborn jet components in 2008--2019}
\label{tab:tb6}
\centering
\begin{tabular}{l l c c c}
\toprule
ID  &  $T_{0}^{1}$  &  $\mu^{2}$  &  $\phi^{3}$  &  $\beta_{\rm app}^{4}$  \\
  &  (MJD)  &  (mas/year)  &  ($^{\circ}$)  &  ($c$)  \\
\midrule
B1  &  $54620\pm15$  &  $0.53\pm0.04$  &  $21\pm22$  &  $10.0\pm0.3$  \\
B2  &  $54854\pm2$  &  $1.40\pm0.03$  &  $-3\pm6$  &  $26.4\pm0.2$  \\
B3  &  $55083\pm10$  &  $1.04\pm0.05$  &  $-9\pm4$  &  $19.6\pm0.3$  \\
B4  &  $55248\pm52$  &  $0.33\pm0.04$  &  $26\pm5$  &  $6.2\pm0.2$  \\
B5  &  $55526\pm32$  &  $0.82\pm0.03$  &  $23\pm2$  &  $15.6\pm0.2$  \\
B6  &  $55856\pm8$  &  $1.24\pm0.05$  &  $23.3\pm1.2$  &  $23.4\pm0.3$  \\
B7  &  $55943\pm79$  &  $1.07\pm0.11$  &  $30\pm8$  &  $20.1\pm0.6$  \\
B8  &  $56213\pm2$  &  $0.83\pm0.13$  &  $34\pm6$  &  $15.6\pm0.7$  \\
B9  &  $56444\pm6$  &  $0.88\pm0.07$  &  $5\pm5$  &  $16.7\pm0.4$  \\
B10  &  $56760\pm61$  &  $0.51\pm0.03$  &  $25\pm2$  &  $9.6\pm0.2$  \\
B11  &  $57653\pm1$  &  $1.04\pm0.12$  &  $8\pm12$  &  $19.6\pm0.7$  \\
B12  &  $57777\pm4$  &  $0.66\pm0.03$  &  $5\pm8$  &  $12.4\pm0.2$  \\
B13  &  $58037\pm2$  &  $0.84\pm0.03$  &  $27\pm4$  &  $15.8\pm0.2$  \\
\bottomrule 
\multicolumn{4}{l}{$^1$ Ejection time.}\\
\multicolumn{4}{l}{$^2$ Proper motion.}\\
\multicolumn{4}{l}{$^3$ Direction of motion (North through East).}\\
\multicolumn{4}{l}{$^4$ Apparent velocity in units of speed of light.}\\
\end{tabular}
\end{table}

The apparent velocities ($\beta_{\rm app}$) of the knots $B1$--$B13$ are located in the range from 6 to 26\,$c$. The direction of the motion varies between $-$9$^{\circ}$ (Western) and 34$^{\circ}$ (Eastern) as measured from North. On average, the knots were ejected from the core every 0.8\,years.  
This is a bit higher than what \citet{2017ApJ...846...98J} expected. However, if we exclude the long quiescent period between $B10$ and $B11$ (i.e., 893\,days), the ejection interval becomes 0.6\,years. 
Two knots were found in each of the $T1$, $T2$, and $T3$ periods: $B2$--$B3$ in $T1$, $B8$--$B9$ in $T2$, and $B11$--$B12$ in $T3$. The timing between ejections and radio/$\gamma$-ray flares differs from knot to knot. However, a consistent behavior can be seen in $T2$: each ejection coincides with the decaying phase of a $\gamma$-ray flare and the rising phase of a radio flare.
In the case of $T3$, the ejection of $B11$ seems not to be correlated with any significant events at both radio and $\gamma$-rays. However, strong radio and $\gamma$-ray flares started rising just after the ejection of $B12$ almost simultaneously.

\begin{figure}
\centering
\includegraphics[angle=0, width=0.45\textwidth, keepaspectratio]{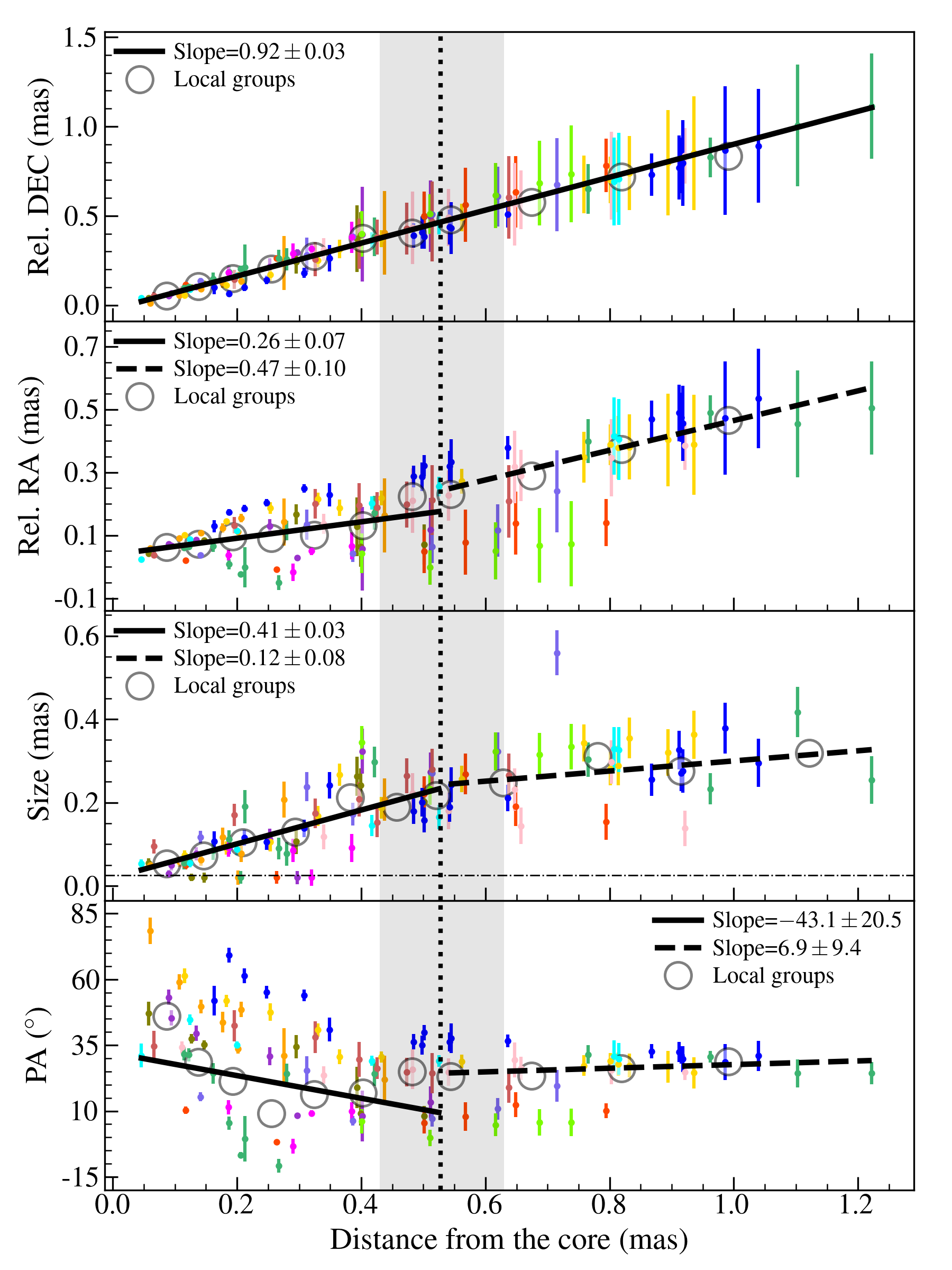} 
\caption{From top to bottom: relative DEC and RA, size, and position angle (PA; North through East, where 0$^{\circ}$ corresponds to North) as function of the distance from the core for the knots $B1$--$B13$ obtained from the model fits to the VLBA images. Data colors are the same as in Figure~\ref{fig:f7}. The black solid/dashed lines indicate linear regressions. For RA, size, and PA, regressions are performed separately for distances within and beyond the mean position of $A2$ (i.e., 0.53\,mas) which is denoted by the vertical black dotted line. The shaded gray color region represents the mean FWHM of $A2$ (i.e., 0.2\,mas) which shows the overall size of the region. The empty gray circles show the mean variations of the parameters by averaging 10 consecutive points each. Lower limits on sizes (third panel from top) are indicated by the horizontal dot-dashed line.}
\label{fig:f8}
\end{figure}

We further investigated the global kinematic properties of the 0716$+$714 jet. Figure~\ref{fig:f8} shows the results of the analysis of coponent motions, component sizes (FWHM), and position angles.
The parameters in the radial distance domain ($r$) show complex behavior with large scatter, except for the motion n declination (see the top panel of Figure~\ref{fig:f8}). We used linear regressions to approximate the global behaviors of the parameters of the knots. 
As was shown above, there is a stationary feature $A2$ at about $\sim$0.53\,mas from the core. This feature could be a transition region or a shock which leads to a change in the jet physical conditions \citep[e.g.,][]{2008ApJ...680..867K, 2013ApJ...768...40L, 2018A&A...610A..32B}. Thus, we performed the analysis for two groups of data separately: one with $r < 0.53$\,mas and the other with $r \geq 0.53$\,mas. The angular resolution of our data is insufficient to apply the analysis on smaller scales (e.g., less than 0.2\,mas). Thus, the stationary feature $A1$ was excluded from this analysis.

The jet components move in norther direction. We quantify their collective motion by assuming the relationship $d_{xy} \propto k r$, where $d_{xy}$ is either the relative RA ($x$) or DEC ($y$) of the knots, $r$ is the radial distance from the core, and $k$ is the slope. The linear slope of the motion in the $y$ direction is $k\sim$0.9, i.e. an almost one-to-one match with the radial distance. However, the slope of the motion along the $x$-axis, which is approximately perpendicular to the jet axis, is much smaller, meaning that the motion is primarily in the $y$ direction. Interestingly, at least one knot shows a curved trajectory, which indicates a bend in the jet \citep[see also][]{2013ApJ...768...40L, 2015A&A...578A.123R}. Indeed, from calculating $d_{x}$ we found that there are two separate linear trends in the transverse motion: one with $k \sim 0.26$ for $r < 0.53$\,mas, and one with $k \sim 0.47$ for $r \geq 0.53$\,mas. This indicates that there is a change in the jet direction at $A2$. 

The position angles show large variations out to $A2$, indicating a bending of the jet. Due to the large scatter, the position angle data are not well described by a linear trend. However, the outer region (i.e., downstream from $A2$) shows a clear linear trend. It should be noted that the uncertainties of the slopes are quite large. This implies that the angle variations are complicated and linear trend lines cannot fully describe the behavior of the position angle as function of core distance. We form groups by averaging over 10 consecutive points (total 110 model jet components) in the $r$ domain; as for the component size, seven lower limits on the size were rejected from the analysis, and the last bin contains three data points. 
The grouped position angles indicate that the jet axis moves from East to North (from about 50$^{\circ}$ to 10$^{\circ}$) up to a radial distance of $\sim$0.3\,mas, after which the jet bends back toward the eastern direction. However, the variation in the position angle is relatively weak beyond $\sim$0.3\,mas from the core before converging at about 30$^{\circ}$.

\begin{figure}
\centering
\includegraphics[angle=0, width=0.45\textwidth, keepaspectratio]{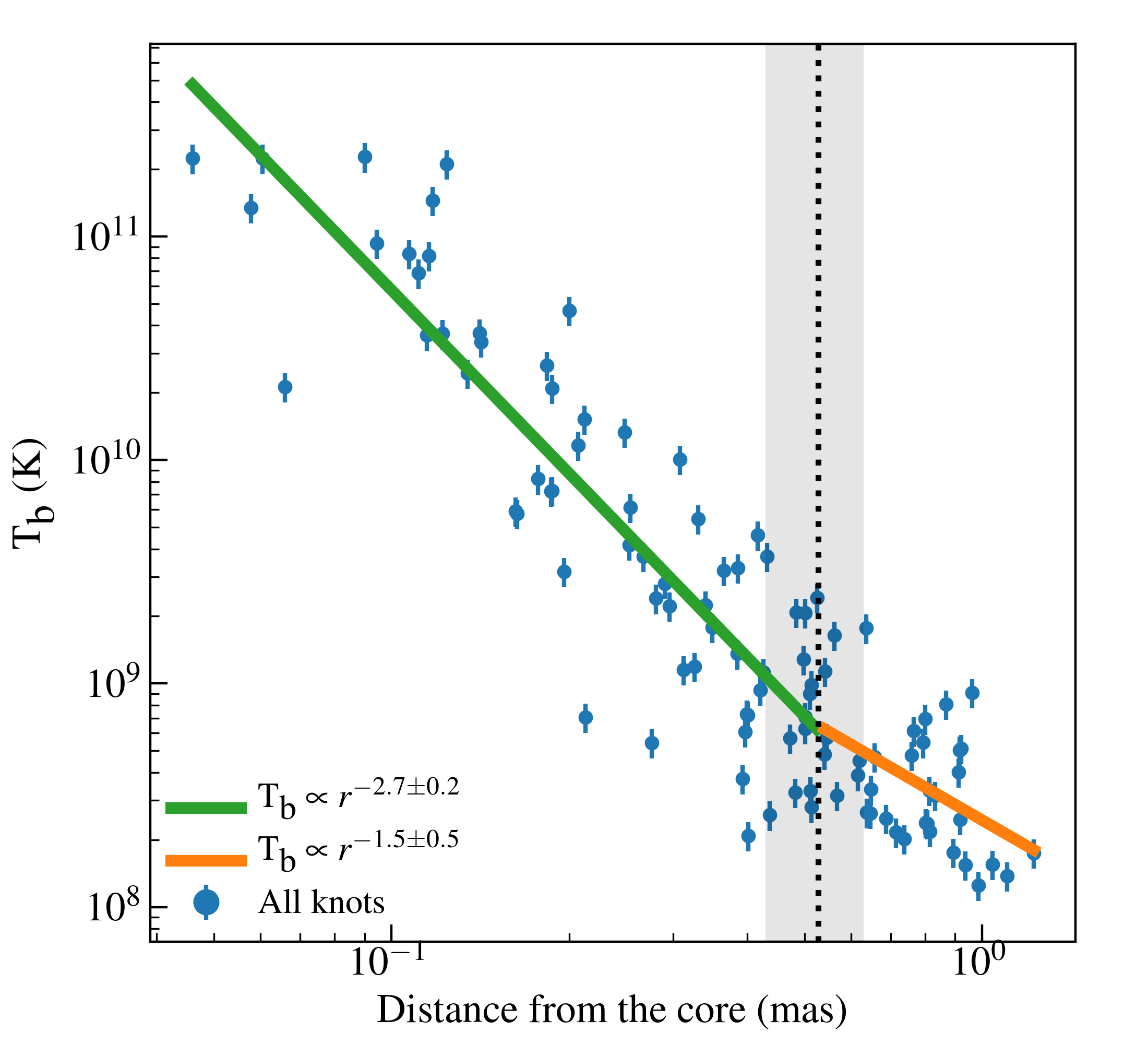} 
\caption{Observed brightness temperatures of all knots ($B1$--$B13$) as a function of the distance from the core. The analysis is the same as in Figure~\ref{fig:f8}, but using a power-law (T$_{\rm b} \propto r^{\epsilon}$). The uncertainties of the brightness temperatures were assumed to be 15\% \citep{2018A&A...610A..32B}.
}
\label{fig:f9}
\end{figure}

Figure~\ref{fig:f9} shows the distribution of the observed brightness temperatures versus radial distance  \citep{2005AJ....130.2473K, 2017ApJ...846...98J}. Seven outliers, which correspond to the unresolved jet components shown in Figure~\ref{fig:f8}, were excluded from the analysis. 
We fit power-law models (i.e., T$_{\rm b} \propto r^{\epsilon}$) to the brightness temperature data. 
To investigate any transition at the position of $A2$, we again divided the distribution in the same manner as in Figure~\ref{fig:f8}. The resultant power-law indices are $\epsilon = -2.7\pm0.2$ and $\epsilon =-1.5\pm0.5$ in the inner and outer regions, respectively. 
The power-law index within the inner region is consistent with the values obtained from the parsec-scale jets of other blazars \citep[e.g.,][]{2008ApJ...680..867K, 2015A&A...578A.123R, 2016MNRAS.462.2747K}. The notable change in the brightness temperature gradient at the location of $A2$ is an indication of a sudden change in the physical properties of the jet \citep[e.g.,][]{2005ApJ...631..169B, 2013AA...551A..32F, 2018A&A...610A..32B}.

\begin{figure}[!t]
\centering
\includegraphics[angle=0, width=0.45\textwidth, keepaspectratio]{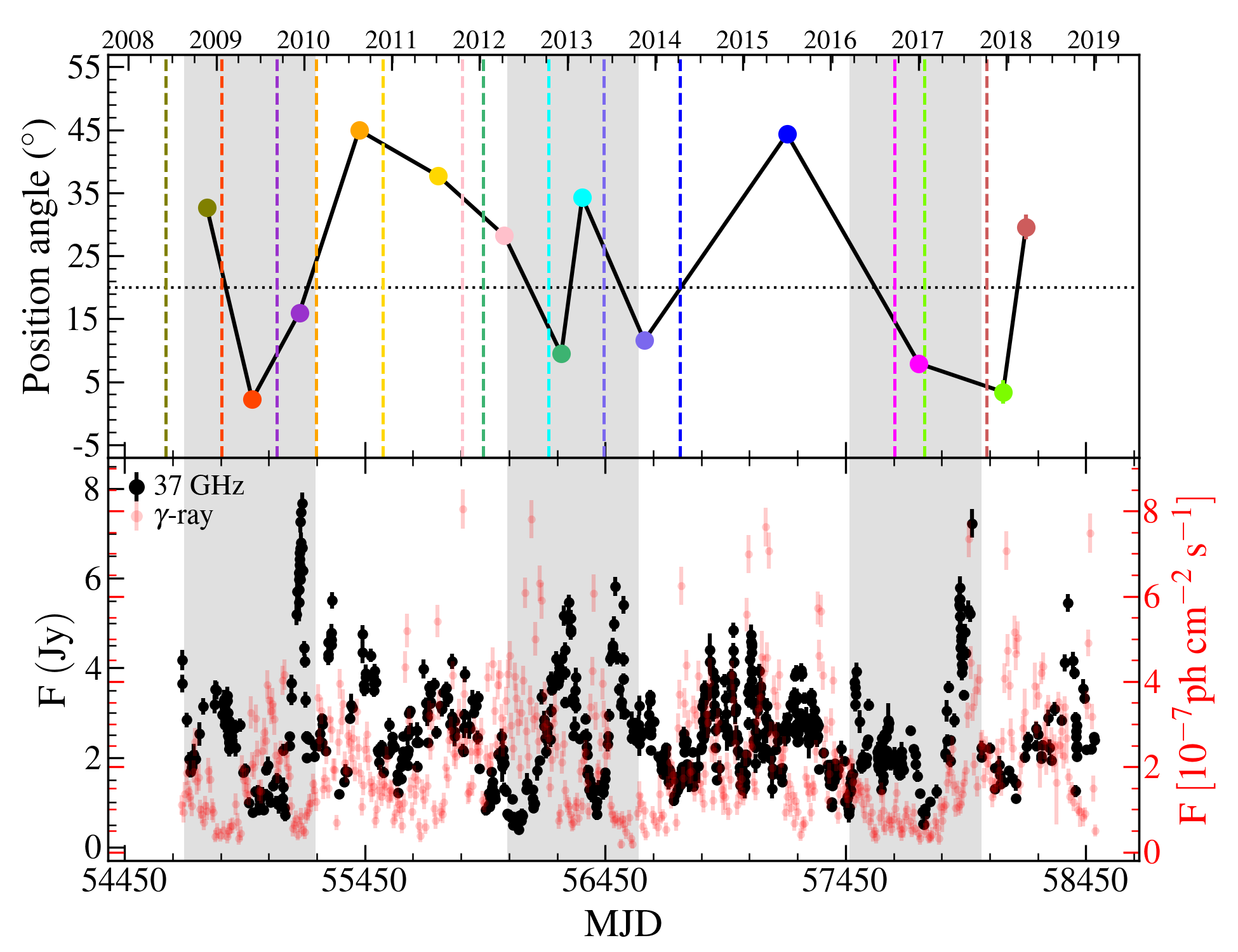} 
\caption{\textit{Top}: mean position angles of the jet knots at their mean epochs. Colors are the same as in Figure~\ref{fig:f7}. The vertical dashed lines indicate the ejection times of the knots. The horizontal dotted line marks 20$^{\circ}$. The gray shaded regions indicate the periods $T1$, $T2$, and $T3$, respectively. \textit{Bottom}: the observed radio and $\gamma$-ray light curves.}
\label{fig:f11}
\end{figure}

\citet{2014A&A...571L...2R} suggested a connection between the $\gamma$-ray emission and the inner jet position angle (near the core).
We investigated the variations in the mean position angle of the knots as shown in Figure~\ref{fig:f11}. For each knot ($B1$--$B13$), the mean position angle and its location in the time domain were determined by calculating the average values of the position angles and observing epochs (see the model-fit results in Table~\ref{app:buparam}).
Interestingly, the three periods with significant radio-to-$\gamma$-ray correlations (i.e., $T1$, $T2$, and $T3$) correspond to epochs of smaller mean position angles (below 20$^{\circ}$). This indicates that the global jet direction was more aligned toward North during those periods. Changes in the jet orientation can be due to propagation of jet components at different position angles \citep[][]{2011ApJ...733...11G, 2015ApJ...808..162C, 2017MNRAS.468.4992P}. This might determine the association of the knot with a $\gamma$-ray flare by changing the Doppler factor and viewing angle of the jet \citep[e.g.,][]{2015ApJ...813...51C}. Thus, we suggest that the northernmost position angles corresponds to closer alignment of the jet with the line of sight, with the corresponding beaming effects in turn leading to the correlated radio and $\gamma$-ray flares \citep[see also][for discussions of different beaming effects between the radio and $\gamma$-ray emitting regions]{2001ApJS..134..181J, 2014A&A...571L...2R, 2018ApJ...866..137L}.


\section{Discussion} \label{sec:dis}

\subsection{Interpretation of radio--\texorpdfstring{$\gamma$}{}-ray correlations and time lags} \label{sec:dis1}
The occasional occurrence of a significant correlation and/or anti-correlation between the radio and \emph{Fermi}-LAT 
$\gamma$-ray activity might indicate that the variability in these bands is initiated by the same physical process. 
However, causality arguments in combination with the typical $\lesssim$~day scale $\gamma$-ray variability time scales 
constrain the $\gamma$-ray emitting region to a size of $\lesssim 10^{16}$~cm. Such a compact emission region is generally 
found to be optically thick to synchrotron self-absorption (SSA) at radio wavelengths. Therefore, the $\gamma$-ray 
emission is likely to be produced up-stream from the 37~GHz radio core. This is in agreement with previous observations 
of $\gamma$-ray bright blazars which suggest that a time lag between $\gamma$-ray and radio emission is due to the
opacity effect \citep{2010ApJ...722L...7P, 2013A&A...552A..11R, 2016MNRAS.456..171R, 2018MNRAS.480.5517L}. In general, 
the $\gamma$-rays precede the radio emission, thus suggesting an upstream production of the $\gamma$-rays with respect 
to the radio emitting site \citep{2014MNRAS.441.1899F, 2014MNRAS.445..428M}.

Using VLBA observations, we found that all $\gamma$-ray flares which were correlated with radio flares, were 
accompanied by the ejection of a VLBI jet component: $B2$ and $B3$ in $T1$, $B8$ and $B9$ in $T2$, and $B12$ (or $B13$) 
in $T3$. The calculated ejection epochs ($T_{0}$) are expected to coincide with the time when the knot passes through 
the radio core. Thus, this could explain the estimated $T_{0}$ locations of the knots relative to the evolutionary 
stages of the 43\,GHz core flares.

We find that the ejection times of $B3$, $B8$, and $B9$ are located at the rising 
stage of a radio flare, whereas that of $B2$ is near the peak of a radio flare. However, the actual peaks of the radio 
flares can be a product of superposed underlying components \citep[e.g.,][]{2011A&A...532A.146L}, and thus care should 
be taken in the comparison. The position of $T_{0}$ in the radio light curves might be affected by variable beaming 
effects \citep[e.g.,][]{2018ApJ...866..137L} and/or complex trajectories of the knots. \citet{2013ApJ...768...40L} 
suggested that a shock wave moving down a helical 
path along the jet can explain the observed flaring features in S5~0716$+$714. This might play a crucial role for the 
individual knots, thus resulting in less accurate estimates of $T_{0}$ or distortion of the observed flaring features 
\citep[see also][for the significantly bent structure even at $\mu$as scales]{2020ApJ...893...68K}. Furthermore, there 
could be a substantial change in the velocity of the knots due to the interaction with the standing shocks 
\citep[e.g.,][]{2012A&A...537A..70S, 2018A&A...616A..93L} or a bent-jet structure \citep[e.g.,][]{2015A&A...578A.123R}. 
This suggests a transition from smaller velocities in a region near the radio core to higher velocities beyond this 
region. The knot $B2$ has the highest apparent velocity ($26\,c$) in our observations, and this might affect the 
estimation of its $T_{0}$ making it appear further downstream from its true ejection epoch.
Given the overall trend in $T_{0}$ of the knots (i.e., a position between the $\gamma$-ray and radio flares), however,
we suggest that these knots likely caused the notable radio flares in $T1$ and $T2$ by interacting with the radio core.

We therefore propose the following scenario to explain the marginal positve radio--$\gamma$-ray correlations with
leading $\gamma$-ray emission as well as the more significant anti-correlation with radio leading features: A new
jet component is being ejected from the central engine, accompanied by efficient relativistic particle acceleration,
possibly mediated by internal shocks
\citep[e.g.,][]{Spada01,Sokolov04,Sokolov05,Graff08,2010ApJ...711..445B,Chen11,Joshi11,Zhang15,Baring17,BB19}. 
Both high-frequency (mm--IR--optical) synchrotron radiation as well as external (e.g., IR from the dust torus) radiation fields are intense in that region and serve as efficient targets
for Compton scattering to produce a bright $\gamma$-ray flare, while the emission region is still opaque to SSA.
Subsequent radio flares are produced by the passage of the disturbance through the radio cores at the various
frequencies, with radio flares occurring earlier at the higher radio frequencies.

Visual inspection of the radio and $\gamma$-ray light curves suggests that the significant radio--$\gamma$-ray 
anti-correlations in $T1$ and $T2$ are due to long-lasting dips in the $\gamma$-ray light curves following radio
flares by a few -- few tens of days. We here suggest that this may be either due to efficient radiative cooling
of $\gamma$-ray emitting electrons or due to the emission region leaving the influence of dense target radiation
fields for Compton scattering.

The best-fit time delays may be used to estimate the corresponding propagation distances ($d_{\gamma \rm r}$) in terms 
of $\beta_{\rm app}$ (Table~\ref{tab:tb6}) for each period and frequency. We considered the mean $\beta_{\rm app}$ 
values of the knots ejected in the $T1$ and $T2$ periods separately: $B2$--$B3$ for $T1$ and $B8$--$B9$ for $T2$. 
In the case of $T3$ ($B11$--$B12$), $B13$ was used instead of $B11$ due to its closer timing to the flares. 
Using the relation $\Gamma \approx (1 + \beta_{\rm app}^{2})^{1/2}$ \citep[e.g.,][]{2012A&A...545A.113P, 2016A&A...586A..60K}, 
we obtained the following set of parameters: $\beta_{\rm app} = 23$ and $\theta_{\rm obs} = 2.5^{\circ}$ for $T1$, 
$\beta_{\rm app} = 16$ and $\theta_{\rm obs} = 3.6^{\circ}$ for $T2$, and $\beta_{\rm app} = 14$ and $\theta_{\rm obs} 
= 4.1^{\circ}$ for $T3$. With these estimates, we calculated the $d_{\gamma \rm r}$ values (see Table~\ref{tab:tb7}) 
using the relation \citep[e.g.,][]{2014MNRAS.445.1636R}: $d_{\gamma \rm r} = \frac{\beta_{\rm app} \, c \, 
\Delta t_{\rm r\gamma}}{\sin\theta \, (1+z)}$. The resulting propagation distances are listed in Table~\ref{tab:tb7}.

\begin{table}
\caption{The $d_{\gamma \rm r}$ values obtained from the delays of the significant correlations in $T1$, $T2$, and $T3$.
}
\label{tab:tb7}
\centering
\begin{tabular}{l c c c}   
\toprule
  &  $T1^{1}$  &  $T2^{1}$  &  $T3^{2}$  \\
\midrule
$d_{\gamma , \rm 15GHz}$ (pc)  &  $0.1\pm0.2$  &  $3.6\pm0.2$  &  $1.4\pm0.2$  \\
$d_{\gamma , \rm 37GHz}$ (pc)  &  $2.7\pm0.4$  &  $4.7\pm0.2$  &  $1.7\pm0.4$  \\
$d_{\gamma , \rm 230GHz}$ (pc)  &  $3.6\pm0.4$  &  $< 6.5^{3}$  &  -  \\
\bottomrule 
\multicolumn{4}{l}{$^1$ Negative correlation with the radio leading feature.}\\
\multicolumn{4}{l}{$^2$ Positive correlation with the $\gamma$-ray leading feature.}\\
\multicolumn{4}{l}{$^3$ considered to be overestimated due to the sparse sampling.}\\
\end{tabular}
\end{table}

For the flaring episode in $T3$, we can estimate the location of the $\gamma$-ray production site 
by assuming the location of the 15\,GHz radio core from the central engine to be 7--8\,pc as reported by 
\citet{2012A&A...545A.113P} and \citet{2018ARep...62..654B}. With the $d_{\gamma , \rm 15 \, GHz}$ value of 
1.5\,pc, the $\gamma$-ray site can be estimated to be located 5.5--6.5\,pc from the central engine. 
\citet{2015A&A...578A.123R} reported the location of the 43\,GHz core to be at $\sim$6.5\,pc. This might 
be explaining the occurrence of the radio and $\gamma$-ray flares in close proximity to each other in $T3$.

For $T1$ and $T2$, the delays of the positive correlations, with the $\gamma$-rays leading, decrease with higher frequencies, 
as expected. The delays in $T1$ and $T2$ at 15\,GHz are roughly $-$140 and $-$120\,days, respectively. From these 
delays, $d_{\gamma \rm r}$ can be estimated as 47 and 20 \,pc for $T1$ and $T2$, respectively. Given the locations 
of the 15 and 43\,GHz cores suggested by previous studies (i.e., 6--8\,pc), such large values seem to be highly 
overestimated. We suggest that this could be attributed to a complicated path (e.g., helices) of the moving 
disturbances and/or significant changes in $\beta_{\rm app}$ of the knots. Assuming the viewing angles found 
in Section~\ref{sec:dis1}, we found that the $d_{\gamma \rm r}$ values are comparable to the expected 
locations of the radio cores with the $\beta_{\rm app}$ values around 3.5 for $T1$ and 6.0 for $T2$. 
Thus, we suggest that strong moving disturbances produce the $\gamma$-ray flares in $T1$ and $T2$ at 
subparsec scales from the jet apex and undergo significant velocity changes (possibly with variable 
viewing angles) in their motions by the time they reach the radio cores \citep[see also][for a 
subparsec-scale origin of $\gamma$-rays in S5~0716$+$714]{2020ApJ...904...67G}.

Further physical insight will be gained by comparing the observed time delays to the radiative cooling
time scales of electrons emitting synchrotron at GHz frequencies ($\gamma \sim 100$) and Compton emission
at GeV $\gamma$-ray energies ($\gamma \sim 10^4$--$10^5$). These can be estimated as $t^{\rm obs}_{\rm rad} 
\approx 1.2 \times 10^5 \, B_{-1}^{-2} \, \delta_1^{-1} \, (1 + C)^{-1} \, \gamma^{-1}$~days, 
where $B = 0.1 \, B_{-1}$~G is the magnetic field, $\delta = 10 \, \delta_1$ the Doppler factor, 
and $C \equiv \nu F_{\nu}^C / \nu F_{\nu}^{sy} \sim 1$ 
is the Compton dominance parameter. This corresponds to observed cooling time scales of $\sim 60 \, B_{-1}^{-2}
\, \delta_1^{-1}$~days for radio-emitting electrons and $\lesssim 15 \, B_{-1}^{-2} \, \delta_1^{-1}$~hours for 
$\gamma$-ray emitting electrons. The latter time scale is well consistent with the $\sim$day scale 
$\gamma$-ray variability and suggests that there must be on-going particle acceleration during the flaring
periods that last several days.  
37-GHz-radio-emitting electrons (with Lorentz factor $\gamma'_{37} \approx 94 \, B_{-1}^{-1/2} \, \delta_1^{-1/2}$) 
that may have been co-accelerated with the $\gamma$-ray emitting electrons in the $\gamma$-ray emission region, 
will not have had time to cool significantly by the time they reach the radio emitting zone. Hence, if a $\gamma$-ray 
flaring event produces excess electrons both low ($\gamma \sim 100$) and high ($\gamma \gtrsim 10^4$) 
energies simultaneously, these excess radio emitting electrons will not have been affected by radiative
cooling by the time the emission region reaches the radio core (becoming optically thin to SSA), thus 
leading to positively correlated variability among the radio and $\gamma$-ray bands with
leading $\gamma$-ray activity and delay times decreasing with increasing frequencies.

The significant anti-correlations between radio and $\gamma$-ray activity, with radio flares preceding
dips in the $\gamma$-ray light curve by several -- several tens of days and time delays increasing with increasing 
frequencies, may be a consequence of efficient radiative cooling of $\gamma$-ray emitting electrons. 
The delay between the radio flares and the $\gamma$-ray dips could possibly originate from synchrotron self-Compton (SSC) 
cooling on the light crossing time scale through the radio core: as the radio flare evolves, the synchrotron photon field 
in the radio core builds up, leading to SSC emission (and cooling) of the high-energy electrons. Thus, if the observed
time delay represents the light-crossing time scale through the emission region, it would indicate a size of the 
radio core of $R_{\rm radio} \sim 2 \times 10^{17} \, \delta_1 \, \Delta t_1$~cm, where $\Delta t = 10 \, \Delta t_1$~days 
is the time delay. The resulting SSC emission 
is expected to emerge in X-rays, so this scenario would predict that the radio flare(s) should be correlated with X-ray 
flares. There have been a number of previous studies 
\citep[e.g.,][]{2009ApJ...706.1433V, 2014ApJ...783...83L, 2015A&A...578A.123R, 2018A&A...619A..45M} that report correlated 
radio/X-ray emission in S5~0716$+$714. Given sparse sampling of the X-rays \citep[e.g.,][]{2013A&A...552A..11R, 2015MNRAS.452L..11W} 
and the concave X-ray spectrum generated by both synchrotron and IC \citep[e.g.,][]{2014ApJ...783...83L, 2018A&A...619A..45M}, 
however, more detailed hard X-ray observations would be necessary for clarity. 
It is worth noting that the 230\,GHz data suffers relatively sparse sampling in $T2$. This prevents us from measuring the 
delay of the anti-correlation accurately in this period. Thus, we consider that the $d_{\gamma , \rm 230GHz}$ of $T2$ is 
significantly overestimated and might actually be smaller (e.g., around 5.0--5.5\,pc) given the opacity effect in the jet.

Alternatively, the motion of the emission region out of the core region of the AGN might result in the energy densities of
target radiation fields for Compton scattering dropping off rapidly. With the location of the 37~GHz radio core assumed to
be $\sim$7--8\,pc from the black hole, the radio--$\gamma$-ray delay then implies a distance of $\sim$10\,pc from the
central engine, where the dips in the $\gamma$-ray light curve are produced. This is a characteristic size of dusty, 
infrared-emitting tori of AGN, which have often been invoked as the source
of the dominant target photon field for Compton scattering in low- and intermediate-frequency peaked BL Lac objects \citep[e.g.,][]{2011ApJ...726...43A, 2012ApJ...751..159A, 2013ApJ...768...54B}. It is therefore quite plausible that the reason for the delayed $\gamma$-ray dips is the $\gamma$-ray emission region leaving the sphere of influence of the dust-torus radiation field.

\begin{figure}
\centering
\includegraphics[angle=0, width=0.45\textwidth, keepaspectratio]{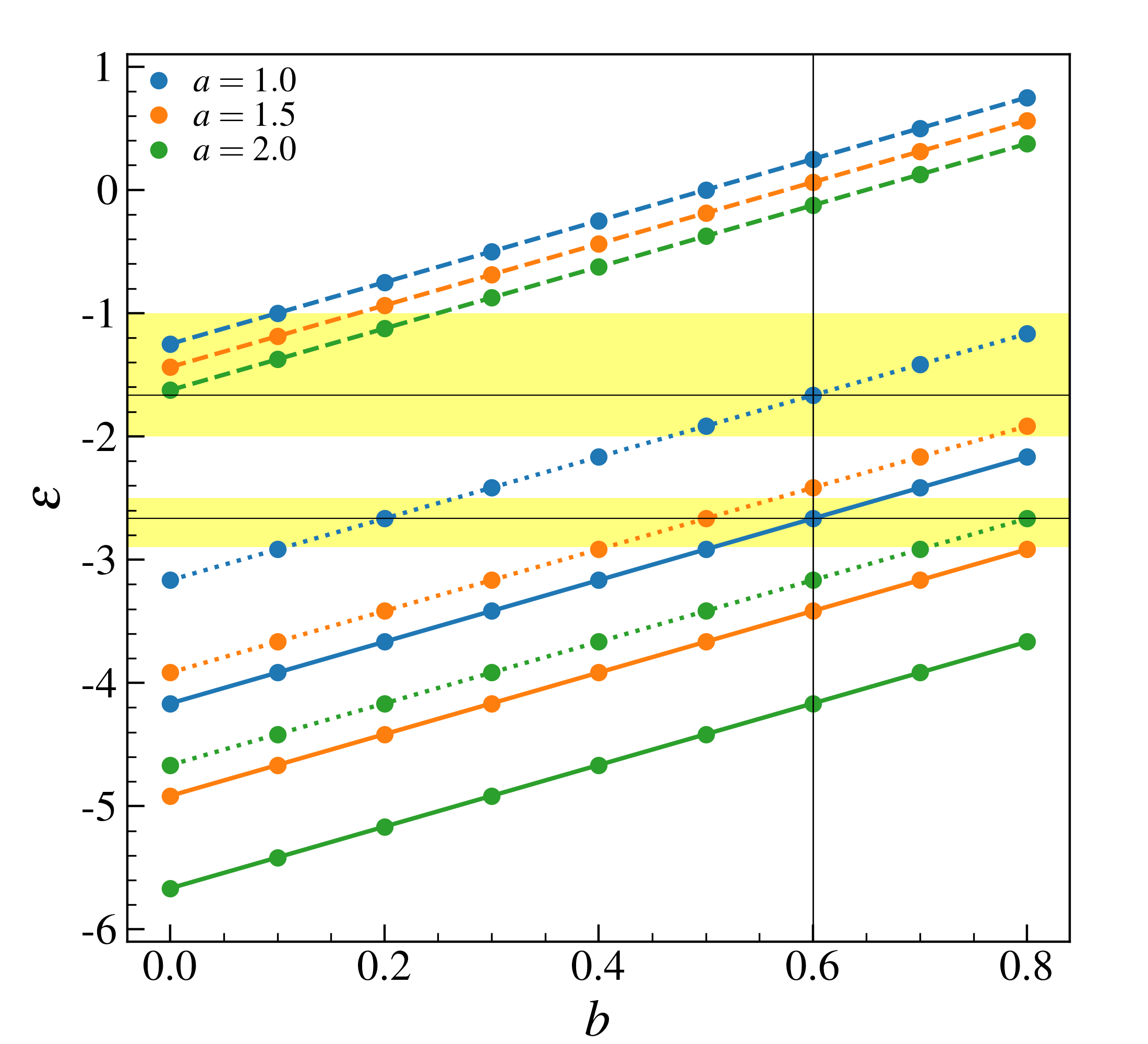} 
\caption{Calculated $\epsilon$ as a function of $b$. $a$ depends on the dominant magnetic field component: from 1 (toroidal) to 2 (poloidal). The dashed, dotted, and solid lines indicate $\epsilon_{c}$, $\epsilon_{a}$, and $\epsilon_{s}$ respectively. The yellow regions denote the power-law dependencies of the inner and outer regions with their uncertainties. The most probable estimates ($\epsilon_{s} = -2.7$ and $\epsilon_{a} = -1.7$) are marked by the vertical and horizontal lines.
}
\label{fig:f10}
\end{figure}

\subsection{Evolution of the parsec scale jet} \label{sec:dis4}
Considering the three primary evolutionary stages in the energy losses of the travelling shocks: Compton, synchrotron, and adiabatic \citep{1992vob..conf...85M}, \citet{2015A&A...578A.123R} derived the three power-law dependencies of the brightness temperature on the radial distance from the core \citep{1999ApJ...521..509L} in the 0716$+$714 jet: $\epsilon_{c} = [11-s-a\,(s+1)]/8 \,+\, [b\,(s+3)]/2 \,-\, 2$ for Compton loss, $\epsilon_{s} = -[4\,(s-1) + 3\,a\,(s+1)]/6 \,+\, [b\,(s+3)]/2 \,-\, 2$ for synchrotron loss, and $\epsilon_{a} = [2\,(5-2\,s) - 3\,a\,(s+1)]/6 \,+\, [b\,(s+3)]/2 \,-\, 2$ for adiabatic loss, where $a$, $b$, and $s$ are the power-law scaling indices of the magnetic field $B \propto r^{-a}$, Doppler factor $\delta \propto r^{b}$, and electron distribution $N(\gamma) \propto \gamma^{-s}$, respectively \citep[see also][for the case of the quasar 3C\,345]{2012A&A...537A..70S}. We employed this approach to interpret our results on the observed brightness temperature. $a$ can be set to be 1 or 2 for toroidal and poloidal magnetic fields, accordingly. However, we also considered an intermediate value of 1.5 for $s$ \citep{1999ApJ...521..509L}. Given the possible variations in Doppler factor \citep[e.g.,][]{2012A&A...537A..70S, 2015A&A...578A.123R}, we used $b$ with values between 0 and 1. Figure~\ref{fig:f10} shows the calculated $\epsilon$ with the typical value of $s=2$ (optically thin spectral index $\alpha \sim -0.5$ from $S_{\nu} \propto \nu^{\alpha}$). We find that the combination of $a=1$ and $b=0.6$ nicely corresponds to our observed power-law dependencies: $\epsilon_{s} = -2.7$ and $\epsilon_{a} = -1.7$ for the inner and outer regions, respectively. This implies that the propagating shocks suffer strong radiative cooling via synchrotron losses at first, then adiabatic losses become prominent as the shocks passes the standing shock $A2$.

Such changes in the brightness temperature gradient could be caused by a recollimation shock (RCS) and compression of the plasma flow, thus suggesting the stationary feature $A2$ being a RCS \citep{2008ApJ...680..867K}. One can expect an increase in the expansion rate of the propagating blobs beyond the RCS, which is supposed to be an overpressured region with respect to the nearby environment \citep[e.g.,][]{2018A&A...610A..32B}. This is consistent with our estimation above: $\epsilon \sim \epsilon_{a}$, after the $A2$ region. The variations in the component size shown in Figure~\ref{fig:f8}, however, indicate that the expansion became less pronounced after $A2$. This might be indicating that $A2$ is a standing feature induced by a bend, rather than recollimation \citep[e.g.,][]{2001ApJS..134..181J, 2015A&A...578A.123R}. The notable changes in the transverse motion and position angle beyond $A2$ in Figure~\ref{fig:f8}, support such a standing oblique shock scenario \citep[see][]{2001ApJS..134..181J, 2014MNRAS.445.1636R}. \citet{2013ApJ...768...40L} also concluded the stationary feature (their knot $K2$) belongs to a bending point accompanied by kinematic changes of a moving knot around the standing component. We suggest that collision with the ambient medium or magneto-hydrodynamical instabilities resulted in the formation of the structure.


\section{Summary} \label{sec:sum}
We have analyzed the long-term cm- and mm-radio and $\gamma$-ray light curves of the blazar 0716$+$714 to explore the connection between high- and low-energy radiation and to pinpoint the production site of the high-energy emission. By using VLBA observations, we were able to probe the parsec-scale jet activity in detail. Our primary conclusions are the following:

\begin{enumerate}

\item We found three significant correlation signals between the radio (15, 37, and 230\,GHz) and $\gamma$-ray light curves in 2008--2019. To identify the epochs from which these signals originate, we split the light curves into segments and found three  1.5\,yr-periods of interest: two ($T1$ and $T2$) with an anti-correlation and one ($T3$) with a positive correlation. All these correlations exceed significance levels of 99.9\% and peak at the time delays within 40\,days (i.e. radio leads $\gamma$-ray emission) in case of an negative correlation and $-$14\,days ($\gamma$-ray leads radio emission) in case of a positive correlation.


\item Using VLBA maps, we found three stationary jet features including the core, which dominates in the radio emission. The kinematic analysis revealed 14 jet components (including $B14$ which is excluded from the analysis) moving downstream the jet. Their apparent velocities range from 6 to 26$c$, and the average ejection rate is estimated of $\sim$0.8 years. Each of the three significant radio/$\gamma$-ray correlation signals was accompanied by two (for $T1$ and $T2$) or one (for $T3$) moving jet components. We found that these components emerged from the core when radio flares were growing (i.e., onset/rising/peaking).

\item The observed radio/$\gamma$-ray correlations can be attributed to jet components that are being newly ejected from the central engine. During their downstream propagation along the jet, efficient relativistic particle acceleration occurs, possibly mediated by internal shocks. Internal/external radiation fields are sufficient to produce a $\gamma$-ray flare. Subsequent flares at radio frequencies are expected to occur as the moving disturbance passes through the radio cores.

\item In each of the $T1$ and $T2$ periods, a marginal positive ($\gamma$-ray leading) radio/$\gamma$-ray correlation as well as a more significant anti-correlation (radio leading) were found. We suggest that these two correlation signals are connected: a $\gamma$-ray flare at subpc-scales and a subsequent radio flare/$\gamma$-ray dip at pc-scales, produced by a strong moving jet component. The $\gamma$-ray dip can be explained by efficient radiative cooling of $\gamma$-ray emitting electrons ($\gamma \sim 10^4$--$10^5$), possibly due to shock-compressed, stronger magnetic field in the emission region, or the emission region leaving the influence of dense radiation fields required for IC.

\item For the significant radio/$\gamma$-ray correlation found in $T3$, our estimates ($d_{\gamma \rm r}$) point toward a $\gamma$-ray production site located 5.5--6.5\,pc from the jet apex. Positively correlated variability between radio and $\gamma$-ray bands indicates that the radio emitting electrons have not enough time to cool down significantly by the time they reach the radio core where the jet becomes transparent at radio frequencies.

\item The overall motions of the jet components are complicated and show highly variable position angles, particularly within $\sim$0.4\,mas from the core. The observed brightness temperature decreases along the jet following a power-law. The power-law index is initially $-2.7\pm0.2$ and changes to $-1.5\pm0.5$ at the position of the stationary feature $A2$ ($\sim$0.53\,mas from the core). This indicates a change in the energy losses from being synchrotron-dominated to adiabatically-dominated.

\item We found that during the periods of significant radio/$\gamma$-ray correlations (i.e., $T1$/$T2$/$T3$) the average position angle of the jet components was smaller than about 20$^{\circ}$. This is consistent with a closer alignment of the jet with the line of sight when the jets points toward the north, leading to an enhancement of beaming effects. 

\end{enumerate}

\acknowledgments
DK acknowledges the support from the National Research Foundation of Korea (NRF) through the next generation fellowship 2019R1A6A3A13095962. ST and DK acknowledge the support from the NRF grant 2019R1F1A1059721.
The work of EVK is supported in the framework of the state project ``Science'' by the Ministry of Science and Higher Education of the Russian Federation under the contract 075-15-2020-778.
The work of MB is supported by the South African Research Chairs Initiative (SARChI) of the Department of Science and Innovation and the National Research Foundation\footnote{Any opinion, finding and conclusion or recommendation expressed in this material is that of the authors, and the NRF does not accept any liability in this regard.} of South Africa through SARChI grant UID 64789.
JLG acknowledges the support of the Spanish Ministerio de Econom\'{\i}a y Competitividad (grants AYA2016-80889-P, PID2019-108995GB-C21), the Consejer\'{\i}a de Econom\'{\i}a, Conocimiento, Empresas y Universidad of the Junta de Andaluc\'{\i}a (grant P18-FR-1769), the Consejo Superior de Investigaciones Cient\'{\i}ficas (grant 2019AEP112), and the State Agency for Research of the Spanish MCIU through the Center of Excellence Severo Ochoa award for the Instituto de Astrof\'{\i}sica de Andaluc\'{\i}a (SEV-2017-0709).
VR acknowledges the support from the FONDECYT postdoctoral grant 3190878.
The Submillimeter Array is a joint project between the Smithsonian Astrophysical Observatory and the Academia Sinica Institute of Astronomy and Astrophysics and is funded by the Smithsonian Institution and the Academia Sinica.
This publication makes use of facilities and data obtained at the Mets\"{a}hovi Radio Observatory, operated by Aalto University, Finland
This research has made use of data from the OVRO 40-m monitoring program \citep{richards_etal11}, supported by private funding from the California Insitute of Technology and the Max Planck Institute for Radio Astronomy, and by NASA grants NNX08AW31G, NNX11A043G, and NNX14AQ89G and NSF grants AST-0808050 and AST- 1109911.
This study makes use of 43 GHz VLBA data from the VLBA-BU Blazar Monitoring Program (VLBA-BU-BLAZAR;
http://www.bu.edu/blazars/VLBAproject.html), funded by NASA through the Fermi Guest Investigator Program. The VLBA is an instrument of the National Radio Astronomy Observatory. The National Radio Astronomy Observatory is a facility of the National Science Foundation operated by Associated Universities, Inc.
This work makes use of public Fermi data obtained from \textit{Fermi} Science Support Center (FSSC).
%
%
%
\textit{facilities}: OVRO, Mets\"{a}hovi, SMA, VLBA, \textit{Fermi}-LAT
\textit{software}: AIPS, Difmap, \textit{Fermi} Science Tools, Python


%
%

\appendix

\section{VLBI jet components}                  

\begin{longtable}{ccccccccccc}
\caption{Gaussian model-fit parameters of the BU-43\,GHz knots} \\
\hline
\hline
Date & MJD & Comp. & Flux & Distance$^{a}$ & Position angle$^{a}$ & Rel. RA & Rel. DEC & FWHM & T$_{\rm{b}}$ & UL$_{\rm T_{b}} ^{b}$ \\
 &  &  & (Jy) & (mas) & ($^{\circ}$) & (mas) & (mas) & (mas) & (K) & \\

\hline
\endfirsthead

\caption{Continued.} \\
\hline
\hline
Date & MJD & Comp. & Flux & Distance$^{a}$ & Position angle$^{a}$ & Rel. RA & Rel. DEC & FWHM & T$_{\rm{b}}$ & UL$_{\rm T_{b}} ^{b}$ \\
 &  &  & (Jy) & (mas) & ($^{\circ}$) & (mas) & (mas) & (mas) & (K) & \\
\hline
\endhead

2008.45 & 54629 & A0 & 1.23$\pm$0.06 &  &  & 0.000$\pm$0.005 & 0.00$\pm$0.01 & 0.039$\pm$0.007 & 6.1\,$\times$\,10$^{11}$ & F  \\
 &  & A1 & 0.06$\pm$0.01 & 0.12$\pm$0.01 & 38.1$\pm$2.0 & 0.076$\pm$0.006 & 0.10$\pm$0.01 & 0.020$\pm$0.011 & 1.0\,$\times$\,10$^{11}$ & T  \\
 &  & A2 & 0.05$\pm$0.01 & 0.57$\pm$0.09 & 11.5$\pm$2.5 & 0.115$\pm$0.048 & 0.56$\pm$0.10 & 0.181$\pm$0.035 & 1.2\,$\times$\,10$^{9}$ & F  \\ 
2008.51 & 54653 & A0 & 2.33$\pm$0.12 &  &  & 0.000$\pm$0.005 & 0.00$\pm$0.01 & 0.020$\pm$0.004 & 4.4\,$\times$\,10$^{12}$ & T  \\
 &  & B1 & 0.54$\pm$0.03 & 0.06$\pm$0.01 & 47.0$\pm$4.6 & 0.042$\pm$0.006 & 0.04$\pm$0.01 & 0.055$\pm$0.011 & 1.3\,$\times$\,10$^{11}$ & F  \\
 &  & A2 & 0.07$\pm$0.01 & 0.51$\pm$0.11 & 17.1$\pm$3.7 & 0.150$\pm$0.058 & 0.49$\pm$0.12 & 0.249$\pm$0.038 & 8.3\,$\times$\,10$^{8}$ & F  \\
2008.62 & 54693 & A0 & 3.85$\pm$0.19 &  &  & 0.000$\pm$0.005 & 0.00$\pm$0.01 & 0.047$\pm$0.006 & 1.3\,$\times$\,10$^{12}$ & F  \\
 &  & B1 & 0.12$\pm$0.01 & 0.13$\pm$0.01 & 37.4$\pm$1.8 & 0.077$\pm$0.005 & 0.10$\pm$0.01 & 0.020$\pm$0.009 & 2.3\,$\times$\,10$^{11}$ & T  \\
\end{longtable}

\bibliography{main}{}       

\begin{thebibliography}{}
\expandafter\ifx\csname natexlab\endcsname\relax\def\natexlab#1{#1}\fi
\providecommand{\url}[1]{\href{#1}{#1}}
\providecommand{\dodoi}[1]{doi:~\href{http://doi.org/#1}{\nolinkurl{#1}}}
\providecommand{\doeprint}[1]{\href{http://ascl.net/#1}{\nolinkurl{http://ascl.net/#1}}}
\providecommand{\doarXiv}[1]{\href{https://arxiv.org/abs/#1}{\nolinkurl{https://arxiv.org/abs/#1}}}

\bibitem[{{Abdo} {et~al.}(2010{\natexlab{a}}){Abdo}, {Ackermann}, {Ajello},
  {Allafort}, {Antolini}, {Atwood}, {Axelsson}, {Baldini}, {Ballet},
  {Barbiellini}, {Bastieri}, {Baughman}, {Bechtol}, {Bellazzini}, {Berenji},
  {Blandford}, {Bloom}, {Bogart}, {Bonamente}, {Borgland }, {Bouvier},
  {Bregeon}, {Brez}, {Brigida}, {Bruel}, {Buehler}, {Burnett}, {Buson},
  {Caliandro}, {Cameron}, {Cannon}, {Caraveo}, {Carrigan}, {Casand jian},
  {Cavazzuti}, {Cecchi}, {{\c{C}}elik}, {Celotti}, {Charles}, {Chekhtman},
  {Chen}, {Cheung}, {Chiang}, {Ciprini}, {Claus}, {Cohen-Tanugi}, {Conrad},
  {Costamante}, {Cotter}, {Cutini}, {D'Elia}, {Dermer}, {de Angelis}, {de
  Palma}, {De Rosa}, {Digel}, {Silva}, {Drell}, {Dubois}, {Dumora}, {Escande},
  {Farnier}, {Favuzzi}, {Fegan}, {Ferrara}, {Focke}, {Fortin}, {Frailis},
  {Fukazawa}, {Funk}, {Fusco}, {Gargano}, {Gasparrini}, {Gehrels}, {Germani},
  {Giebels}, {Giglietto}, {Giommi}, {Giordano}, {Giroletti}, {Glanzman},
  {Godfrey}, {Grandi}, {Grenier}, {Grondin}, {Grove}, {Guiriec}, {Hadasch},
  {Harding}, {Hayashida}, {Hays}, {Healey}, {Hill}, {Horan}, {Hughes},
  {Iafrate}, {Itoh}, {J{\'o}hannesson}, {Johnson}, {Johnson}, {Johnson},
  {Johnson}, {Kamae}, {Katagiri}, {Kataoka}, {Kawai}, {Kerr}, {Kn{\"o}dlseder},
  {Kuss}, {Lande}, {Latronico}, {Lavalley}, {Lemoine-Goumard}, {Llena Garde},
  {Longo}, {Loparco}, {Lott}, {Lovellette}, {Lubrano}, {Madejski}, {Makeev},
  {Malaguti}, {Massaro}, {Mazziotta}, {McConville}, {McEnery}, {McGlynn},
  {Michelson}, {Mitthumsiri}, {Mizuno}, {Moiseev}, {Monte}, {Monzani},
  {Morselli}, {Moskalenko}, {Murgia}, {Nolan}, {Norris}, {Nuss}, {Ohno},
  {Ohsugi}, {Omodei}, {Orlando}, {Ormes}, {Ozaki}, {Paneque}, {Panetta},
  {Parent}, {Pelassa}, {Pepe}, {Pesce-Rollins}, {Piranomonte}, {Piron},
  {Porter}, {Rain{\`o}}, {Rando}, {Razzano}, {Reimer}, {Reimer}, {Reposeur},
  {Ripken}, {Ritz}, {Rodriguez}, {Romani}, {Roth}, {Ryde}, {Sadrozinski},
  {Sanchez}, {Sander}, {Saz Parkinson}, {Scargle}, {Sgr{\`o}}, {Shaw},
  {Siskind}, {Smith}, {Spand re}, {Spinelli}, {Starck}, {Stawarz}, {Strickman},
  {Suson}, {Tajima}, {Takahashi}, {Takahashi}, {Tanaka}, {Taylor}, {Thayer},
  {Thayer}, {Thompson}, {Tibaldo}, {Torres}, {Tosti}, {Tramacere}, {Ubertini},
  {Uchiyama}, {Usher}, {Vasileiou}, {Vilchez}, {Villata}, {Vitale}, {Waite},
  {Wallace}, {Wang}, {Winer}, {Wood}, {Yang}, {Ylinen}, \&
  {Ziegler}}]{2010ApJ...715..429A}
{Abdo}, A.~A., {Ackermann}, M., {Ajello}, M., {et~al.} 2010{\natexlab{a}},
  \apj, 715, 429, \dodoi{10.1088/0004-637X/715/1/429}

\bibitem[{{Abdo} {et~al.}(2010{\natexlab{b}}){Abdo}, {Ackermann}, {Ajello},
  {Antolini}, {Baldini}, {Ballet}, {Barbiellini}, {Bastieri}, {Bechtol},
  {Bellazzini}, {Berenji}, {Blandford}, {Bloom}, {Bonamente}, {Borgland},
  {Bouvier}, {Bregeon}, {Brez}, {Brigida}, {Bruel}, {Buehler}, {Burnett},
  {Buson}, {Caliand ro}, {Cameron}, {Caraveo}, {Carrigan}, {Casandjian},
  {Cavazzuti}, {Cecchi}, {{\c{C}}elik}, {Chekhtman}, {Cheung}, {Chiang},
  {Ciprini}, {Claus}, {Cohen-Tanugi}, {Cominsky}, {Conrad}, {Costamante},
  {Cutini}, {Dermer}, {de Angelis}, {de Palma}, {Silva}, {Drell}, {Dubois},
  {Dumora}, {Farnier}, {Favuzzi}, {Fegan}, {Focke}, {Fortin}, {Frailis},
  {Fukazawa}, {Funk}, {Fusco}, {Gargano}, {Gasparrini}, {Gehrels}, {Germani},
  {Giebels}, {Giglietto}, {Giommi}, {Giordano}, {Glanzman}, {Godfrey},
  {Grenier}, {Grondin}, {Grove}, {Guiriec}, {Hadasch}, {Hayashida}, {Hays},
  {Healey}, {Horan}, {Hughes}, {Itoh}, {J{\'o}hannesson}, {Johnson}, {Johnson},
  {Kamae}, {Katagiri}, {Kataoka}, {Kawai}, {Kn{\"o}dlseder}, {Kuss}, {Lande},
  {Larsson}, {Latronico}, {Lemoine-Goumard}, {Longo}, {Loparco}, {Lott},
  {Lovellette}, {Lubrano}, {Madejski}, {Makeev}, {Massaro}, {Mazziotta},
  {McEnery}, {Michelson}, {Mitthumsiri}, {Mizuno}, {Moiseev}, {Monte},
  {Monzani}, {Morselli}, {Moskalenko}, {Mueller}, {Murgia}, {Nolan}, {Norris},
  {Nuss}, {Ohno}, {Ohsugi}, {Omodei}, {Orlando}, {Ormes}, {Ozaki}, {Panetta},
  {Parent}, {Pelassa}, {Pepe}, {Pesce-Rollins}, {Piron}, {Porter}, {Rain{\`o}},
  {Rando}, {Razzano}, {Reimer}, {Reimer}, {Ritz}, {Rodriguez}, {Romani},
  {Roth}, {Ryde}, {Sadrozinski}, {Sand er}, {Scargle}, {Sgr{\`o}}, {Shaw},
  {Smith}, {Spandre}, {Spinelli}, {Starck}, {Strickman}, {Suson}, {Takahashi},
  {Takahashi}, {Tanaka}, {Thayer}, {Thayer}, {Thompson}, {Tibaldo}, {Torres},
  {Tosti}, {Tramacere}, {Uchiyama}, {Usher}, {Vasileiou}, {Vilchez}, {Vitale},
  {Waite}, {Wallace}, {Wang}, {Winer}, {Wood}, {Yang}, {Ylinen}, \&
  {Ziegler}}]{2010ApJ...722..520A}
---. 2010{\natexlab{b}}, \apj, 722, 520, \dodoi{10.1088/0004-637X/722/1/520}

\bibitem[{{Abdo} {et~al.}(2011{\natexlab{a}}){Abdo}, {Ackermann}, {Ajello},
  {Antolini}, {Baldini}, {Ballet}, {Barbiellini}, {Bastieri}, {Bechtol},
  {Bellazzini}, {Berenji}, {Blandford}, {Bonamente}, {Borgland}, {Bregeon},
  {Brez}, {Brigida}, {Bruel}, {Buehler}, {Buson}, {Caliandro}, {Cameron},
  {Cannon}, {Caraveo}, {Carrigan}, {Casandjian}, {Cecchi}, {{\c C}elik},
  {Charles}, {Chekhtman}, {Cheung}, {Chiang}, {Ciprini}, {Claus},
  {Cohen-Tanugi}, {Conrad}, {Costamante}, {Cutini}, {Dermer}, {de Palma},
  {Donato}, {Silva}, {Drell}, {Dubois}, {Escande}, {Favuzzi}, {Fegan}, {Finke},
  {Focke}, {Fortin}, {Frailis}, {Fukazawa}, {Funk}, {Fusco}, {Gargano},
  {Gasparrini}, {Gehrels}, {Germani}, {Giglietto}, {Giordano}, {Giroletti},
  {Glanzman}, {Godfrey}, {Grenier}, {Guiriec}, {Hadasch}, {Hayashida}, {Hays},
  {Hughes}, {Itoh}, {J{\'o}hannesson}, {Johnson}, {Johnson}, {Kamae},
  {Katagiri}, {Kataoka}, {Kn{\"o}dlseder}, {Kuss}, {Lande}, {Larsson},
  {Latronico}, {Lee}, {Llena Garde}, {Longo}, {Loparco}, {Lott}, {Lovellette},
  {Lubrano}, {Makeev}, {Mazziotta}, {McEnery}, {Mehault}, {Michelson},
  {Mizuno}, {Monte}, {Monzani}, {Morselli}, {Moskalenko}, {Murgia}, {Nakamori},
  {Naumann-Godo}, {Nishino}, {Nolan}, {Norris}, {Nuss}, {Ohsugi}, {Okumura},
  {Omodei}, {Orlando}, {Ormes}, {Ozaki}, {Paneque}, {Panetta}, {Parent},
  {Pelassa}, {Pepe}, {Pesce-Rollins}, {Piron}, {Porter}, {Rain{\`o}}, {Rando},
  {Razzano}, {Reimer}, {Reimer}, {Ritz}, {Roth}, {Sadrozinski}, {Sanchez},
  {Sander}, {Schinzel}, {Sgr{\`o}}, {Siskind}, {Smith}, {Sokolovsky},
  {Spandre}, {Spinelli}, {Strickman}, {Suson}, {Takahashi}, {Tanaka}, {Thayer},
  {Thayer}, {Thompson}, {Tibaldo}, {Torres}, {Tosti}, {Tramacere}, {Uehara},
  {Usher}, {Vandenbroucke}, {Vasileiou}, {Vilchez}, {Vitale}, {Waite},
  {Wallace}, {Wang}, {Winer}, {Wood}, {Yang}, {Ylinen}, {Ziegler}, {Berdyugin},
  {Boettcher}, {Carrami{\~n}ana}, {Carrasco}, {de la Fuente}, {Diltz},
  {Hovatta}, {Kadenius}, {Kovalev}, {L{\"a}hteenm{\"a}ki}, {Lindfors},
  {Marscher}, {Nilsson}, {Pereira}, {Reinthal}, {Roustazadeh}, {Savolainen},
  {Sillanp{\"a}{\"a}}, {Takalo}, \& {Tornikoski}}]{2011ApJ...730..101A}
---. 2011{\natexlab{a}}, \apj, 730, 101, \dodoi{10.1088/0004-637X/730/2/101}

\bibitem[{{Abdo} {et~al.}(2011{\natexlab{b}}){Abdo}, {Ackermann}, {Ajello},
  {Baldini}, {Ballet}, {Barbiellini}, {Bastieri}, {Bechtol}, {Bellazzini},
  {Berenji}, {Blandford}, {Bonamente}, {Borgland}, {Bouvier}, {Bregeon},
  {Brez}, {Brigida}, {Bruel}, {Buehler}, {Buson}, {Caliandro}, {Cameron},
  {Caraveo}, {Carrigan}, {Casandjian}, {Cavazzuti}, {Cecchi}, {{\c{C}}elik},
  {Charles}, {Chekhtman}, {Cheung}, {Chiang}, {Ciprini}, {Claus},
  {Cohen-Tanugi}, {Conrad}, {Costamante}, {Cutini}, {Davis}, {Dermer}, {de
  Palma}, {Digel}, {do Couto e Silva}, {Drell}, {Dubois}, {Dumora}, {Favuzzi},
  {Fegan}, {Fortin}, {Frailis}, {Fuhrmann}, {Fukazawa}, {Funk}, {Fusco},
  {Gargano}, {Gasparrini}, {Gehrels}, {Germani}, {Giglietto}, {Giommi},
  {Giordano}, {Giroletti}, {Glanzman}, {Godfrey}, {Grenier}, {Grove},
  {Guillemot}, {Guiriec}, {Hadasch}, {Hayashida}, {Hays}, {Horan}, {Hughes},
  {Itoh}, {J{\'o}hannesson}, {Johnson}, {Johnson}, {Johnson}, {Kamae},
  {Katagiri}, {Kataoka}, {Kn{\"o}dlseder}, {Kuss}, {Lande}, {Latronico}, {Lee},
  {Longo}, {Loparco}, {Lott}, {Lovellette}, {Lubrano}, {Makeev}, {Mazziotta},
  {McEnery}, {Mehault}, {Michelson}, {Mizuno}, {Moiseev}, {Monte}, {Monzani},
  {Morselli}, {Moskalenko}, {Murgia}, {Nakamori}, {Naumann-Godo}, {Nestoras},
  {Nolan}, {Norris}, {Nuss}, {Ohsugi}, {Okumura}, {Omodei}, {Orlando}, {Ormes},
  {Ozaki}, {Paneque}, {Panetta}, {Parent}, {Pelassa}, {Pepe}, {Pesce-Rollins},
  {Piron}, {Porter}, {Rain{\`o}}, {Rando}, {Razzano}, {Reimer}, {Reimer},
  {Reyes}, {Ripken}, {Ritz}, {Romani}, {Roth}, {Sadrozinski}, {Sanchez},
  {Sander}, {Scargle}, {Sgr{\`o}}, {Shaw}, {Smith}, {Spandre}, {Spinelli},
  {Strickman}, {Suson}, {Takahashi}, {Tanaka}, {Thayer}, {Thayer}, {Thompson},
  {Tibaldo}, {Torres}, {Tosti}, {Tramacere}, {Usher}, {Vandenbroucke},
  {Vasileiou}, {Vilchez}, {Vitale}, {Waite}, {Wang}, {Winer}, {Wood}, {Yang},
  {Ylinen}, {Ziegler}, {Acciari}, {Aliu}, {Arlen}, {Aune}, {Beilicke},
  {Benbow}, {B{\"o}ttcher}, {Boltuch}, {Bradbury}, {Buckley}, {Bugaev},
  {Byrum}, {Cannon}, {Cesarini}, {Christiansen}, {Ciupik}, {Cui}, {de la Calle
  Perez}, {Dickherber}, {Errando}, {Falcone}, {Finley}, {Finnegan}, {Fortson},
  {Furniss}, {Galante}, {Gall}, {Gillanders}, {Godambe}, {Grube}, {Guenette},
  {Gyuk}, {Hanna}, {Holder}, {Hui}, {Humensky}, {Imran}, {Kaaret}, {Karlsson},
  {Kertzman}, {Kieda}, {Konopelko}, {Krawczynski}, {Krennrich}, {Lang},
  {LeBohec}, {Maier}, {McArthur}, {McCann}, {McCutcheon}, {Moriarty},
  {Mukherjee}, {Ong}, {Otte}, {Pandel}, {Perkins}, {Pichel}, {Pohl}, {Quinn},
  {Ragan}, {Reynolds}, {Roache}, {Rose}, {Schroedter}, {Sembroski}, {Senturk},
  {Smith}, {Steele}, {Swordy}, {Te{\v{s}}i{\'c}}, {Theiling}, {Thibadeau},
  {Varlotta}, {Vassiliev}, {Vincent}, {Wakely}, {Ward}, {Weekes}, {Weinstein},
  {Weisgarber}, {Williams}, {Wissel}, {Wood}, {Villata}, {Raiteri}, {Gurwell},
  {Larionov}, {Kurtanidze}, {Aller}, {L{\"a}hteenm{\"a}ki}, {Chen},
  {Berduygin}, {Agudo}, {Aller}, {Arkharov}, {Bach}, {Bachev}, {Beltrame},
  {Ben{\'\i}tez}, {Buemi}, {Dashti}, {Calcidese}, {Capezzali}, {Carosati}, {Da
  Rio}, {Di Paola}, {Diltz}, {Dolci}, {Dultzin}, {Forn{\'e}}, {G{\'o}mez},
  {Hagen-Thorn}, {Halkola}, {Heidt}, {Hiriart}, {Hovatta}, {Hsiao}, {Jorstad},
  {Kimeridze}, {Konstantinova}, {Kopatskaya}, {Koptelova}, {Leto}, {Ligustri},
  {Lindfors}, {Lopez}, {Marscher}, {Mommert}, {Mujica}, {Nikolashvili},
  {Nilsson}, {Palma}, {Pasanen}, {Roca-Sogorb}, {Ros}, {Roustazadeh}, {Sadun},
  {Saino}, {Sigua}, {Sillan{\"a}{\"a}}, {Sorcia}, {Takalo}, {Tornikoski},
  {Trigilio}, {Turchetti}, {Umana}, {Belloni}, {Blake}, {Bloom}, {Angelakis},
  {Fumagalli}, {Hauser}, {Prochaska}, {Riquelme}, {Sievers}, {Starr},
  {Tagliaferri}, {Ungerechts}, {Wagner}, {Zensus}, {Fermi LAT Collaboration},
  {VERITAS Collaboration}, \& {GASP-WEBT Consortium}}]{2011ApJ...726...43A}
---. 2011{\natexlab{b}}, \apj, 726, 43, \dodoi{10.1088/0004-637X/726/1/43}

\bibitem[{{Acero} {et~al.}(2015){Acero}, {Ackermann}, {Ajello}, {Albert},
  {Atwood}, {Axelsson}, {Baldini}, {Ballet}, {Barbiellini}, {Bastieri},
  {Belfiore}, {Bellazzini}, {Bissaldi}, {Blandford}, {Bloom}, {Bogart},
  {Bonino}, {Bottacini}, {Bregeon}, {Britto}, {Bruel}, {Buehler}, {Burnett},
  {Buson}, {Caliandro}, {Cameron}, {Caputo}, {Caragiulo}, {Caraveo},
  {Casandjian}, {Cavazzuti}, {Charles}, {Chaves}, {Chekhtman}, {Cheung},
  {Chiang}, {Chiaro}, {Ciprini}, {Claus}, {Cohen-Tanugi}, {Cominsky}, {Conrad},
  {Cutini}, {D'Ammando}, {de Angelis}, {DeKlotz}, {de Palma}, {Desiante},
  {Digel}, {Di Venere}, {Drell}, {Dubois}, {Dumora}, {Favuzzi}, {Fegan},
  {Ferrara}, {Finke}, {Franckowiak}, {Fukazawa}, {Funk}, {Fusco}, {Gargano},
  {Gasparrini}, {Giebels}, {Giglietto}, {Giommi}, {Giordano}, {Giroletti},
  {Glanzman}, {Godfrey}, {Grenier}, {Grondin}, {Grove}, {Guillemot}, {Guiriec},
  {Hadasch}, {Harding}, {Hays}, {Hewitt}, {Hill}, {Horan}, {Iafrate}, {Jogler},
  {J{\'o}hannesson}, {Johnson}, {Johnson}, {Johnson}, {Johnson}, {Kamae},
  {Kataoka}, {Katsuta}, {Kuss}, {La Mura}, {Landriu}, {Larsson}, {Latronico},
  {Lemoine-Goumard}, {Li}, {Li}, {Longo}, {Loparco}, {Lott}, {Lovellette},
  {Lubrano}, {Madejski}, {Massaro}, {Mayer}, {Mazziotta}, {McEnery},
  {Michelson}, {Mirabal}, {Mizuno}, {Moiseev}, {Mongelli}, {Monzani},
  {Morselli}, {Moskalenko}, {Murgia}, {Nuss}, {Ohno}, {Ohsugi}, {Omodei},
  {Orienti}, {Orlando}, {Ormes}, {Paneque}, {Panetta}, {Perkins},
  {Pesce-Rollins}, {Piron}, {Pivato}, {Porter}, {Racusin}, {Rando}, {Razzano},
  {Razzaque}, {Reimer}, {Reimer}, {Reposeur}, {Rochester}, {Romani},
  {Salvetti}, {S{\'a}nchez-Conde}, {Saz Parkinson}, {Schulz}, {Siskind},
  {Smith}, {Spada}, {Spandre}, {Spinelli}, {Stephens}, {Strong}, {Suson},
  {Takahashi}, {Takahashi}, {Tanaka}, {Thayer}, {Thayer}, {Thompson},
  {Tibaldo}, {Tibolla}, {Torres}, {Torresi}, {Tosti}, {Troja}, {Van Klaveren},
  {Vianello}, {Winer}, {Wood}, {Wood}, {Zimmer}, \& {Fermi-LAT
  Collaboration}}]{2015ApJS..218...23A}
{Acero}, F., {Ackermann}, M., {Ajello}, M., {et~al.} 2015, \apjs, 218, 23,
  \dodoi{10.1088/0067-0049/218/2/23}

\bibitem[{{Ackermann} {et~al.}(2012){Ackermann}, {Ajello}, {Ballet},
  {Barbiellini}, {Bastieri}, {Bellazzini}, {Blandford}, {Bloom}, {Bonamente},
  {Borgland}, {Bottacini}, {Bregeon}, {Brigida}, {Bruel}, {Buehler}, {Buson},
  {Caliandro}, {Cameron}, {Caraveo}, {Casandjian}, {Cavazzuti}, {Cecchi},
  {Charles}, {Chekhtman}, {Chiang}, {Ciprini}, {Claus}, {Cohen-Tanugi},
  {Cutini}, {D'Ammando}, {de Palma}, {Dermer}, {Silva}, {Drell},
  {Drlica-Wagner}, {Dubois}, {Favuzzi}, {Fegan}, {Ferrara}, {Focke}, {Fortin},
  {Fuhrmann}, {Fukazawa}, {Fusco}, {Gargano}, {Gasparrini}, {Gehrels},
  {Germani}, {Giglietto}, {Giommi}, {Giordano}, {Giroletti}, {Glanzman},
  {Godfrey}, {Grenier}, {Guiriec}, {Hadasch}, {Hayashida}, {Hughes}, {Itoh},
  {J{\'o}hannesson}, {Johnson}, {Katagiri}, {Kataoka}, {Kn{\"o}dlseder},
  {Kuss}, {Lande}, {Larsson}, {Lee}, {Longo}, {Loparco}, {Lott}, {Lovellette},
  {Lubrano}, {Madejski}, {Mazziotta}, {McEnery}, {Mehault}, {Michelson},
  {Mitthumsiri}, {Mizuno}, {Monte}, {Monzani}, {Morselli}, {Moskalenko},
  {Murgia}, {Naumann-Godo}, {Nishino}, {Norris}, {Nuss}, {Ohsugi}, {Okumura},
  {Omodei}, {Orlando}, {Ozaki}, {Paneque}, {Panetta}, {Pelassa},
  {Pesce-Rollins}, {Pierbattista}, {Piron}, {Pivato}, {Porter}, {Rain{\`o}},
  {Rando}, {Rastawicki}, {Razzano}, {Readhead}, {Reimer}, {Reimer}, {Reyes},
  {Richards}, {Sbarra}, {Sgr{\`o}}, {Siskind}, {Spandre}, {Spinelli},
  {Szostek}, {Takahashi}, {Tanaka}, {Thayer}, {Thayer}, {Thompson},
  {Tinivella}, {Torres}, {Tosti}, {Troja}, {Usher}, {Vandenbroucke},
  {Vasileiou}, {Vianello}, {Vitale}, {Waite}, {Winer}, {Wood}, {Yang},
  {Zimmer}, {Fermi-LAT Collaboration}, {Moderski}, {Nalewajko}, {Sikora},
  {Villata}, {Raiteri}, {Aller}, {Aller}, {Arkharov}, {Ben{\'\i}tez},
  {Berdyugin}, {Blinov}, {Boettcher}, {Bravo Calle}, {Buemi}, {Carosati},
  {Chen}, {Diltz}, {Di Paola}, {Dolci}, {Efimova}, {Forn{\'e}}, {Gurwell},
  {Heidt}, {Hiriart}, {Jordan}, {Kimeridze}, {Konstantinova}, {Kopatskaya},
  {Koptelova}, {Kurtanidze}, {L{\"a}hteenm{\"a}ki}, {Larionova}, {Larionova},
  {Larionov}, {Leto}, {Lindfors}, {Lin}, {Morozova}, {Nikolashvili}, {Nilsson},
  {Oksman}, {Roustazadeh}, {Sievers}, {Sigua}, {Sillanp{\"a}{\"a}},
  {Takahashi}, {Takalo}, {Tornikoski}, {Trigilio}, {Troitsky}, {Umana},
  {GASP-WEBT Consortium}, {Angelakis}, {Krichbaum}, {Nestoras}, {Riquelme},
  {F-GAMMA}, {Krips}, {Trippe}, {Iram-PdBI}, {Arai}, {Kawabata}, {Sakimoto},
  {Sasada}, {Sato}, {Uemura}, {Yamanaka}, {Yoshida}, {Kanata}, {Belloni},
  {Tagliaferri}, {RXTE}, {Bonning}, {Isler}, {Urry}, {SMARTS}, {Hoversten},
  {Falcone}, {Pagani}, {Stroh}, \& {(Swift-XRT}}]{2012ApJ...751..159A}
{Ackermann}, M., {Ajello}, M., {Ballet}, J., {et~al.} 2012, \apj, 751, 159,
  \dodoi{10.1088/0004-637X/751/2/159}

\bibitem[{{Angel} \& {Stockman}(1980)}]{1980ARA&A..18..321A}
{Angel}, J.~R.~P., \& {Stockman}, H.~S. 1980, \araa, 18, 321,
  \dodoi{10.1146/annurev.aa.18.090180.001541}

\bibitem[{{Angelakis} {et~al.}(2019){Angelakis}, {Fuhrmann}, {Myserlis},
  {Zensus}, {Nestoras}, {Karamanavis}, {Marchili}, {Krichbaum}, {Kraus}, \&
  {Rachen}}]{2019A&A...626A..60A}
{Angelakis}, E., {Fuhrmann}, L., {Myserlis}, I., {et~al.} 2019, \aap, 626, A60,
  \dodoi{10.1051/0004-6361/201834363}

\bibitem[{{Bach} {et~al.}(2006){Bach}, {Krichbaum}, {Kraus}, {Witzel}, \&
  {Zensus}}]{2006AA...452...83B}
{Bach}, U., {Krichbaum}, T.~P., {Kraus}, A., {Witzel}, A., \& {Zensus}, J.~A.
  2006, \aap, 452, 83, \dodoi{10.1051/0004-6361:20053943}

\bibitem[{{Bach} {et~al.}(2005){Bach}, {Krichbaum}, {Ros}, {Britzen}, {Tian},
  {Kraus}, {Witzel}, \& {Zensus}}]{2005AA...433..815B}
{Bach}, U., {Krichbaum}, T.~P., {Ros}, E., {et~al.} 2005, \aap, 433, 815,
  \dodoi{10.1051/0004-6361:20040388}

\bibitem[{{Baring} {et~al.}(2017){Baring}, {B{\"o}ttcher}, \&
  {Summerlin}}]{Baring17}
{Baring}, M.~G., {B{\"o}ttcher}, M., \& {Summerlin}, E.~J. 2017, \mnras, 464,
  4875, \dodoi{10.1093/mnras/stw2344}

\bibitem[{{Bennett} {et~al.}(2014){Bennett}, {Larson}, {Weiland}, \&
  {Hinshaw}}]{2014ApJ...794..135B}
{Bennett}, C.~L., {Larson}, D., {Weiland}, J.~L., \& {Hinshaw}, G. 2014, \apj,
  794, 135, \dodoi{10.1088/0004-637X/794/2/135}

\bibitem[{{Berton} {et~al.}(2018){Berton}, {Liao}, {La Mura},
  {J{\"a}rvel{\"a}}, {Congiu}, {Foschini}, {Frezzato}, {Ramakrishnan}, {Fan},
  {L{\"a}hteenm{\"a}ki}, {Pursimo}, {Abate}, {Bai}, {Calcidese}, {Ciroi},
  {Chen}, {Cracco}, {Li}, {Tornikoski}, \& {Rafanelli}}]{2018A&A...614A.148B}
{Berton}, M., {Liao}, N.~H., {La Mura}, G., {et~al.} 2018, \aap, 614, A148,
  \dodoi{10.1051/0004-6361/201731625}

\bibitem[{{Beuchert} {et~al.}(2018){Beuchert}, {Kadler}, {Perucho},
  {Gro{\ss}berger}, {Schulz}, {Agudo}, {Casadio}, {G{\'o}mez}, {Gurwell},
  {Homan}, {Kovalev}, {Lister}, {Markoff}, {Molina}, {Pushkarev}, {Ros},
  {Savolainen}, {Steinbring}, {Thum}, \& {Wilms}}]{2018A&A...610A..32B}
{Beuchert}, T., {Kadler}, M., {Perucho}, M., {et~al.} 2018, \aap, 610, A32,
  \dodoi{10.1051/0004-6361/201731952}

\bibitem[{{Bhatta} {et~al.}(2013){Bhatta}, {Webb}, {Hollingsworth}, {Dhalla},
  {Khanuja}, {Bachev}, {Blinov}, {B{\"o}ttcher}, {Bravo Calle}, {Calcidese},
  {Capezzali}, {Carosati}, {Chigladze}, {Collins}, {Coloma}, {Efimov}, {Gupta},
  {Hu}, {Kurtanidze}, {Lamerato}, {Larionov}, {Lee}, {Lindfors}, {Murphy},
  {Nilsson}, {Ohlert}, {Oksanen}, {P{\"a}{\"a}kk{\"o}nen}, {Pollock}, {Rani},
  {Reinthal}, {Rodriguez}, {Ros}, {Roustazadeh}, {Sagar}, {Sanchez}, {Shastri},
  {Sillanp{\"a}{\"a}}, {Strigachev}, {Takalo}, {Vennes}, {Villata},
  {Villforth}, {Wu}, \& {Zhou}}]{2013A&A...558A..92B}
{Bhatta}, G., {Webb}, J.~R., {Hollingsworth}, H., {et~al.} 2013, \aap, 558,
  A92, \dodoi{10.1051/0004-6361/201220236}

\bibitem[{{B{\"o}ttcher} \& {Baring}(2019)}]{BB19}
{B{\"o}ttcher}, M., \& {Baring}, M.~G. 2019, \apj, 887, 133,
  \dodoi{10.3847/1538-4357/ab552a}

\bibitem[{{B{\"o}ttcher} \& {Chiang}(2002)}]{2002ApJ...581..127B}
{B{\"o}ttcher}, M., \& {Chiang}, J. 2002, \apj, 581, 127,
  \dodoi{10.1086/344155}

\bibitem[{{B{\"o}ttcher} \& {Dermer}(2010)}]{2010ApJ...711..445B}
{B{\"o}ttcher}, M., \& {Dermer}, C.~D. 2010, \apj, 711, 445,
  \dodoi{10.1088/0004-637X/711/1/445}

\bibitem[{{B{\"o}ttcher} {et~al.}(2013){B{\"o}ttcher}, {Reimer}, {Sweeney}, \&
  {Prakash}}]{2013ApJ...768...54B}
{B{\"o}ttcher}, M., {Reimer}, A., {Sweeney}, K., \& {Prakash}, A. 2013, \apj,
  768, 54, \dodoi{10.1088/0004-637X/768/1/54}

\bibitem[{{B{\"o}ttcher} {et~al.}(2005){B{\"o}ttcher}, {Harvey}, {Joshi},
  {Villata}, {Raiteri}, {Bramel}, {Mukherjee}, {Savolainen}, {Cui}, {Fossati},
  {Smith}, {Able}, {Aller}, {Aller}, {Arkharov}, {Augusteijn}, {Baliyan},
  {Barnaby}, {Berdyugin}, {Ben{\'\i}tez}, {Boltwood}, {Carini}, {Carosati},
  {Ciprini}, {Coloma}, {Crapanzano}, {de Diego}, {Di Paola}, {Dolci}, {Fan},
  {Frasca}, {Hagen-Thorn}, {Horan}, {Ibrahimov}, {Kimeridze}, {Kovalev},
  {Kovalev}, {Kurtanidze}, {L{\"a}hteenm{\"a}ki}, {Lanteri}, {Larionov},
  {Larionova}, {Lindfors}, {Marilli}, {Mirabal}, {Nikolashvili}, {Nilsson},
  {Ohlert}, {Ohnishi}, {Oksanen}, {Ostorero}, {Oyer}, {Papadakis}, {Pasanen},
  {Poteet}, {Pursimo}, {Sadakane}, {Sigua}, {Takalo}, {Tartar},
  {Ter{\"a}sranta}, {Tosti}, {Walters}, {Wiik}, {Wilking}, {Wills}, {Xilouris},
  {Fletcher}, {Gu}, {Lee}, {Pak}, \& {Yim}}]{2005ApJ...631..169B}
{B{\"o}ttcher}, M., {Harvey}, J., {Joshi}, M., {et~al.} 2005, \apj, 631, 169,
  \dodoi{10.1086/432609}

\bibitem[{{Butuzova}(2018)}]{2018ARep...62..654B}
{Butuzova}, M.~S. 2018, Astronomy Reports, 62, 654,
  \dodoi{10.1134/S1063772918100037}

\bibitem[{{Casadio} {et~al.}(2015{\natexlab{a}}){Casadio}, {G{\'o}mez},
  {Grandi}, {Jorstad}, {Marscher}, {Lister}, {Kovalev}, {Savolainen}, \&
  {Pushkarev}}]{2015ApJ...808..162C}
{Casadio}, C., {G{\'o}mez}, J.~L., {Grandi}, P., {et~al.} 2015{\natexlab{a}},
  \apj, 808, 162, \dodoi{10.1088/0004-637X/808/2/162}

\bibitem[{{Casadio} {et~al.}(2015{\natexlab{b}}){Casadio}, {G{\'o}mez},
  {Jorstad}, {Marscher}, {Larionov}, {Smith}, {Gurwell}, {L{\"a}hteenm{\"a}ki},
  {Agudo}, {Molina}, {Bala}, {Joshi}, {Taylor}, {Williamson}, {Arkharov},
  {Blinov}, {Borman}, {Di Paola}, {Grishina}, {Hagen-Thorn}, {Itoh},
  {Kopatskaya}, {Larionova}, {Larionova}, {Morozova}, {Rastorgueva-Foi},
  {Sergeev}, {Tornikoski}, {Troitsky}, {Thum}, \&
  {Wiesemeyer}}]{2015ApJ...813...51C}
{Casadio}, C., {G{\'o}mez}, J.~L., {Jorstad}, S.~G., {et~al.}
  2015{\natexlab{b}}, \apj, 813, 51, \dodoi{10.1088/0004-637X/813/1/51}

\bibitem[{{Chen} {et~al.}(2011){Chen}, {Fossati}, {Liang}, \&
  {B{\"o}ttcher}}]{Chen11}
{Chen}, X., {Fossati}, G., {Liang}, E.~P., \& {B{\"o}ttcher}, M. 2011, \mnras,
  416, 2368, \dodoi{10.1111/j.1365-2966.2011.19215.x}

\bibitem[{{Danforth} {et~al.}(2013){Danforth}, {Nalewajko}, {France}, \&
  {Keeney}}]{2013ApJ...764...57D}
{Danforth}, C.~W., {Nalewajko}, K., {France}, K., \& {Keeney}, B.~A. 2013,
  \apj, 764, 57, \dodoi{10.1088/0004-637X/764/1/57}

\bibitem[{{Edelson} \& {Krolik}(1988)}]{1988ApJ...333..646E}
{Edelson}, R.~A., \& {Krolik}, J.~H. 1988, \apj, 333, 646,
  \dodoi{10.1086/166773}

\bibitem[{{Emmanoulopoulos} {et~al.}(2013){Emmanoulopoulos}, {McHardy}, \&
  {Papadakis}}]{2013MNRAS.433..907E}
{Emmanoulopoulos}, D., {McHardy}, I.~M., \& {Papadakis}, I.~E. 2013, \mnras,
  433, 907, \dodoi{10.1093/mnras/stt764}

\bibitem[{{Fromm} {et~al.}(2013){Fromm}, {Ros}, {Perucho}, {Savolainen},
  {Mimica}, {Kadler}, {Lobanov}, {Lister}, {Kovalev}, \&
  {Zensus}}]{2013AA...551A..32F}
{Fromm}, C.~M., {Ros}, E., {Perucho}, M., {et~al.} 2013, \aap, 551, A32,
  \dodoi{10.1051/0004-6361/201219913}

\bibitem[{{Fuhrmann} {et~al.}(2014){Fuhrmann}, {Larsson}, {Chiang},
  {Angelakis}, {Zensus}, {Nestoras}, {Krichbaum}, {Ungerechts}, {Sievers},
  {Pavlidou}, {Readhead}, {Max-Moerbeck}, \& {Pearson}}]{2014MNRAS.441.1899F}
{Fuhrmann}, L., {Larsson}, S., {Chiang}, J., {et~al.} 2014, \mnras, 441, 1899,
  \dodoi{10.1093/mnras/stu540}

\bibitem[{{Geng} {et~al.}(2020){Geng}, {Zeng}, {Rani}, {Britto}, {Zhang},
  {Wen}, {Hu}, {Larsson}, {Thompson}, {Yang}, {Cao}, \&
  {Dai}}]{2020ApJ...904...67G}
{Geng}, X., {Zeng}, W., {Rani}, B., {et~al.} 2020, \apj, 904, 67,
  \dodoi{10.3847/1538-4357/abb603}

\bibitem[{{G{\'o}mez} {et~al.}(2011){G{\'o}mez}, {Roca-Sogorb}, {Agudo},
  {Marscher}, \& {Jorstad}}]{2011ApJ...733...11G}
{G{\'o}mez}, J.~L., {Roca-Sogorb}, M., {Agudo}, I., {Marscher}, A.~P., \&
  {Jorstad}, S.~G. 2011, \apj, 733, 11, \dodoi{10.1088/0004-637X/733/1/11}

\bibitem[{{Graff} {et~al.}(2008){Graff}, {Georganopoulos}, {Perlman}, \&
  {Kazanas}}]{Graff08}
{Graff}, P.~B., {Georganopoulos}, M., {Perlman}, E.~S., \& {Kazanas}, D. 2008,
  \apj, 689, 68, \dodoi{10.1086/592427}

\bibitem[{{Gurwell} {et~al.}(2007){Gurwell}, {Peck}, {Hostler}, {Darrah}, \&
  {Katz}}]{2007ASPC..375..234G}
{Gurwell}, M.~A., {Peck}, A.~B., {Hostler}, S.~R., {Darrah}, M.~R., \& {Katz},
  C.~A. 2007, in Astronomical Society of the Pacific Conference Series, Vol.
  375, From Z-Machines to ALMA: (Sub)Millimeter Spectroscopy of Galaxies, ed.
  A.~J. {Baker}, J.~{Glenn}, A.~I. {Harris}, J.~G. {Mangum}, \& M.~S. {Yun},
  234

\bibitem[{{Hovatta} {et~al.}(2008){Hovatta}, {Lehto}, \&
  {Tornikoski}}]{2008A&A...488..897H}
{Hovatta}, T., {Lehto}, H.~J., \& {Tornikoski}, M. 2008, \aap, 488, 897,
  \dodoi{10.1051/0004-6361:200810200}

\bibitem[{{Hovatta} {et~al.}(2007){Hovatta}, {Tornikoski}, {Lainela}, {Lehto},
  {Valtaoja}, {Torniainen}, {Aller}, \& {Aller}}]{2007A&A...469..899H}
{Hovatta}, T., {Tornikoski}, M., {Lainela}, M., {et~al.} 2007, \aap, 469, 899,
  \dodoi{10.1051/0004-6361:20077529}

\bibitem[{{Hovatta} {et~al.}(2009{\natexlab{a}}){Hovatta}, {Valtaoja},
  {Tornikoski}, \& {L{\"a}hteenm{\"a}ki}}]{2009A&A...494..527H}
{Hovatta}, T., {Valtaoja}, E., {Tornikoski}, M., \& {L{\"a}hteenm{\"a}ki}, A.
  2009{\natexlab{a}}, \aap, 494, 527, \dodoi{10.1051/0004-6361:200811150}

\bibitem[{{Hovatta} {et~al.}(2009{\natexlab{b}}){Hovatta}, {Valtaoja},
  {Tornikoski}, \& {L{\"a}hteenm{\"a}ki}}]{2009AA...494..527H}
---. 2009{\natexlab{b}}, \aap, 494, 527, \dodoi{10.1051/0004-6361:200811150}

\bibitem[{{Jorstad} {et~al.}(2001){Jorstad}, {Marscher}, {Mattox}, {Wehrle},
  {Bloom}, \& {Yurchenko}}]{2001ApJS..134..181J}
{Jorstad}, S.~G., {Marscher}, A.~P., {Mattox}, J.~R., {et~al.} 2001, \apjs,
  134, 181, \dodoi{10.1086/320858}

\bibitem[{{Jorstad} {et~al.}(2017){Jorstad}, {Marscher}, {Morozova},
  {Troitsky}, {Agudo}, {Casadio}, {Foord}, {G{\'o}mez}, {MacDonald}, {Molina},
  {L{\"a}hteenm{\"a}ki}, {Tammi}, \& {Tornikoski}}]{2017ApJ...846...98J}
{Jorstad}, S.~G., {Marscher}, A.~P., {Morozova}, D.~A., {et~al.} 2017, \apj,
  846, 98, \dodoi{10.3847/1538-4357/aa8407}

\bibitem[{{Joshi} \& {B{\"o}ttcher}(2011)}]{Joshi11}
{Joshi}, M., \& {B{\"o}ttcher}, M. 2011, \apj, 727, 21,
  \dodoi{10.1088/0004-637X/727/1/21}

\bibitem[{{Kadler} {et~al.}(2008){Kadler}, {Ros}, {Perucho}, {Kovalev},
  {Homan}, {Agudo}, {Kellermann}, {Aller}, {Aller}, {Lister}, \&
  {Zensus}}]{2008ApJ...680..867K}
{Kadler}, M., {Ros}, E., {Perucho}, M., {et~al.} 2008, \apj, 680, 867,
  \dodoi{10.1086/529539}

\bibitem[{{Karamanavis} {et~al.}(2016){Karamanavis}, {Fuhrmann}, {Krichbaum},
  {Angelakis}, {Hodgson}, {Nestoras}, {Myserlis}, {Zensus}, {Sievers}, \&
  {Ciprini}}]{2016A&A...586A..60K}
{Karamanavis}, V., {Fuhrmann}, L., {Krichbaum}, T.~P., {et~al.} 2016, \aap,
  586, A60, \dodoi{10.1051/0004-6361/201527225}

\bibitem[{{Kardashev} {et~al.}(2013){Kardashev}, {Khartov}, {Abramov},
  {Avdeev}, {Alakoz}, {Aleksandrov}, {Ananthakrishnan}, {Andreyanov},
  {Andrianov}, {Antonov}, {Artyukhov}, {Arkhipov}, {Baan}, {Babakin},
  {Babyshkin}, {Bartel'}, {Belousov}, {Belyaev}, {Berulis}, {Burke},
  {Biryukov}, {Bubnov}, {Burgin}, {Busca}, {Bykadorov}, {Bychkova},
  {Vasil'kov}, {Wellington}, {Vinogradov}, {Wietfeldt}, {Voitsik},
  {Gvamichava}, {Girin}, {Gurvits}, {Dagkesamanskii}, {D'Addario},
  {Giovannini}, {Jauncey}, {Dewdney}, {D'yakov}, {Zharov}, {Zhuravlev},
  {Zaslavskii}, {Zakhvatkin}, {Zinov'ev}, {Ilinen}, {Ipatov}, {Kanevskii},
  {Knorin}, {Casse}, {Kellermann}, {Kovalev}, {Kovalev}, {Kovalenko}, {Kogan},
  {Komaev}, {Konovalenko}, {Kopelyanskii}, {Korneev}, {Kostenko}, {Kotik},
  {Kreisman}, {Kukushkin}, {Kulishenko}, {Cooper}, {Kut'kin}, {Cannon},
  {Larionov}, {Lisakov}, {Litvinenko}, {Likhachev}, {Likhacheva}, {Lobanov},
  {Logvinenko}, {Langston}, {McCracken}, {Medvedev}, {Melekhin}, {Menderov},
  {Murphy}, {Mizyakina}, {Mozgovoi}, {Nikolaev}, {Novikov}, {Novikov},
  {Oreshko}, {Pavlenko}, {Pashchenko}, {Ponomarev}, {Popov}, {Pravin-Kumar},
  {Preston}, {Pyshnov}, {Rakhimov}, {Rozhkov}, {Romney}, {Rocha}, {Rudakov},
  {R{\"a}is{\"a}nen}, {Sazankov}, {Sakharov}, {Semenov}, {Serebrennikov},
  {Schilizzi}, {Skulachev}, {Slysh}, {Smirnov}, {Smith}, {Soglasnov},
  {Sokolovskii}, {Sondaar}, {Stepan'yants}, {Turygin}, {Turygin}, {Tuchin},
  {Urpo}, {Fedorchuk}, {Finkel'shtein}, {Fomalont}, {Fejes}, {Fomina},
  {Khapin}, {Tsarevskii}, {Zensus}, {Chuprikov}, {Shatskaya}, {Shapirovskaya},
  {Sheikhet}, {Shirshakov}, {Schmidt}, {Shnyreva}, {Shpilevskii}, {Ekers}, \&
  {Yakimov}}]{2013ARep...57..153K}
{Kardashev}, N.~S., {Khartov}, V.~V., {Abramov}, V.~V., {et~al.} 2013,
  Astronomy Reports, 57, 153, \dodoi{10.1134/S1063772913030025}

\bibitem[{{Kim} {et~al.}(2020){Kim}, {Trippe}, \&
  {Kravchenko}}]{2020A&A...636A..62K}
{Kim}, D.-W., {Trippe}, S., \& {Kravchenko}, E.~V. 2020, \aap, 636, A62,
  \dodoi{10.1051/0004-6361/202037474}

\bibitem[{{Kim} {et~al.}(2018){Kim}, {Trippe}, {Lee}, {Kim}, {Algaba},
  {Hodgson}, {Park}, {Kino}, {Zhao}, {Wajima}, {Lee}, \&
  {Kang}}]{2018MNRAS.480.2324K}
{Kim}, D.-W., {Trippe}, S., {Lee}, S.-S., {et~al.} 2018, \mnras, 480, 2324,
  \dodoi{10.1093/mnras/sty1993}

\bibitem[{{Koay} {et~al.}(2019){Koay}, {Jauncey}, {Hovatta}, {Kiehlmann},
  {Bignall}, {Max-Moerbeck}, {Pearson}, {Readhead}, {Reeves}, {Reynolds}, \&
  {Vedantham}}]{2019MNRAS.489.5365K}
{Koay}, J.~Y., {Jauncey}, D.~L., {Hovatta}, T., {et~al.} 2019, \mnras, 489,
  5365, \dodoi{10.1093/mnras/stz2488}

\bibitem[{{Kovalev} {et~al.}(2005){Kovalev}, {Kellermann}, {Lister}, {Homan},
  {Vermeulen}, {Cohen}, {Ros}, {Kadler}, {Lobanov}, {Zensus}, {Kardashev},
  {Gurvits}, {Aller}, \& {Aller}}]{2005AJ....130.2473K}
{Kovalev}, Y.~Y., {Kellermann}, K.~I., {Lister}, M.~L., {et~al.} 2005, \aj,
  130, 2473, \dodoi{10.1086/497430}

\bibitem[{{Kovalev} {et~al.}(2009){Kovalev}, {Aller}, {Aller}, {Homan},
  {Kadler}, {Kellermann}, {Kovalev}, {Lister}, {McCormick}, {Pushkarev}, {Ros},
  \& {Zensus}}]{2009ApJ...696L..17K}
{Kovalev}, Y.~Y., {Aller}, H.~D., {Aller}, M.~F., {et~al.} 2009, \apjl, 696,
  L17, \dodoi{10.1088/0004-637X/696/1/L17}

\bibitem[{{Kravchenko} {et~al.}(2016){Kravchenko}, {Kovalev}, {Hovatta}, \&
  {Ramakrishnan}}]{2016MNRAS.462.2747K}
{Kravchenko}, E.~V., {Kovalev}, Y.~Y., {Hovatta}, T., \& {Ramakrishnan}, V.
  2016, \mnras, 462, 2747, \dodoi{10.1093/mnras/stw1776}

\bibitem[{{Kravchenko} {et~al.}(2020){Kravchenko}, {G{\'o}mez}, {Kovalev},
  {Lobanov}, {Savolainen}, {Bruni}, {Fuentes}, {Anderson}, {Jorstad},
  {Marscher}, {Tornikoski}, {L{\"a}hteenm{\"a}ki}, \&
  {Lisakov}}]{2020ApJ...893...68K}
{Kravchenko}, E.~V., {G{\'o}mez}, J.~L., {Kovalev}, Y.~Y., {et~al.} 2020, \apj,
  893, 68, \dodoi{10.3847/1538-4357/ab7dae}

\bibitem[{{Larionov} {et~al.}(2013){Larionov}, {Jorstad}, {Marscher},
  {Morozova}, {Blinov}, {Hagen-Thorn}, {Konstantinova}, {Kopatskaya},
  {Larionova}, {Larionova}, \& {Troitsky}}]{2013ApJ...768...40L}
{Larionov}, V.~M., {Jorstad}, S.~G., {Marscher}, A.~P., {et~al.} 2013, \apj,
  768, 40, \dodoi{10.1088/0004-637X/768/1/40}

\bibitem[{{Lee} {et~al.}(2017){Lee}, {Lee}, {Hodgson}, {Kim}, {Algaba}, {Kang},
  {Kang}, \& {Kim}}]{2017ApJ...841..119L}
{Lee}, J.~W., {Lee}, S.-S., {Hodgson}, J.~A., {et~al.} 2017, \apj, 841, 119,
  \dodoi{10.3847/1538-4357/aa72f7}

\bibitem[{{Le{\'o}n-Tavares} {et~al.}(2011){Le{\'o}n-Tavares}, {Valtaoja},
  {Tornikoski}, {L{\"a}hteenm{\"a}ki}, \& {Nieppola}}]{2011A&A...532A.146L}
{Le{\'o}n-Tavares}, J., {Valtaoja}, E., {Tornikoski}, M.,
  {L{\"a}hteenm{\"a}ki}, A., \& {Nieppola}, E. 2011, \aap, 532, A146,
  \dodoi{10.1051/0004-6361/201116664}

\bibitem[{{Li} \& {Kusunose}(2000)}]{2000ApJ...536..729L}
{Li}, H., \& {Kusunose}, M. 2000, \apj, 536, 729, \dodoi{10.1086/308960}

\bibitem[{{Liao} {et~al.}(2014){Liao}, {Bai}, {Liu}, {Weng}, {Chen}, \&
  {Li}}]{2014ApJ...783...83L}
{Liao}, N.~H., {Bai}, J.~M., {Liu}, H.~T., {et~al.} 2014, \apj, 783, 83,
  \dodoi{10.1088/0004-637X/783/2/83}

\bibitem[{{Liodakis}(2018)}]{2018A&A...616A..93L}
{Liodakis}, I. 2018, \aap, 616, A93, \dodoi{10.1051/0004-6361/201832766}

\bibitem[{{Liodakis} {et~al.}(2018{\natexlab{a}}){Liodakis}, {Hovatta},
  {Huppenkothen}, {Kiehlmann}, {Max-Moerbeck}, \&
  {Readhead}}]{2018ApJ...866..137L}
{Liodakis}, I., {Hovatta}, T., {Huppenkothen}, D., {et~al.} 2018{\natexlab{a}},
  \apj, 866, 137, \dodoi{10.3847/1538-4357/aae2b7}

\bibitem[{{Liodakis} {et~al.}(2018{\natexlab{b}}){Liodakis}, {Romani},
  {Filippenko}, {Kiehlmann}, {Max-Moerbeck}, {Readhead}, \&
  {Zheng}}]{2018MNRAS.480.5517L}
{Liodakis}, I., {Romani}, R.~W., {Filippenko}, A.~V., {et~al.}
  2018{\natexlab{b}}, \mnras, 480, 5517, \dodoi{10.1093/mnras/sty2264}

\bibitem[{{Lisakov} {et~al.}(2017){Lisakov}, {Kovalev}, {Savolainen},
  {Hovatta}, \& {Kutkin}}]{2017MNRAS.468.4478L}
{Lisakov}, M.~M., {Kovalev}, Y.~Y., {Savolainen}, T., {Hovatta}, T., \&
  {Kutkin}, A.~M. 2017, \mnras, 468, 4478, \dodoi{10.1093/mnras/stx710}

\bibitem[{{Lobanov} \& {Zensus}(1999)}]{1999ApJ...521..509L}
{Lobanov}, A.~P., \& {Zensus}, J.~A. 1999, \apj, 521, 509,
  \dodoi{10.1086/307555}

\bibitem[{{MAGIC Collaboration} {et~al.}(2018){MAGIC Collaboration}, {Ahnen},
  {Ansoldi}, {Antonelli}, {Arcaro}, {Baack}, {Babi{\'c}}, {Banerjee},
  {Bangale}, {Barres de Almeida}, {Barrio}, {Becerra Gonz{\'a}lez}, {Bednarek},
  {Bernardini}, {Ch Berse}, {Berti}, {Bhattacharyya}, {Biland }, {Blanch},
  {Bonnoli}, {Carosi}, {Carosi}, {Ceribella}, {Chatterjee}, {Colak}, {Colin},
  {Colombo}, {Contreras}, {Cortina}, {Covino}, {Cumani}, {da Vela}, {Dazzi},
  {de Angelis}, {de Lotto}, {Delfino}, {Delgado}, {di Pierro},
  {Dom{\'\i}nguez}, {Dominis Prester}, {Dorner}, {Doro}, {Einecke},
  {Elsaesser}, {Fallah Ramazani}, {Fern{\'a}ndez-Barral}, {Fidalgo}, {Fonseca},
  {Font}, {Fruck}, {Galindo}, {Gallozzi}, {Garc{\'\i}a L{\'o}pez},
  {Garczarczyk}, {Gaug}, {Giammaria}, {Godinovi{\'c}}, {Gora}, {Guberman},
  {Hadasch}, {Hahn}, {Hassan}, {Hayashida}, {Herrera}, {Hose}, {Hrupec},
  {Ishio}, {Konno}, {Kubo}, {Kushida}, {Kuve{\v{z}}di{\'c}}, {Lelas},
  {Lindfors}, {Lombardi}, {Longo}, {L{\'o}pez}, {Maggio}, {Majumdar},
  {Makariev}, {Maneva}, {Manganaro}, {Mannheim}, {Maraschi}, {Mariotti},
  {Mart{\'\i}nez}, {Masuda}, {Mazin}, {Mielke}, {Minev}, {Miranda}, {Mirzoyan},
  {Moralejo}, {Moreno}, {Moretti}, {Nagayoshi}, {Neustroev}, {Niedzwiecki},
  {Nievas Rosillo}, {Nigro}, {Nilsson}, {Ninci}, {Nishijima}, {Noda},
  {Nogu{\'e}s}, {Paiano}, {Palacio}, {Paneque}, {Paoletti}, {Paredes},
  {Pedaletti}, {Peresano}, {Persic}, {Prada Moroni}, {Prand ini}, {Puljak},
  {Garcia}, {Reichardt}, {Rhode}, {Rib{\'o}}, {Rico}, {Righi}, {Rugliancich},
  {Saito}, {Satalecka}, {Schweizer}, {Sitarek}, {{\v{S}}nidari{\'c}},
  {Sobczynska}, {Stamerra}, {Strzys}, {Suri{\'c}}, {Takahashi}, {Takalo},
  {Tavecchio}, {Temnikov}, {Terzi{\'c}}, {Teshima}, {Torres-Alb{\`a}},
  {Treves}, {Tsujimoto}, {Vanzo}, {Vazquez Acosta}, {Vovk}, {Ward}, {Will},
  {Zari{\'c}}, {Fermi-Lat Collaboration}, {Bastieri}, {Gasparrini}, {Lott},
  {Rani}, {Thompson}, {MWL Collaborators}, {Agudo}, {Angelakis}, {Borman},
  {Casadio}, {Grishina}, {Gurwell}, {Hovatta}, {Itoh}, {J{\"a}rvel{\"a}},
  {Jermak}, {Jorstad}, {Kopatskaya}, {Kraus}, {Krichbaum}, {Kuin},
  {L{\"a}hteenm{\"a}ki}, {Larionov}, {Larionova}, {Lien}, {Madejski},
  {Marscher}, {Myserlis}, {Max-Moerbeck}, {Molina}, {Morozova}, {Nalewajko},
  {Pearson}, {Ramakrishnan}, {Readhead}, {Reeves}, {Savchenko}, {Steele},
  {Tornikoski}, {Troitskaya}, {Troitsky}, {Vasilyev}, \&
  {Zensus}}]{2018A&A...619A..45M}
{MAGIC Collaboration}, {Ahnen}, M.~L., {Ansoldi}, S., {et~al.} 2018, \aap, 619,
  A45, \dodoi{10.1051/0004-6361/201832677}

\bibitem[{{Mannheim}(1998)}]{1998Sci...279..684M}
{Mannheim}, K. 1998, Science, 279, 684, \dodoi{10.1126/science.279.5351.684}

\bibitem[{{Marscher} {et~al.}(1992){Marscher}, {Gear}, \&
  {Travis}}]{1992vob..conf...85M}
{Marscher}, A.~P., {Gear}, W.~K., \& {Travis}, J.~P. 1992, in Variability of
  Blazars, ed. E.~{Valtaoja} \& M.~{Valtonen}, 85

\bibitem[{{Mastichiadis} \& {Kirk}(1997)}]{1997A&A...320...19M}
{Mastichiadis}, A., \& {Kirk}, J.~G. 1997, \aap, 320, 19.
\newblock \doarXiv{astro-ph/9610058}

\bibitem[{{Mattox} {et~al.}(1996){Mattox}, {Bertsch}, {Chiang}, {Dingus},
  {Digel}, {Esposito}, {Fierro}, {Hartman}, {Hunter}, {Kanbach}, {Kniffen},
  {Lin}, {Macomb}, {Mayer-Hasselwander}, {Michelson}, {von Montigny},
  {Mukherjee}, {Nolan}, {Ramanamurthy}, {Schneid}, {Sreekumar}, {Thompson}, \&
  {Willis}}]{1996ApJ...461..396M}
{Mattox}, J.~R., {Bertsch}, D.~L., {Chiang}, J., {et~al.} 1996, \apj, 461, 396,
  \dodoi{10.1086/177068}

\bibitem[{{Max-Moerbeck} {et~al.}(2014{\natexlab{a}}){Max-Moerbeck},
  {Richards}, {Hovatta}, {Pavlidou}, {Pearson}, \&
  {Readhead}}]{2014MNRAS.445..437M}
{Max-Moerbeck}, W., {Richards}, J.~L., {Hovatta}, T., {et~al.}
  2014{\natexlab{a}}, \mnras, 445, 437, \dodoi{10.1093/mnras/stu1707}

\bibitem[{{Max-Moerbeck} {et~al.}(2014{\natexlab{b}}){Max-Moerbeck}, {Hovatta},
  {Richards}, {King}, {Pearson}, {Readhead}, {Reeves}, {Shepherd}, {Stevenson},
  {Angelakis}, {Fuhrmann}, {Grainge}, {Pavlidou}, {Romani}, \&
  {Zensus}}]{2014MNRAS.445..428M}
{Max-Moerbeck}, W., {Hovatta}, T., {Richards}, J.~L., {et~al.}
  2014{\natexlab{b}}, \mnras, 445, 428, \dodoi{10.1093/mnras/stu1749}

\bibitem[{{Nilsson} {et~al.}(2008){Nilsson}, {Pursimo}, {Sillanp{\"a}{\"a}},
  {Takalo}, \& {Lindfors}}]{2008A&A...487L..29N}
{Nilsson}, K., {Pursimo}, T., {Sillanp{\"a}{\"a}}, A., {Takalo}, L.~O., \&
  {Lindfors}, E. 2008, \aap, 487, L29, \dodoi{10.1051/0004-6361:200810310}

\bibitem[{{Planck Collaboration} {et~al.}(2016){Planck Collaboration}, {Ade},
  {Aghanim}, {Arnaud}, {Ashdown}, {Aumont}, {Baccigalupi}, {Banday},
  {Barreiro}, {Bartlett}, \& et~al.}]{2016AA...594A..13P}
{Planck Collaboration}, {Ade}, P.~A.~R., {Aghanim}, N., {et~al.} 2016, \aap,
  594, A13, \dodoi{10.1051/0004-6361/201525830}

\bibitem[{{Pushkarev} {et~al.}(2019){Pushkarev}, {Butuzova}, {Kovalev}, \&
  {Hovatta}}]{2019MNRAS.482.2336P}
{Pushkarev}, A.~B., {Butuzova}, M.~S., {Kovalev}, Y.~Y., \& {Hovatta}, T. 2019,
  \mnras, 482, 2336, \dodoi{10.1093/mnras/sty2724}

\bibitem[{{Pushkarev} {et~al.}(2012){Pushkarev}, {Hovatta}, {Kovalev},
  {Lister}, {Lobanov}, {Savolainen}, \& {Zensus}}]{2012A&A...545A.113P}
{Pushkarev}, A.~B., {Hovatta}, T., {Kovalev}, Y.~Y., {et~al.} 2012, \aap, 545,
  A113, \dodoi{10.1051/0004-6361/201219173}

\bibitem[{{Pushkarev} {et~al.}(2010){Pushkarev}, {Kovalev}, \&
  {Lister}}]{2010ApJ...722L...7P}
{Pushkarev}, A.~B., {Kovalev}, Y.~Y., \& {Lister}, M.~L. 2010, \apjl, 722, L7,
  \dodoi{10.1088/2041-8205/722/1/L7}

\bibitem[{{Pushkarev} {et~al.}(2017){Pushkarev}, {Kovalev}, {Lister}, \&
  {Savolainen}}]{2017MNRAS.468.4992P}
{Pushkarev}, A.~B., {Kovalev}, Y.~Y., {Lister}, M.~L., \& {Savolainen}, T.
  2017, \mnras, 468, 4992, \dodoi{10.1093/mnras/stx854}

\bibitem[{{Ramakrishnan} {et~al.}(2015){Ramakrishnan}, {Hovatta}, {Nieppola},
  {Tornikoski}, {L{\"a}hteenm{\"a}ki}, \& {Valtaoja}}]{2015MNRAS.452.1280R}
{Ramakrishnan}, V., {Hovatta}, T., {Nieppola}, E., {et~al.} 2015, \mnras, 452,
  1280, \dodoi{10.1093/mnras/stv321}

\bibitem[{{Ramakrishnan} {et~al.}(2014){Ramakrishnan}, {Le{\'o}n-Tavares},
  {Rastorgueva-Foi}, {Wiik}, {Jorstad}, {Marscher}, {Tornikoski}, {Agudo},
  {L{\"a}hteenm{\"a}ki}, {Valtaoja}, {Aller}, {Blinov}, {Casadio}, {Efimova},
  {Gurwell}, {G{\'o}mez}, {Hagen-Thorn}, {Joshi}, {J{\"a}rvel{\"a}},
  {Konstantinova}, {Kopatskaya}, {Larionov}, {Larionova}, {Larionova},
  {Lavonen}, {MacDonald}, {McHardy}, {Molina}, {Morozova}, {Nieppola}, {Tammi},
  {Taylor}, \& {Troitsky}}]{2014MNRAS.445.1636R}
{Ramakrishnan}, V., {Le{\'o}n-Tavares}, J., {Rastorgueva-Foi}, E.~A., {et~al.}
  2014, \mnras, 445, 1636, \dodoi{10.1093/mnras/stu1873}

\bibitem[{{Ramakrishnan} {et~al.}(2016){Ramakrishnan}, {Hovatta}, {Tornikoski},
  {Nilsson}, {Lindfors}, {Balokovi{\'c}}, {L{\"a}hteenm{\"a}ki}, {Reinthal}, \&
  {Takalo}}]{2016MNRAS.456..171R}
{Ramakrishnan}, V., {Hovatta}, T., {Tornikoski}, M., {et~al.} 2016, \mnras,
  456, 171, \dodoi{10.1093/mnras/stv2653}

\bibitem[{{Rani} {et~al.}(2015){Rani}, {Krichbaum}, {Marscher}, {Hodgson},
  {Fuhrmann}, {Angelakis}, {Britzen}, \& {Zensus}}]{2015A&A...578A.123R}
{Rani}, B., {Krichbaum}, T.~P., {Marscher}, A.~P., {et~al.} 2015, \aap, 578,
  A123, \dodoi{10.1051/0004-6361/201525608}

\bibitem[{{Rani} {et~al.}(2014){Rani}, {Krichbaum}, {Marscher}, {Jorstad},
  {Hodgson}, {Fuhrmann}, \& {Zensus}}]{2014A&A...571L...2R}
---. 2014, \aap, 571, L2, \dodoi{10.1051/0004-6361/201424796}

\bibitem[{{Rani} {et~al.}(2013){Rani}, {Krichbaum}, {Fuhrmann}, {B{\"o}ttcher},
  {Lott}, {Aller}, {Aller}, {Angelakis}, {Bach}, {Bastieri}, {Falcone},
  {Fukazawa}, {Gabanyi}, {Gupta}, {Gurwell}, {Itoh}, {Kawabata}, {Krips},
  {L{\"a}hteenm{\"a}ki}, {Liu}, {Marchili}, {Max-Moerbeck}, {Nestoras},
  {Nieppola}, {Quintana-Lacaci}, {Readhead}, {Richards}, {Sasada}, {Sievers},
  {Sokolovsky}, {Stroh}, {Tammi}, {Tornikoski}, {Uemura}, {Ungerechts},
  {Urano}, \& {Zensus}}]{2013A&A...552A..11R}
{Rani}, B., {Krichbaum}, T.~P., {Fuhrmann}, L., {et~al.} 2013, \aap, 552, A11,
  \dodoi{10.1051/0004-6361/201321058}

\bibitem[{{Richards} {et~al.}(2011){Richards}, {Max-Moerbeck}, {Pavlidou},
  {King}, {Pearson}, {Readhead}, {Reeves}, {Shepherd}, {Stevenson},
  {Weintraub}, {Fuhrmann}, {Angelakis}, {Zensus}, {Healey}, {Romani}, {Shaw},
  {Grainge}, {Birkinshaw}, {Lancaster}, {Worrall}, {Taylor}, {Cotter}, \&
  {Bustos}}]{richards_etal11}
{Richards}, J.~L., {Max-Moerbeck}, W., {Pavlidou}, V., {et~al.} 2011, \apjs,
  194, 29, \dodoi{10.1088/0067-0049/194/2/29}

\bibitem[{{Robertson} {et~al.}(2015){Robertson}, {Gallo}, {Zoghbi}, \&
  {Fabian}}]{2015MNRAS.453.3455R}
{Robertson}, D.~R.~S., {Gallo}, L.~C., {Zoghbi}, A., \& {Fabian}, A.~C. 2015,
  \mnras, 453, 3455, \dodoi{10.1093/mnras/stv1575}

\bibitem[{{Schinzel} {et~al.}(2012){Schinzel}, {Lobanov}, {Taylor}, {Jorstad},
  {Marscher}, \& {Zensus}}]{2012A&A...537A..70S}
{Schinzel}, F.~K., {Lobanov}, A.~P., {Taylor}, G.~B., {et~al.} 2012, \aap, 537,
  A70, \dodoi{10.1051/0004-6361/201117705}

\bibitem[{{Shepherd}(1997)}]{1997ASPC..125...77S}
{Shepherd}, M.~C. 1997, in Astronomical Society of the Pacific Conference
  Series, Vol. 125, Astronomical Data Analysis Software and Systems VI, ed.
  G.~{Hunt} \& H.~{Payne}, 77

\bibitem[{{Sikora} {et~al.}(1994){Sikora}, {Begelman}, \&
  {Rees}}]{1994ApJ...421..153S}
{Sikora}, M., {Begelman}, M.~C., \& {Rees}, M.~J. 1994, \apj, 421, 153,
  \dodoi{10.1086/173633}

\bibitem[{{Sokolov} \& {Marscher}(2005)}]{Sokolov05}
{Sokolov}, A., \& {Marscher}, A.~P. 2005, \apj, 629, 52, \dodoi{10.1086/431321}

\bibitem[{{Sokolov} {et~al.}(2004){Sokolov}, {Marscher}, \&
  {McHardy}}]{Sokolov04}
{Sokolov}, A., {Marscher}, A.~P., \& {McHardy}, I.~M. 2004, \apj, 613, 725,
  \dodoi{10.1086/423165}

\bibitem[{{Spada} {et~al.}(2001){Spada}, {Ghisellini}, {Lazzati}, \&
  {Celotti}}]{Spada01}
{Spada}, M., {Ghisellini}, G., {Lazzati}, D., \& {Celotti}, A. 2001, \mnras,
  325, 1559, \dodoi{10.1046/j.1365-8711.2001.04557.x}

\bibitem[{{Ter{\"a}sranta} {et~al.}(1998){Ter{\"a}sranta}, {Tornikoski},
  {Mujunen}, {Karlamaa}, {Valtonen}, {Henelius}, {Urpo}, {Lainela}, {Pursimo},
  {Nilsson}, {Wiren}, {Laehteenmaeki}, {Korpi}, {Rekola}, {Heinaemaeki},
  {Hanski}, {Nurmi}, {Kokkonen}, {Keinaenen}, {Joutsamo}, {Oksanen},
  {Pietilae}, {Valtaoja}, {Valtonen}, \& {Koenoenen}}]{1998AAS..132..305T}
{Ter{\"a}sranta}, H., {Tornikoski}, M., {Mujunen}, A., {et~al.} 1998, \aaps,
  132, 305, \dodoi{10.1051/aas:1998297}

\bibitem[{{Trippe} {et~al.}(2011){Trippe}, {Krips}, {Pi{\'e}tu}, {Neri},
  {Winters}, {Gueth}, {Bremer}, {Salome}, {Moreno}, {Boissier}, \&
  {Fontani}}]{2011A&A...533A..97T}
{Trippe}, S., {Krips}, M., {Pi{\'e}tu}, V., {et~al.} 2011, \aap, 533, A97,
  \dodoi{10.1051/0004-6361/201015558}

\bibitem[{{Uttley} {et~al.}(2002){Uttley}, {McHardy}, \&
  {Papadakis}}]{2002MNRAS.332..231U}
{Uttley}, P., {McHardy}, I.~M., \& {Papadakis}, I.~E. 2002, \mnras, 332, 231,
  \dodoi{10.1046/j.1365-8711.2002.05298.x}

\bibitem[{{Vittorini} {et~al.}(2009){Vittorini}, {Tavani}, {Paggi},
  {Cavaliere}, {Bulgarelli}, {Chen}, {D'Ammando}, {Donnarumma}, {Giuliani},
  {Longo}, {Pacciani}, {Pucella}, {Vercellone}, {Ferrari}, {Colafrancesco}, \&
  {Giommi}}]{2009ApJ...706.1433V}
{Vittorini}, V., {Tavani}, M., {Paggi}, A., {et~al.} 2009, \apj, 706, 1433,
  \dodoi{10.1088/0004-637X/706/2/1433}

\bibitem[{{Wierzcholska} \& {Siejkowski}(2015)}]{2015MNRAS.452L..11W}
{Wierzcholska}, A., \& {Siejkowski}, H. 2015, \mnras, 452, L11,
  \dodoi{10.1093/mnrasl/slv075}

\bibitem[{{Williamson} {et~al.}(2014){Williamson}, {Jorstad}, {Marscher},
  {Larionov}, {Smith}, {Agudo}, {Arkharov}, {Blinov}, {Casadio}, {Efimova},
  {G{\'o}mez}, {Hagen-Thorn}, {Joshi}, {Konstantinova}, {Kopatskaya},
  {Larionova}, {Larionova}, {Malmrose}, {McHardy}, {Molina}, {Morozova},
  {Schmidt}, {Taylor}, \& {Troitsky}}]{2014ApJ...789..135W}
{Williamson}, K.~E., {Jorstad}, S.~G., {Marscher}, A.~P., {et~al.} 2014, \apj,
  789, 135, \dodoi{10.1088/0004-637X/789/2/135}

\bibitem[{{Zhang} {et~al.}(2015){Zhang}, {Chen}, {B{\"o}ttcher}, {Guo}, \&
  {Li}}]{Zhang15}
{Zhang}, H., {Chen}, X., {B{\"o}ttcher}, M., {Guo}, F., \& {Li}, H. 2015, \apj,
  804, 58, \dodoi{10.1088/0004-637X/804/1/58}

\end{thebibliography}
\bibliographystyle{aasjournal}



\end{document}